\documentclass[longauth]{aa}
\usepackage{graphicx}
\usepackage[varg]{txfonts}
\usepackage{multirow}
\usepackage{array}
\usepackage{silence}
\usepackage{etoolbox} 
\makeatletter
\patchcmd{\AA@hyperref}
  {\AddToHook{begindocument}{showkeys}{before}{nameref}}
  {}{}{}
\makeatother
\usepackage{hyperref} 
\usepackage{ulem} 
\usepackage{mathtools}
\usepackage{amsmath}
\usepackage{booktabs}
\usepackage{longtable}
\usepackage{float}
\usepackage{tabularx}
\usepackage{graphicx}
\usepackage{threeparttable}

\newcommand{\rsun}{\mbox{R$_{\odot}$\,}}

\begin{document}

   \title{SN 2016iog: A fast declining Type II-L supernova with an ultra-faint tail, persistently interacting with circumstellar material}

   \subtitle{}

   \authorrunning{Z.-H. Peng et al.} 
   \titlerunning{SN 2016iog}
   
   \author{
    Z.-H. Peng \orcid{0009-0000-7773-553X}\inst{\ref{inst1}} \and 
    S. Benetti \orcid{0000-0002-3256-0016}\inst{\ref{inst2}} \and 
    Y.-Z. Cai \orcid{0000-0002-7714-493X}\inst{\ref{inst3}, \ref{inst4}}\thanks{Corresponding authors: caiyongzhi@ynao.ac.cn (CYZ)} \and 
    A. Pastorello \orcid{0000-0002-7259-4624}\inst{\ref{inst2}} \and 
    G. Valerin \orcid{0000-0002-3334-4585}\inst{\ref{inst2}} \and 
    A. Reguitti \orcid{0000-0003-4254-2724}\inst{\ref{inst2},\ref{inst20}} \and 
    A. Fiore \orcid{0000-0002-0403-3331}\inst{\ref{inst6},\ref{inst2}} \and 
    Q.-L. Fang \orcid{0000-0002-1161-9592}\inst{\ref{inst8}} \and 
    Z.-Y. Wang \orcid{0000-0002-0025-0179}\inst{\ref{inst9}, \ref{inst10}} \and 
    M. Berton \orcid{0000-0002-1058-9109}\inst{\ref{inst11}}   \and 
    L. Borsato \orcid{0000-0003-0066-9268}\inst{\ref{inst2}}   \and 
    E. Cappellaro \orcid{0000-0001-5008-8619}\inst{\ref{inst2}} \and 
    E. Congiu \orcid{0000-0002-8549-4083}\inst{\ref{inst11}} \and 
    N. Elias-Rosa \orcid{0000-0002-1381-9125}\inst{\ref{inst2},\ref{inst12}}  \and 
    V. Granata \orcid{0000-0002-1425-4541}\inst{\ref{inst13},\ref{inst2}} \and 
    J. Isern  \orcid{0000-0002-0819-9574}\inst{\ref{inst23},\ref{inst12}, \ref{inst14}} \and 
    G. La Mura  \orcid{0000-0001-8553-499X}\inst{\ref{inst21}, \ref{inst22}}    \and 
    P. Ochner \orcid{0000-0001-5578-8614}\inst{\ref{inst2},\ref{inst15}}  \and 
    R. Raddi  \orcid{0000-0002-9090-9191}\inst{\ref{inst16}}  \and 
    G. Terreran \orcid{0000-0003-0794-5982}\inst{\ref{inst17}}   \and 
    L. Tomasella  \orcid{0000-0002-3697-2616}\inst{\ref{inst2}}   \and 
    M. Turatto \orcid{0000-0002-9719-3157}\inst{\ref{inst2}}  \and 
    S.-Y. Yan \orcid{0009-0004-4256-1209}\inst{\ref{inst18}} \and 
    S.-P. Pei \orcid{0000-0002-0851-8045}\inst{\ref{inst19}} \and 
    C.-Y. Wu \orcid{0000-0002-2452-551X}\inst{\ref{inst3}, \ref{inst4}}   \and 
    S. Zha  \orcid{0000-0001-6773-7830}\inst{\ref{inst3}, \ref{inst4}}   \and 
    X.-F. Wang   \orcid{0000-0002-7334-2357}\inst{\ref{inst18},\ref{inst62}}   \and  
    B. Wang \orcid{0000-0002-3231-1167}\inst{\ref{inst3}, \ref{inst4}}\thanks{wangbo@ynao.ac.cn (WB)}  \and  
    Y. Pan \orcid{0000-0001-7261-8297} \inst{\ref{inst1}\thanks{panyu@cqupt.edu.cn (PY)}} 
}

\institute{
\label{inst1}School of Electronic Science and Engineering, Chongqing University of Posts and Telecommunications, Chongqing 400065, P.R. China \and
\label{inst2}INAF - Osservatorio Astronomico di Padova, Vicolo dell'Osservatorio 5, 35122 Padova, Italy \and
\label{inst3}Yunnan Observatories, Chinese Academy of Sciences, Kunming 650216, P.R. China \and
\label{inst4}International Centre of Supernovae, Yunnan Key Laboratory, Kunming 650216, P.R. China \and
\label{inst20}INAF - Osservatorio Astronomico di Brera, Via E. Bianchi 46, 23807 Merate (LC), Italy \and
\label{inst6}INAF - Osservatorio Astronomico d'Abruzzo, Via Mentore Maggini, I-64100 Teramo, Italy \and
\label{inst8}National Astronomical Observatory of Japan, National Institutes of Natural Sciences, 2-21-1 Osawa, Mitaka, Tokyo 181-8588, Japan \and
\label{inst9}School of Astronomy and Space Science, University of Chinese Academy of Sciences, Beijing 100049, P.R. China \and
\label{inst10}National Astronomical Observatories, Chinese Academy of Sciences, Beijing 100101, P.R. China \and
\label{inst11}European Southern Observatory (ESO), Alonso de C\'{o}rdova 3107, Vitacura Santiago, Chile\and
\label{inst12}Institute of Space Sciences (ICE, CSIC), Campus UAB, Carrer de Can Magrans, s/n, E-08193 Barcelona, Spain \and
\label{inst13}CISAS - Centro di Ateneo di Studi e Attivit\`{a} Spaziali ''Giuseppe Colombo'', Universit\`{a} degli Studi di Padova, Via Venezia 1, 35131 Padova, Italy \and
\label{inst23}Fabra Observatory, Royal Academy of Sciences and Arts of Barcelona (RACAB), 08001 Barcelona, Spain \and
\label{inst14}Institute for Space Studies of Catalonia (IEEC), Campus UPC, 08860 Castelldefels (Barcelona), Spain \and
\label{inst21}INAF - Osservatorio Astronomico di Cagliari, Via della Scienza 5, 09047, Selargius (CA), Italy \and
\label{inst22}Laboratory of Instrumentation and Experimental Particle Physics, Av. Prof. Gama Pinto, 2 - 1649-003, Lisboa, Portugal \and
\label{inst15}Department of Physics and Astronomy, University of Padova, Via F. Marzolo 8, I-35131, Padova, Italy \and
\label{inst16}Universitat Polit\`{e}cnica de Catalunya, Departament de F\'{i}sica, c/ Esteve Terrades 5, 08860 Castelldefels, Spain \and
\label{inst17}Adler Planetarium, 1300 S. DuSable Lake Shore Drive, Chicago, IL 60605, USA \and
\label{inst18}Physics Department, Tsinghua University, Beijing 100084, P.R. China \and
\label{inst19}School of Physics and Electrical Engineering, Liupanshui Normal University, Liupanshui, Guizhou, 553004, P.R. China \and
\label{inst62}Purple Mountain Observatory, Chinese Academy of Sciences, Nanjing, 210023, P.R. China
}

    \date{Received ; accepted }
 
  \abstract
  {We present optical photometric and spectroscopic observations of the rapidly declining Type IIL supernova (SN) 2016iog. SN 2016iog reached its peak  $\sim$ 14 days  after explosion, with an absolute magnitude in the $V$ band of $-18.64 \pm 0.15$ mag, followed by a steep decline of $8.85 \pm 0.15$~mag~(100\,d)$^{-1}$ post-peak. Such a high decline rate makes SN~2016iog one of the fastest declining Type~IIL SNe observed to date. 
  The rapid rise in the light curve, combined with the nearly featureless continuum observed in the spectrum at +9.3 days, suggests the presence of interaction.
  In the recombination phase, we observed broad H$\alpha$ lines that persist at all epochs.
  In addition, the prominent double-peaked H$\alpha$ feature observed in the late-time spectrum (+190.8 days) is likely attributable either to significant dust formation within a cool dense shell or to asymmetric circumstellar material.
  These features suggest the presence of sustained interaction around SN~2016iog.
  We propose that the observed characteristics of SN~2016iog can be qualitatively explained by assuming a low-mass H-rich envelope surrounding a red supergiant progenitor star with low-density circumstellar material.}


   \maketitle

\nolinenumbers   
\section{Introduction}
The classification of supernovae (SNe) primarily relies on their observed spectroscopic characteristics, with the presence or absence of hydrogen in their spectra serving as the primary distinguishing factor. This distinction separates H-poor from H-rich SNe. The H-rich Type II SNe category is photometrically heterogeneous, with light curves (LCs) displaying a continuous range of decline rates after the peak.
Type II SNe with LCs exhibiting a `plateau phase' are typically classified as Type IIP SNe, while those with LCs showing a linear decline (in magnitudes) are classified as Type IIL SNe \citep{Levesque1979A&A....72..287B, Anderson2014, Valenti2016MNRAS.459.3939V, Patat1994A&A...282..731P}. 
The different decline rates observed in the LCs is believed to result from progenitors with progressively smaller residual hydrogen envelope masses before the explosion \citep{Georgy2012A&A...538L...8G}.
More recently, several authors have suggested that Type II SNe should be regarded as an heterogeneous class, with their LCs covering a continuous range of properties \citep{Anderson2014bMNRAS.441..671A, Sanders2015ApJ...799..208S, Galbany2016AJ....151...33G, Rubin2016ApJ...828..111R, Valenti2016MNRAS.459.3939V, deJaeger2019MNRAS.490.2799D}.

Different evolutionary scenarios for the progenitors of these Type II SNe have been proposed. \cite{Smartt2009MNRAS.395.1409S} suggest that Type IIP SNe originate from red supergiants (RSGs), with progenitor masses as low as 8 M$_\odot$ and luminosities up to log(L/L$_\odot$) = 5 for the most massive progenitors, equivalent to main-sequence stars with masses of 15-18 M$_\odot$ \citep{Smartt2015PASA...32...16S}. Type IIL SNe are believed to originate from more massive stars, which have partially lost their hydrogen envelopes and possess larger radii (around a few thousand solar radii, as noted by \citealt{Blinnikov1993A&A...273..106B}). Compared with the more compact RSGs, which have smaller masses \citep{Elias2010ApJ...714L.254E, Fraser2010ApJ...714L.280F, Anderson2012MNRAS.424.1372A} and radii (less than 1600 \rsun, according to \citealt{Levesque2005ApJ...628..973L}). However, \cite{Morozova2017ApJ...838...28M} suggest that RSGs surrounded by a dense circumstellar material (CSM) may also exhibit the characteristics observed in Type IIL SNe.

The dense CSM surrounding the progenitor is probably formed during a super-wind phase \citep{Quataert2012MNRAS.423L..92Q, Fuller2017MNRAS.470.1642F}, which occurs shortly before core collapse \citep{Khazov2016ApJ...818....3K, Yaron2017NatPh..13..510Y}.
Alternative scenarios for CSM formation have also been proposed, including continuous mass loss during the RSG phase \citep{Luc2017, Soker2021ApJ...906....1S}, convection within the RSG envelope accompanied by associated instabilities \citep{Goldberg2022ApJ...933..164G, Kozyreva2022ApJ...934L..31K}, or interactions within colliding-wind binaries \citep{Kochanek2019MNRAS.483.3762K}.

A characteristic spectral signature of interaction in SNe is the appearance of narrow, symmetric emission line profiles, as opposed to the Doppler-broadened P-Cygni profiles \citep{Schlegel1990MNRAS.244..269S, Stathakis1991MNRAS.250..786S, Chugai2001MNRAS.326.1448C, Luc2009MNRAS.394...21D}. However, the absence of such persistent narrow lines does not necessarily rule out the possibility of  CSM interaction  \citep{Luc2009MNRAS.394...21D, Chevalier2011ApJ...729L...6C, Moriya2012ApJ...747..118M, Andrews2018MNRAS.477...74A}. CSM can reprocess the radiation from the SN, releasing it on a specific timescale, which results in broad, boxy emission features \citep{Pessi2023MNRAS.523.5315P}. The models in \cite{Luc2022} explicitly demonstrate this process from first principles.
If the CSM density is too low  to be optically-thick to electron scattering, no narrow line with wings broadened by electron scattering will form, and the object will not be classified as a SN IIn, even though the interaction may be strong \citep[ and references therein]{Luc2022}.

The number of published Type II SNe exhibiting early-time spectroscopic evidence of ejecta-CSM interaction has already increased significantly \citep{Jacobson2024ApJ...970..189J}. 
Examples include  SNe 2009bw \citep{Inserra2012MNRAS.422.1122I}, 2013cu \citep{Gal-Yam2014Natur.509..471G}, 2013fs \citep{Yaron2017NatPh..13..510Y, Bullivant2018MNRAS.476.1497B}, 2013fr \citep{Bullivant2018MNRAS.476.1497B}, 2014G \citep{Terreran2016TheMT}, 2018zd \citep{Zhang2020MNRAS.498...84Z, Callis2021arXiv210912943C}, 2020pni \citep{Terreran2022ApJ...926...20T}, 2022tlf \citep{Jacobson2025RNAAS...9....5J}, 2022lxg \citep{Charalampopoulos2025A&A...700A.138C}, 2023ixf \citep{Bostroem2023ApJ...956L...5B,  Jacobson2023ApJ...954L..42J,  Zhang2023SciBu..68.2548Z} and 2024ggi \citep{Zhang2024ApJ...970L..18Z}. \citet{Bruch2023ApJ...952..119B} report that more than 36\% of these Type II SNe events show flash-ionization features at the 95\% confidence level.

In addition to the early-time spectroscopic components, the effect resulting from the interaction between the ejecta and the CSM is also reflected in the luminosity evolution of the SNe \citep{Morozova2017ApJ...838...28M,  F.2018NatAs...2..808F, Khatami2024ApJ...972..140K}. 
Observable consequences of this interaction include a rapid rise to maximum brightness in the LC at early times \citep{F.2018NatAs...2..808F}, interpreted as a signature of shock breakout occurring within a dense stellar wind. Alternatively, an excess luminosity may manifest at early phases, resulting from the conversion of kinetic energy from the ejecta into radiation \citep{Luc2022}.

A notable feature of Type II SN interactions is the persistence of a broad H$\alpha$ line several years after the explosion \citep[e.g. SNe 1993J, 1998S, 2004et, 2017eaw;][]{Matheson2000AJ....120.1487M, Szalai2025A&A...697A.132S, Leonard2000ApJ...536..239L, Shahbandeh2023MNRAS.523.6048S, Weil2020ApJ...900...11W}. In the later stages, if there is an asymmetrically distributed layer of stellar wind material or dust, it may cause the observed broad H$\alpha$ line to become an asymmetric and multi-peaked H$\alpha$ spectral line \citep{Leonard2000ApJ...536..239L, Gerardy2000AJ....119.2968G, Luc2022}.
Remarkably, Type II SNe displaying asymmetric, double-peaked H$\alpha$ features during the nebular phase, such as SNe 1993J \citep{Matheson2000AJ....120.1487M, Szalai2025A&A...697A.132S}, 1998S \citep{Leonard2000ApJ...536..239L}, 2010jp \citep{Smith2012MNRAS.420.1135S}, PTF11iqb  \citep{Smith2015MNRAS.449.1876S}, 2018hfm \citep{Zhang2022MNRAS.509.2013Z} and 2023ufx \citep{Ravi2025ApJ...982...12R}, are rare. Among them, SN~1993J, classified as a Type~IIb event, represents a transitional case between Type~II and Type~Ib SNe and is therefore distinct from typical Type~II explosions.

This paper provides a detailed photometric and spectroscopic analysis of the Type IIL SN 2016iog. The SN was observed to undergo a rapid decline in its LC, with continuous spectral evidence of interaction with the surrounding material.
The organization of the paper is as follows: Section \ref{section:Basic_information} presents the basic information of SN 2016iog. Section \ref{section:data} describes the data reduction procedures. Section \ref{section:photometry} analyses the LC, colour evolution, estimate the ejected $^{56}$Ni mass, and construct a LC model. Section \ref{section:spect} focuses on the spectroscopic analysis and compares the spectral characteristics with those of other Type II SNe. Finally, a discussion is provided in Section \ref{section:sum}, and a summary is given in Section \ref{section:conclusion}.

\section{Basic information for SN 2016iog}
\label{section:Basic_information}

\begin{figure}
    \includegraphics[width=\columnwidth]{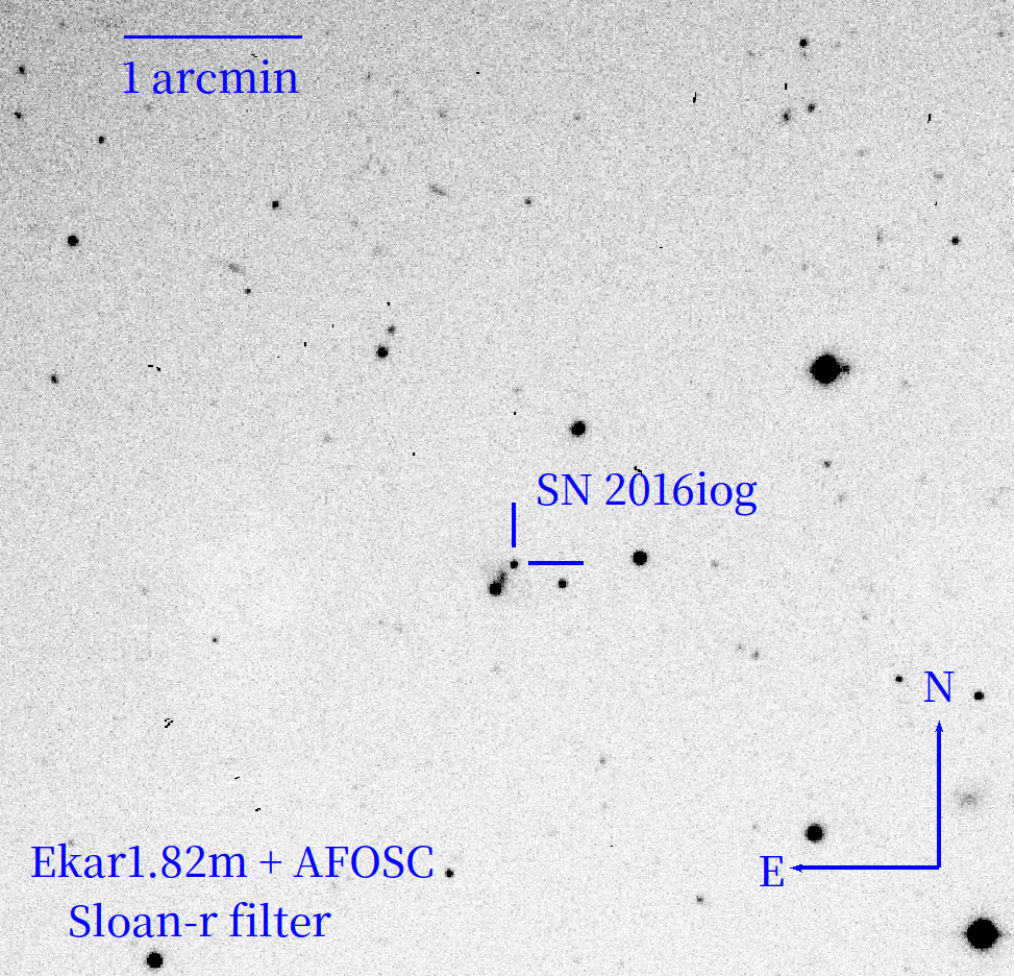}
    \caption{An image showing the location of SN 2016iog, obtained on December 06, 2016, using the $r$-Sloan filter with the Asiago Ekar 1.82-m Copernico Telescope. The orientation and scale are included.}
    \label{fig:location}
\end{figure}

SN 2016iog was discovered on November 27, 2016 (UT dates are used throughout the paper, corresponding to MJD = 57719.56) using the Brutus instrument on the All-Sky Automated Survey for Supernovae (ASAS-SN) telescope by the ASAS-SN team \citep{Shappee2014AAS...22323603S}, with the discovery reported by \cite{Stanek2016TNSTR.963....1S}, and it was named ASASSN-16ns\footnote{\url{https://www.astronomerstelegram.org/?read=9812}} \footnote{also known as Gaia16cfv; \url{http://gsaweb.ast.cam.ac.uk/alerts/alert/Gaia16cfv/}}. The first observation recorded a magnitude of \( V = 17.20 \pm 0.15 \) mag, while the last non-detection on November 20, 2016 (MJD = 57712.59) indicated a limiting magnitude of 17.2. Soon after the discovery, the Padova-Asiago group classified this transient as a Type II SN \citep{Tomasella2016TNSCR1013....1T, Turatto2016ATel.9829....1T}.

The reported coordinates are RA = $10^\mathrm{h}04^\mathrm{m}18^\mathrm{s}.560$ and Dec = $+43^\circ25'28''.45$, placing the SN 2016iog near the possible host galaxy GALEXASC J100418.99+432525.7.
The location of the SN 2016iog is shown in Fig.~\ref{fig:location}. The recessional velocity adopted is \( v = 7819 \pm 14 \, \mathrm{km \, s^{-1}} \) \citep{Mould2000ApJ...529..786M}, corresponding to a redshift of \( z = 0.026 \), determined after applying corrections for the influences of the Virgo Cluster, the Great Attractor, and the Shapley Supercluster. These corrections account for the peculiar motion of the Milky Way and other nearby large structures, which could skew the recessional velocity of the SN 2016iog if not properly taken into account \citep{Marinoni1998ApJ...505..484M}. Using a standard cosmological model with \( H_0 = 73 \pm 5 \, \mathrm{km \, s^{-1} \, Mpc^{-1}} \), \( \Omega_M = 0.27 \), and \( \Omega_\Lambda = 0.73 \) \citep{Spergel2007ApJS..170..377S}, we derive a luminosity distance of \( d_L = 109.00 \pm 7.50 \, \mathrm{Mpc} \) (with \( \mu_L = 35.19 \pm 0.15 \, \mathrm{mag} \)) for SN 2016iog.

For the interstellar reddening, we use \( E(B-V)_{\text{Gal}} = 0.011 \, \mathrm{mag} \) \citep[$A_V$ = 0.034 mag]{Schlafly2011ApJ...737..103S} for the Galactic reddening component, as derived from the NASA/IPAC Extragalactic Database (NED)\footnote{\url{https://ned.ipac.caltech.edu}}, and assume a reddening law with \( R_V = 3.1 \) \citep{Cardelli1989ApJ...345..245C}. However, due to the limitations in the signal-to-noise (S/N) ratios of the spectral data, the extinction from the host galaxy remains uncertain. Consequently, we adopt \( E(B-V)_{\text{tot}} = 0.011 \, \mathrm{mag} \) as the total reddening towards~SN 2016iog, knowing that this is likely a lower limit to the extinction affecting the target.
The very blue continuum observed in the earliest spectra, together with the remote location of SN~2016iog within its host galaxy, both support the assumption of negligible host-galaxy reddening.

\section{Observations and data reductions}
\label{section:data}

\subsection{Photometric data}

\begin{figure*}[ht]
    \centering
    \includegraphics[width=0.8\textwidth]{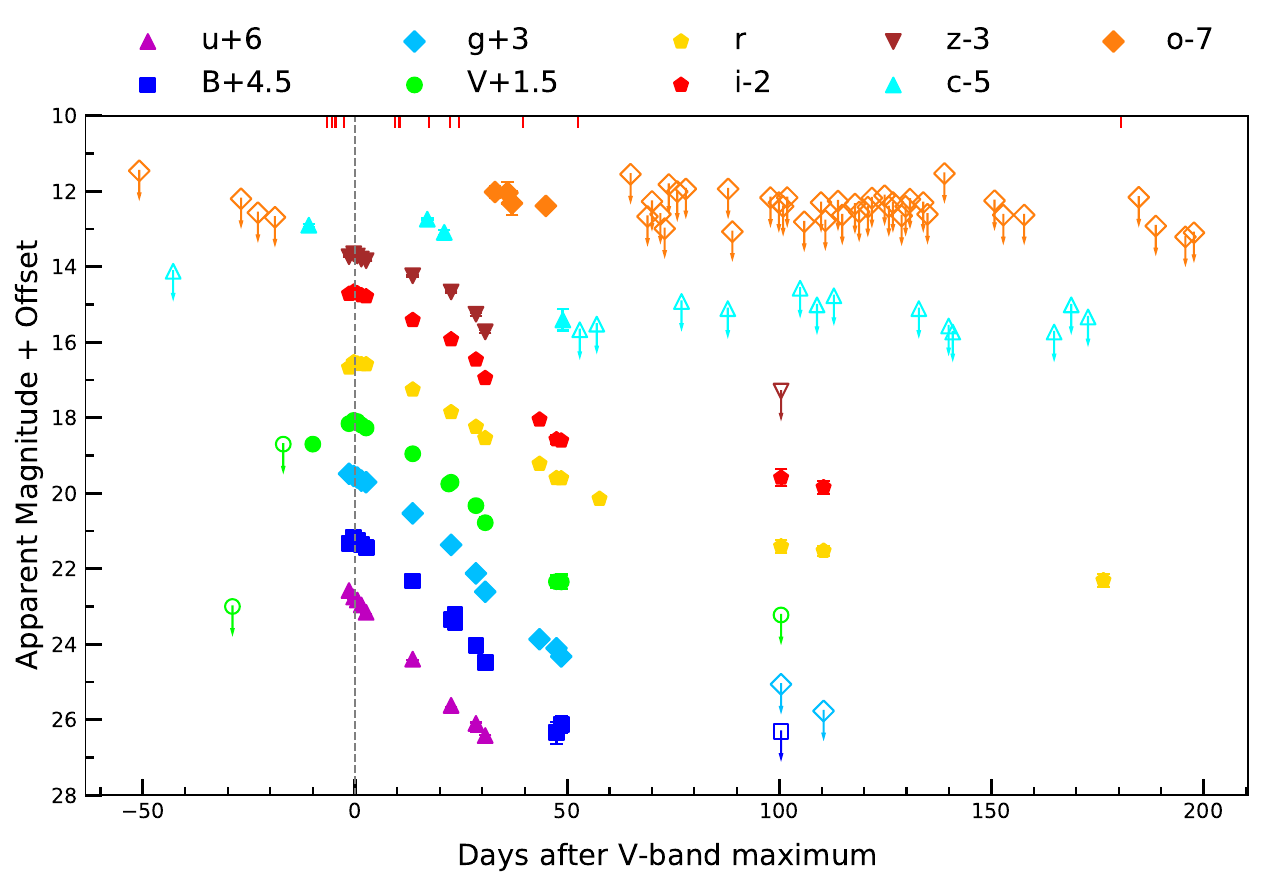} 
    \caption{Multi-band LCs of SN\,2016iog. A dashed vertical line represents the reference epoch, which corresponds to the $V$-band maximum light. The epochs of our spectroscopic observations are marked with solid red vertical lines at the top. Upper limits are denoted by empty symbols with downward arrows. For clarity, the LCs are offset by constant values, as indicated in the legends. In most instances, the uncertainties in the magnitudes are smaller than the size of the plotted symbols.}
    \label{fig:light_curve}
\end{figure*}

We conducted multi-band optical follow-up observations of SN~2016iog, covering the Sloan $ugriz$, Johnson-Cousins $BV$ bands starting shortly after its discovery.
The telescopes and instruments employed in this study were as follows: The 1.82m Copernico Telescope, equipped with the Asiago Faint Object Spectrograph and Camera (AFOSC), operated by the INAF - Padova Astronomical Observatory, located at the Asiago Observatory in Italy.  
The 10.4m Gran Telescopio Canarias (GTC), situated at the Roque de los Muchachos Observatory on La Palma, Spain, equipped with the  Optical System for Imaging and low-Intermediate-Resolution Integrated Spectroscopy (OSIRIS) instrument. 

All raw images were first pre-reduced using standard processed procedures in \textsc{iraf}\footnote{\url{http://iraf.noao.edu/}} \citep{Tody1986SPIE..627..733T, Tody1993ASPC...52..173T}, such as bias, overscan, trimming, and flat-field corrections. When SN\,2016iog was faint, we combined the multiple exposures into one frame to increase the S/N ratio. Photometric measurements data reduction was carried out using the dedicated pipeline {\sl ecsnoopy}\footnote{{\sl ecsnoopy} is a package for SN photometry using PSF fitting and/or template subtraction developed by E. Cappellaro. A package description can be found at \url{http://sngroup.oapd.inaf.it/snoopy.html}.}, which includes several  photometric packages, such as {\sc sextractor}\footnote{\url{www.astromatic.net/software/sextractor/}} \citep[][]{Bertin1996A&AS..117..393B} for source extraction, {\sc daophot}\footnote{\url{http://www.star.bris.ac.uk/~mbt/daophot/}} \citep[][]{Stetson1987PASP...99..191S} for measuring the target magnitude with the point spread function (PSF) fitting, and {\sc hotpants}\footnote{\url{http://www.astro.washington.edu/users/becker/v2.0/hotpants.html}} \citep[][]{Becker2015ascl.soft04004B} for image subtraction with PSF match. We measured the SN instrumental magnitudes via 
the PSF-fitting method, after the subtraction of the sky background. The PSF model was constructed by averaging the profiles of isolated, non-saturated stars in the SN field. Then, the PSF model was removed from the original frames, the local background was estimated again, and the fitting procedure was iterated. In typical runs,  the residuals were visually inspected to evaluate the fit quality. In the case of SN\,2016iog, a straightforward PSF-fitting technique was performed for the Johnson $BV$ images, while the Sloan photometry was reduced after removing the host galaxy contamination using public Sloan Digital Sky Survey (SDSS) templates.

The final photometric calibration was performed using zero points (ZPs) and colour terms (CTs) for each instrument, after the SN instrumental magnitudes were obtained. ZPs and CTs were derived from observations of standard stars during photometric nights. Johnson-Cousins magnitudes were calibrated using the catalogue of \citet{Landolt1992AJ....104..340L}, while Sloan-filter photometry was directly determined from the SDSS DR 18 catalogue \citep{Almeida2023ApJS..267...44A}. In addition, a sequence of local standard stars in the vicinity of the SN field was used to adjust the photometric ZPs obtained on non-photometric nights. This allow us to improve the SN calibration accuracy. 

Instrumental magnitude errors were estimated through artificial star experiments. Several fake stars of known magnitudes were evenly placed in the vicinity of the SN in the PSF-fit residual image. The simulated image was subsequently processed through the PSF fitting procedure. The dispersion of these measurements was taken as an estimate of the instrumental magnitude errors. The final photometric errors were estimated by combining in quadrature the artificial star experiment error, the PSF fit error returned by {\sc daophot}, and the error on the ZP correction. Note that the artificial star experiments were not performed if the template subtraction was applied. Instead, the background uncertainty was derived from the root mean square of the residuals in the background after subtracting the PSF-fitted source.

We also collected photometric data from the public 
Asteroid Terrestrial-impact Last Alert System  \citep[ATLAS;][]{Tonry2018PASP..130f4505T} and ASAS-SN sky surveys, which observed SN 2016iog.
The LCs for the orange ($o$) and cyan ($c$) bands were obtained directly from the ATLAS data release platform\footnote{\url{https://fallingstar-data.com/forcedphot/}} \citep{Shingles2021TNSAN...7....1S}, while some $V$-band data were obtained from the ASAS-SN Sky Patrol\footnote{\url{https://asas-sn.osu.edu}} \citep{Hart2023arXiv230403791H}. 
However, although the ASAS-SN telescopes continued to monitor SN 2016iog, their 14 cm aperture limited effective observations for sources fainter than the 17th magnitude, offering insufficient constraints after the maximum brightness.
The corresponding apparent LCs are presented in Fig.~\ref{fig:light_curve}.

\subsection{Spectroscopic data}

Spectroscopic observations of SN 2016iog were conducted using the following facilities: 
the 2.16m XingLong Telescope (XLT), equipped with the Beijing Faint Object Spectrograph and Camera (BFOSC); the 1.82m Asiago Ekar Telescope, equipped with the AFOSC; and the 10.4m GTC, equipped with the OSIRIS instrument.

All raw spectral data were processed using standard reduction techniques in \textsc{iraf}, or by the \texttt{gtcmos} pipeline for OSIRIS data\footnote{\url{https://www.gtc.iac.es/instruments/osiris/}} \citep{M.2020MNRAS.491..129C}. The initial reduction steps, including bias subtraction, overscan correction, flat-fielding, and trimming, followed the same procedures as those for imaging data. Next, one-dimensional (1D) spectra were extracted from the 2D images.
Wavelength calibration was performed using arc lamps, while flux calibration was carried out with spectrophotometric standard stars observed on the same nights. 
The prominent telluric absorption features, such as O$_2$ and H$_2$O, were removed from the SN spectra using standard star data. Finally, the flux calibration accuracy for all spectra was validated by comparing with the corresponding photometric data. Details of the instrumentation used for the spectroscopic observations are listed in Table \ref{tab:spec} (Appendix \ref{SpecInfo}).
 
\section{Photometry}
\label{section:photometry}

\subsection{Apparent magnitude light curves}

\begin{table}[ht]
\centering
\caption{Decline rates of the LCs of SN 2016iog, along with uncertainties, in units of magnitude $(100 \, \text{d})^{-1}$.}
\label{tab:decay_rates}
\begin{tabular}{cccc}
\hline\hline
\textbf{Filter} & $\gamma_{0-25}$ & $\gamma_{25-50}$ & $\gamma_{50-200}$ \\
\hline
$u$ & $12.37 \pm 0.23$ & $-$ & $-$ \\
$B$ & $9.24 \pm 0.44$ & $10.73 \pm 1.11$ &  $-$ \\
$g$ & $8.07 \pm 0.24$ & $10.19 \pm 0.97$ & $-$\\
$V$ & $7.02 \pm 0.22$ & $9.77 \pm 0.70$ & $-$ \\
$r$ & $5.84 \pm 0.25$ & $6.58  \pm 0.61$ & $2.05 \pm 0.37$ \\
$i$ & $5.53 \pm 0.07$ & $10.31 \pm 0.89$ & $1.96 \pm 0.09$\\
$z$ & $4.20 \pm 0.16$ & $-$  & $-$ \\
\hline
\bottomrule
\end{tabular}
    \label{tab:slope}
\end{table}

We monitored the photometric evolution of SN 2016iog for about 200 days after its discovery. The optical LCs of SN 2016iog are shown in Fig.~\ref{fig:light_curve}.
The explosion epoch of an SN is determined to lie between the last non-detection and the first detection of the event. The last non-detection, $t_l$, occurred on November 20, 2016 (MJD = 57712.59), with a limiting magnitude of $17.2$ Vega-mag in the ASAS-SN $V$ band. The first detection, $t_d$, is dated on November 26, 2016 (MJD = 57718.60), with a detection magnitude of $17.9$ AB-mag in the ATLAS $c$-band. The maximum uncertainty on the explosion epoch is calculated as half of the difference between $t_l$ and $t_d$. Based on this method, the explosion epoch of SN 2016iog is assumed to be derived as MJD = $57715.6 \pm 3.0$, which is adopted as the reference epoch throughout this paper.  However, due to the unconstraining nature of the non-detection, the explosion may have occurred somewhat earlier. To estimate the peak magnitude of SN 2016iog, a third-order polynomial fit was applied to the $V$-band LC data within a two-week period centred at about the magnitude peak. The peak magnitude was determined to be $V = 16.58 \pm 0.02$ mag at MJD = 57729.50 $\pm$ 0.10 for SN 2016iog.

We determined the post-maximum decline rates of SN~2016iog across different bands by applying linear regression to the data after the peak. The findings, presented in Table~\ref{tab:slope}, indicate differences in the decline rates among the various filters.
From 0 to 25 days, the LCs of SN 2016iog show a more rapid decline in the blue bands (e.g. $\gamma_{0-25}$($u$) $\approx 0.12$ mag~d$^{-1}$), whereas in the red bands, the decline is slower (e.g. $\gamma_{0-25}$($r$) $\approx 0.05$ mag~d$^{-1}$). After 25 days, a steeper decline is observed in the LCs (e.g. $\gamma_{25-50}$($r$) $\approx 0.07$ mag~d$^{-1}$). In the later phases, the decline apparently slows down (e.g. $\gamma_{50-200}$($r$) $\approx 0.02$ mag~d$^{-1}$), although the available detection data is insufficient to provide precise measurements. However, this trend can be inferred from the tail of Fig.~\ref{fig:light_curve}.

\subsection{Absolute magnitude light curves}
\begin{figure*}[ht]
    \centering
    \includegraphics[width=0.94\textwidth]{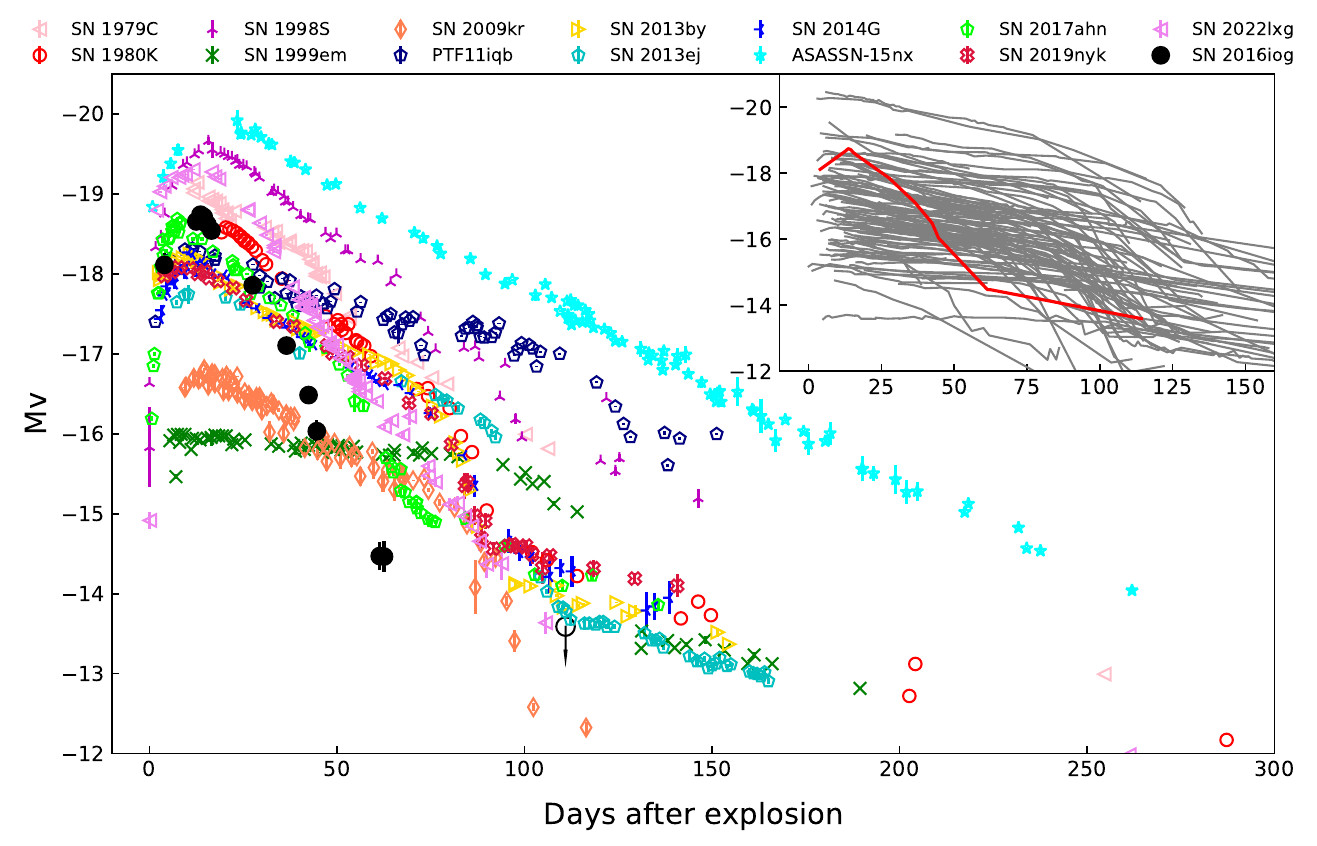} 
    \caption{Comparison of the $V$-band absolute light curves of SN~2016iog and other Type II SNe. For PTF11iqb and SN~2022lxg, the $r$-band light curve is used instead. The upper-right panel shows the absolute \(V\)-band light curve of SN 2016iog (red line), compared with the 104 Type II SNe of the sample presented by \cite{Anderson2014}. All absolute $V/r$-band light curves are corrected for reddening.}
    \label{fig:Absolute_magnitude}
\end{figure*}

\begin{figure*}[ht]
    \centering
    \includegraphics[width=0.49\textwidth]{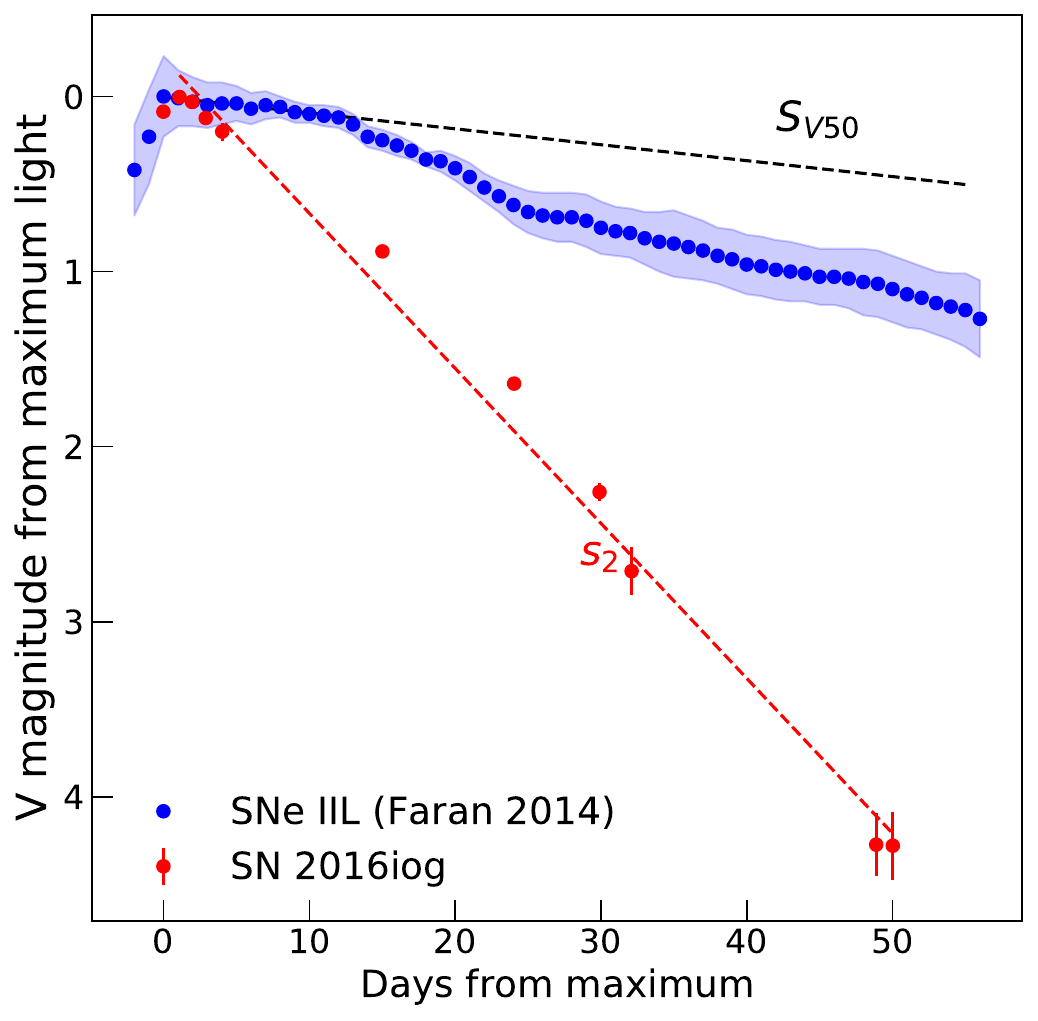} 
    \includegraphics[width=0.5\textwidth]{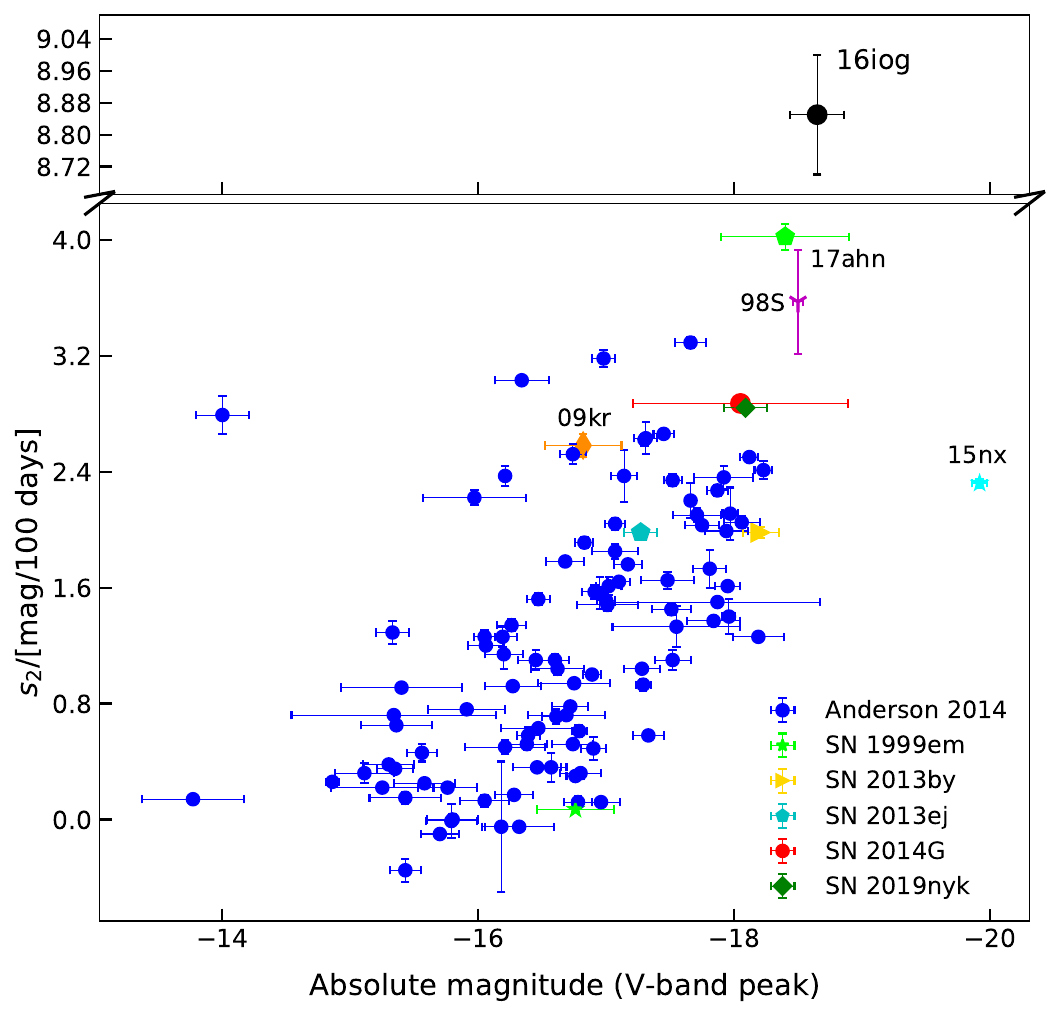}
\caption{Comparison of the $V$-band decline slope of SN~2016iog with other Type II SNe.
    \textit{Left panel}: Absolute $V$-band LC of SN\,2016iog compared to SNe IIL template LCs from \cite{Faran2014ASO}.  $S_{V50}$ = 0.5 mag 50~d$^{-1}$ in the $V$-band. 
    \textit{Right panel}: Absolute $V$-band magnitude of SN~2016iog vs. $s_2$, compared to objects from \cite{Anderson2014}, and other SNe in Table~\ref{tab:s2} have been also added for comparison.
}

    \label{fig:compare}
\end{figure*}

\begin{table*}[ht]
\centering
\caption{Parameters for the comparison sample of SNe II.}
\label{tab:s2}
\resizebox{\textwidth}{!}{
\begin{tabular}{ccccccccc}
\hline\hline
SN        & $s_1$ (mag/100d) & $s_2$ (mag/100d) & $s_3$ (mag/100d)  & $M^{50d}_{V}$ (mag)&$M^{\text{peak}}_V$(mag)&$L_{peak}$ (erg s$^{-1}$)&$M_{\text{Ni}}(M_{\odot})$&References \\ 
\hline
1979C     & --               & $3.55 \pm 0.08$  & --              & $-18.31\pm0.11$&$ \leq -19.85$      &$\geq 9.75$ $\times 10 ^{42}$& --                      & 1\\ 
1980K     & $ 4.22 \pm 0.10$ & $3.44\pm 0.39$  & $1.17 \pm 0.17$  &$-17.30\pm0.50$ & $ \leq -18.58$     &$-$                          &$-$                      &2\\ 
1998S     & $3.67 \pm 0.06$ & $3.57 \pm 0.36$  & --               &$-18.49\pm0.16$ & $-19.66 \pm 0.05$  &1.50 $\times 10 ^{43}$       &$0.067\pm0.034$          & 3\\
1999em    & $0.86 \pm 0.11$ & $0.30 \pm 0.02$  & $0.88 \pm 0.05$  &$-16.48\pm0.52$ & $-16.76 \pm 0.07$  &1.23 $\times 10 ^{42}$       &$0.042^{+0.027}_{-0.019}$& 4\\
2009kr    & --              & $2.58 \pm 0.08$  & --               & $-15.74\pm0.08$& $-16.82 \pm 0.30$  &8.30 $\times 10 ^{41}$       &$0.009\pm0.004$          &5\\
PTF11iqb$^a$    & $1.61 \pm 0.10$ & $0.83 \pm 0.17$   &--        &  $-17.73\pm0.04$   &  $-18.32\pm0.01$   & --& --   &6\\
2013by    & $3.43 \pm 0.02$ & $1.98 \pm 0.04$  & $1.20 \pm 0.06$  & $-17.40\pm0.44$& $-18.05 \pm 0.15$  &2.50  $\times 10 ^{42}$      &$0.033\pm0.012$          &7\\ 
2013ej    & $3.64 \pm 0.26$ & $2.04 \pm 0.02$  & $1.65 \pm 0.08$  &$-16.79\pm0.16$ & $-17.80 \pm 0.13$  &2.89  $\times 10 ^{42}$      &$0.020\pm0.002$          &8\\ 
2014G     & --              & $2.87 \pm 0.05$  & $1.68 \pm 0.18$  & $-17.18\pm0.47$& $-18.05 \pm 0.84$  &4.14 $\times 10 ^{42}$       &$0.073\pm0.013$          &9\\ 
ASASSN-15nx& $2.32 \pm 0.03$ & $2.32 \pm 0.03$  & --              &$-19.13\pm0.05$ & $-19.92 \pm 0.06$  &1.34 $\times 10 ^{43}$       &$1.3\pm0.2$              &10\\
2017ahn   & --              & $4.02 \pm 0.09$  & $1.90 \pm 0.07$  &$-16.75\pm0.46$ & $-18.40 \pm 0.50$  &5.56 $\times 10 ^{42}$       &$0.051\pm0.016$          &11\\
2019nyk   & --              & $2.84 \pm 0.03$  & $1.20 \pm 0.20$  & $-16.95\pm0.17$& $-18.09 \pm 0.17$  &3.61 $\times 10 ^{42}$       &$0.050\pm0.010$          &12\\
2022lxg$^a$& $6.53 \pm 0.07$         & $6.53 \pm 0.07$   &--         &  $-17.10\pm0.02$   &  $-19.31 \pm 0.02$  & 6.40 $\times 10 ^{43}$& $\geq 0.003 \pm 0.002 $   &13\\
\hline
2016iog   & $8.85 \pm 0.15$ & $8.85 \pm 0.15$  & --               & $-15.54\pm0.21$& $-18.64\pm 0.15$  &6.31  $\times 10 ^{42}$      &$0.014\pm0.007$          &14\\
\hline\hline
\end{tabular}
}
\raggedright
\footnotesize
$^a$ The data for PTF11iqb and SN 2022lxg is obtained using the $r$-band as a substitute.
\newline

References: 1 = \cite{Fesen1999AJ....117..725F}, 2 = \cite{Fesen1999AJ....117..725F,Zsiros2023SerendipitousDO}, 3 = \cite{Fassia2000OpticalAI}, 4 = \cite{Elmhamdi2002PhotometryAS,Leonard2002AAS...201.2303L}, 5 = \cite{Elias2010ApJ...714L.254E}, 6 = \cite{Smith2015MNRAS.449.1876S}, 7= \cite{Valenti2015Supernova2A}, 8 = \cite{Huang2015SN2I}, 9 = \cite{Terreran2016TheMT,Bose2016MNRAS.455.2712B}, 10 = \cite{Bose2018ASASSN15nxAL}, 11 = \cite{Tartaglia2020TheED}, 12 = \cite{Dastidar2024SN2A},13 = \cite{Charalampopoulos2025A&A...700A.138C}, 14 = This work.
\end{table*}

In this work, we assembled a sample of Type II SNe to serve as a reference for the analysis of the LC and spectral properties of SN~2016iog, as presented in Table~\ref{tab:s2}. Specifically, we selected SNe classified as Type II with decline rates exceeding 1.9 mag/100d, along with the Type II SNe 1999em and PTF11iqb for comparison,  in order to compare their LCs and spectral properties. A comparison of the absolute $V$-band magnitudes for the selected Type II SNe, is shown in Fig.~\ref{fig:Absolute_magnitude}.

Considering the distance and extinction values provided in Sect. \ref{section:Basic_information}, we determine the absolute $V$-band magnitude at peak for SN\,2016iog to be $M_V = -18.64 \pm 0.15$ mag. All SNe reach their peak luminosity within $\leq$ 3 weeks, with peak magnitudes spanning a wide range between $-16.8 \pm 0.3$ and $-19.9 \pm 0.1$ mag, and an average $M = -18.45 \pm 0.20$ mag. The peak luminosity of SN 2016iog in the selected sample lies in the middle range, but it is relatively bright compared to other Type II SNe (see Fig. \ref{fig:Absolute_magnitude}). Among the selected sample, only SNe 1979C ($M_V \leq -19.85$ mag), 1998S ($M_V =-19.66 \pm 0.05$ mag), 2022lxg ($M_r = -19.31 \pm 0.02$ mag) and ASASSN-15nx ($M_V =-19.91 \pm 0.06$ mag) have higher peak luminosities than SN 2016iog. However,  SNe 1999em ($M_V =-16.76 \pm 0.07$ mag) and 2009kr ($M_V =-16.82 \pm 0.30$ mag) exhibit a different behaviour compared to the other SNe in the comparison, showing notably dimmer luminosities.

To better understand the position of SN~2016iog relative to other SNe II in the literature, we quantify the post-maximum decline rate. Comprehensive statistical studies on brightness decline rates have revealed a correlation with the absolute magnitude at the peak, as shown by \cite{Li2010NearbySR}, \cite{Anderson2014}, and \cite{Valenti2015Supernova2A}. Following the method outlined in \cite{Valenti2015Supernova2A}, we measure the decline rate of the shallower slope in the $V$-band LC of SN 2016iog, denoted as $s_2$.
Specifically, \(s_1\) represents the initial rapid decline rate in the $V$-band after peak brightness, \(s_2\) refers to the decline rate in the $V$-band during the slow-fading, linear phase, and \(s_3\) denotes the radioactive tail decline rate in the $V$-band following the plateau, dominated by \(^{56}\)Co decay \citep{Anderson2014}.
To maintain consistency with definitions in the literature, we express $s_2$ in units of mag/100d as described in \cite{Anderson2014}.
After peak, the luminosities of SNe 1979C, 1980K, 1998S, ASASSN-15nx, 2017ahn, 2022lxg, and 2016iog  all decline in a manner similar to a linear decay until the radioactive decay phase, showing very short plateau phases.
SNe 2009kr, 2013by, 2013ej, 2014G, and 2019nyk exhibit a slower decline after reaching peak luminosity, followed by a brief plateau before quickly declining. The corresponding $s_2$ values for these SNe are listed in Table~\ref{tab:s2}.

SN\,2016iog lies below the \( S_{V50} = 0.5 \, \text{mag}/50d \), which is the standard value for Type IIL SNe (see the left panel of Fig. \ref{fig:compare},  e.g. the comparison with the Type IIL SN templates of \cite{Faran2014ASO}), indicating that SN 2016iog evolves faster than a typical Type IIL SN.
For SN\,2016iog, the best-fit slope for the early decline is $s_1=s_2$ = 8.85 $\pm$ 0.15 mag/100d (see red dashed line in left panel of Fig.~\ref{fig:compare}). This represents the fastest early decline compared to the sample of \cite{Anderson2014} and Table~\ref{tab:s2}.
Most SNe II are clustered in the region where \(s_2 < 3.0\) mag/100d and \(M_V > -18\) mag.
The high $M_{V}$ and exceptionally large $s_2$ of SN~2016iog place it in a previously unexplored region of parameter space.
We present this comparison in the right panel of Fig.~\ref{fig:compare}, where we combine the sample data from \citet{Anderson2014}.

Several parameters have been proposed in the literature for quantitatively distinguishing Type IIP and IIL SNe. These parameters rely on different phase intervals and are used to measure the average decline rate in various optical bands. For example, \cite{Patat1994A&A...282..731P} introduced a decline rate of 3.5 mag per 100 days in the B~band (\(B_{100}\)), while \cite{Faran2014ASO} proposed a decline rate of 0.5 mag per 50 days in the V~band (\(V_{50}\)).
Using the decline rates measured according to these different criteria, we find that SN 2016iog has  $B_{100} \approx 10.55$ mag and $V_{50} \approx 4.48$ mag.  Based on these values, it can be confirmed that SN 2016iog fits the characteristics of a Type IIL SN. Furthermore, SN 2016iog stands out for its fast decline, compared to the 104 Type II SNe sample from \cite{Anderson2014}, as shown in the upper-right panel of Fig.~\ref{fig:Absolute_magnitude}. 

\subsection{Colour evolution}
\label{section:Colour_evolution}

\begin{figure*}[ht]
    \centering
    \includegraphics[width=0.9\textwidth]{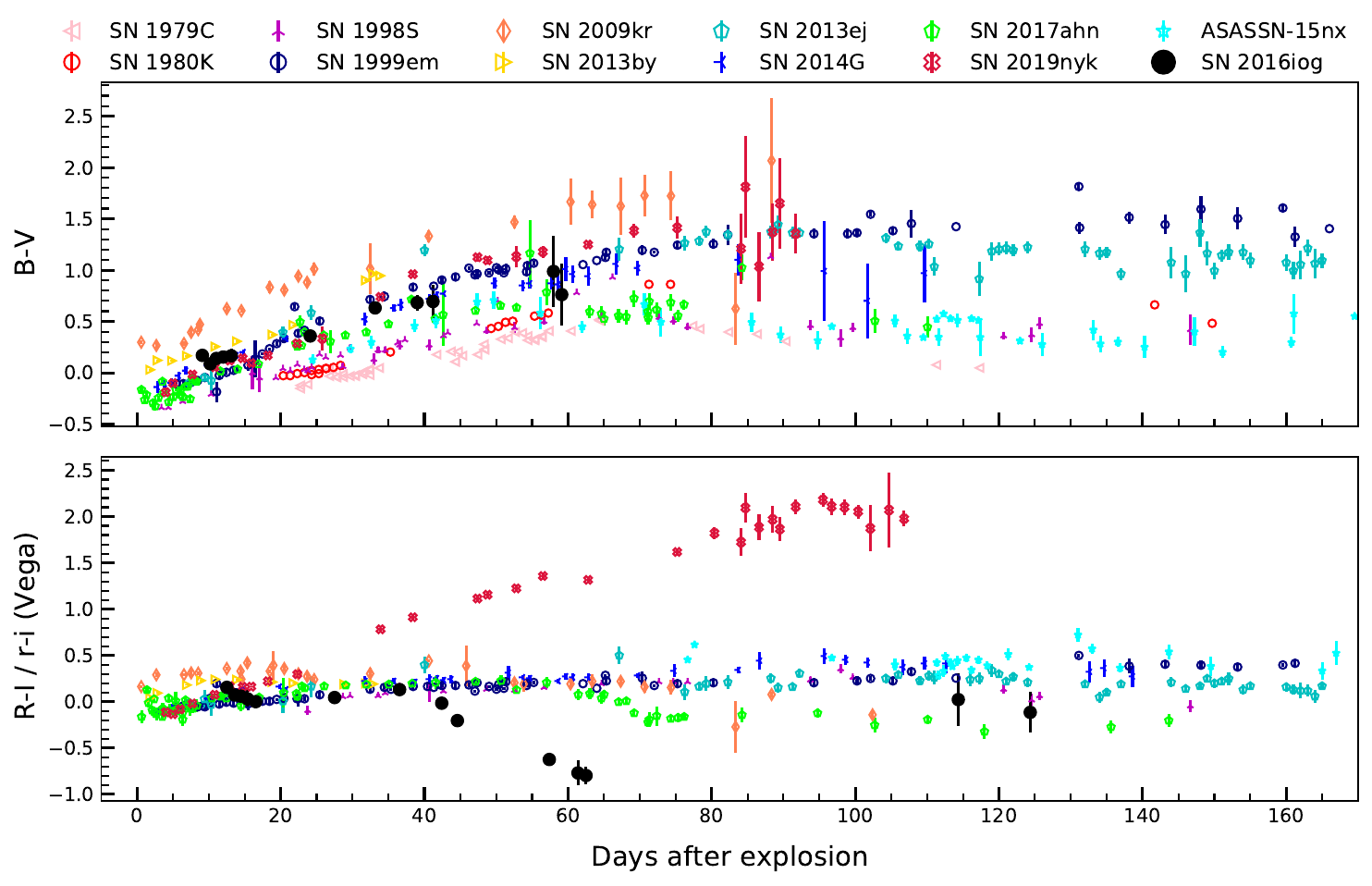} 
    \caption{Colour evolution of SN\,2016iog, compared to a sample of SNe II.
    Upper panel: $B~-~V$ colour evolution; Lower panel: ($R~-~I$) or ($r~-~i$) colour evolution.
    The colour curves are corrected for both Galactic and host galaxy extinction.}
    \label{fig:colour_evolution}
\end{figure*}

The intrinsic colour evolution of SN~2016iog, compared with the selected Type II SNe listed in Table~\ref{tab:s2}, is shown in Fig.~\ref{fig:colour_evolution}.
During the first +15 days past explosion, the $B-V$ colour of SN 2016iog remains fairly stable at approximately 0.16 mag ($\pm~0.02$ mag), with only minor fluctuations.
Between +15 and +40 days, the $B-V$ colour rises, reaching 0.69 mag. Around +60 days, the $B-V$ colour peaks at 0.98 mag, indicating a substantial decrease in the temperature of the ejecta. Subsequently, the $B-V$ colour exhibits a decline, although with considerable uncertainty.  Overall, the evolution of the $B-V$ colour of SN 2016iog closely follows the trends observed in other Type II SNe studied to date (see the top panel in Fig.~\ref{fig:colour_evolution}), showing an increase in the $B-V$ colour, consistent with expectations from the expansion of the SN envelope. Due to the limited observational data, we do not have $B-V$ measurements for the later phases of SN 2016iog.

In terms of the $r-i$ colour evolution, SN 2016iog behaves somewhat differently (see the bottom panel in Fig~\ref{fig:colour_evolution}).
From +10 to +20 days post-explosion, the $r-i$ colour decreases from +0.16 mag to nearly 0, then between +20 and +40 days it begins to increase, peaking at 0.13 mag at +36 days. Afterward, between +40 and +60 days, the $r-i$ colour becomes rapidly bluer, decreasing from $-$0.02 mag and reaching $-$0.80 mag at +62 days. 
In the later stages, the $r-i$ colour begins to redden, returning to 0. The evolution of SN 2016iog is similar to that of SN 2017ahn, except that in the intermediate phase, when the $r-i$ colour of SN 2016iog is bluer. 
During the intermediate phase, considering that strong H$\alpha$ emission, the enhanced brightness in the $r$ band relative to other bands leads to a bluer $r-i$ colour evolution. Circumstellar interaction is characterized by strong H$\alpha$ emission, which suggests that SN~2016iog may also have experienced interaction with CSM at this stage.
SN 2019nyk represents an exceptional case, where its $r-i$ evolution behaves differently from other SNe, showing a linear fast trend from 0 days, until it begins to flatten around 90 days.

\subsection{Pseudo-bolometric light curves} 
\begin{figure*}[ht]
    \centering
    \includegraphics[width=0.8\textwidth]{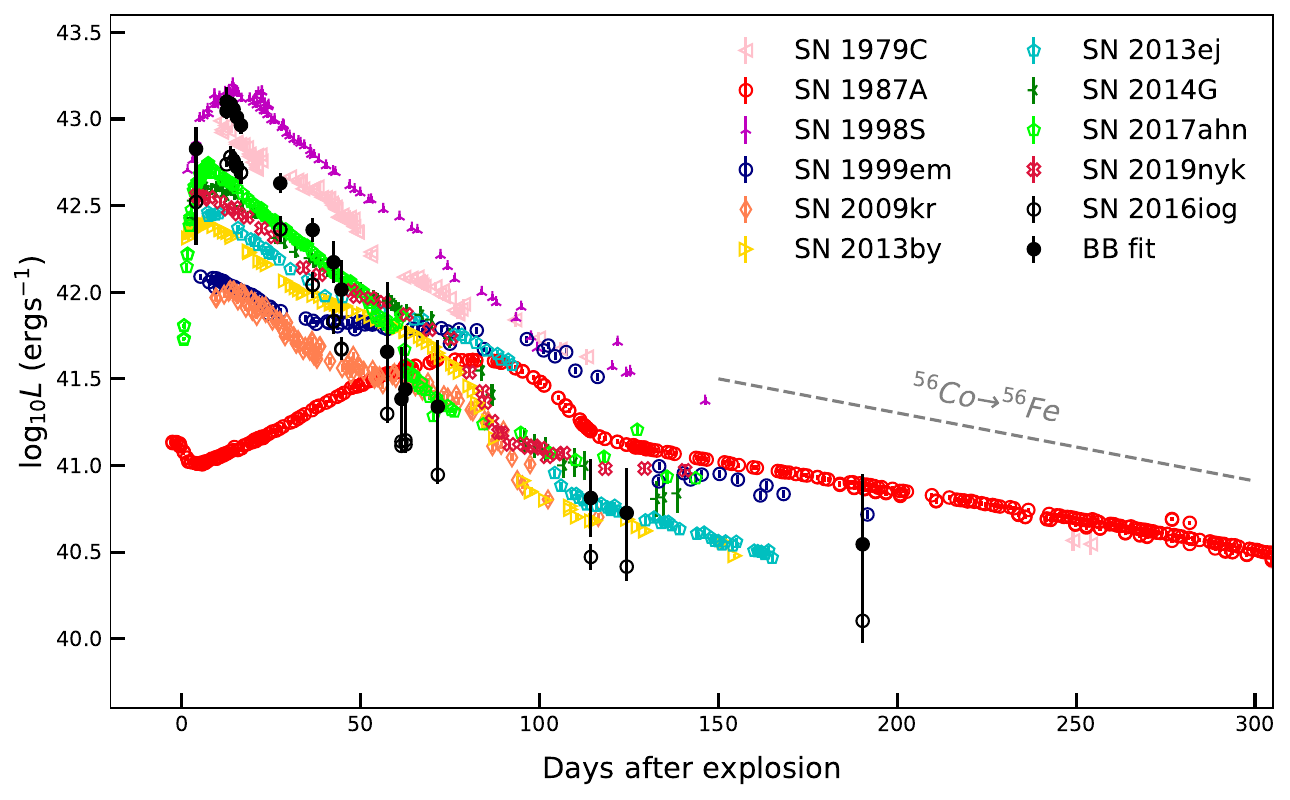} 
    \caption{Comparison of the pseudo-bolometric LC of SN 2016iog with those of other Type II SNe. The pseudo-bolometric LCs are limited to the observed U/u through I/i bands.  The black solid dot represent the blackbody-fit LC of SN~2016iog. The grey dashed line illustrates the expected slope of the LC under the assumption that all energy from $^{56}$Co decay is fully thermalized by the ejecta.}
    \label{fig:Pseudobolometric_light_curves}
\end{figure*}

With the optical photometric data and applying the \texttt{SuperBol}\footnote{\url{https://github.com/mnicholl/superbol}} program \citep{Nicholl_2018}, we calculated the pseudo-bolometric LC of SN~2016iog based on the optical bands.
When photometric data were missing for specific epochs in certain filters, we relied on the available bands and used an extrapolation to estimate the missing values.
For a meaningful comparison, we selected the $U/u$ through $I/i$ bands to derive the pseudo-bolometric LCs for other SNe\footnote{The pseudo-bolometric LCs for SNe 1979C, 1998S, 1999em, 2009kr, and 2013et were constructed using the $UBVRI$ bands; for SN 2013by, the $UBVri$ bands were used; for SN 2014G, the $UBVrRiI$ bands were utilized; and for SNe 2017ahn and 2019nyk, the $UBVgri$ bands were taken into account.}, excluding bands with insufficient available data from the fit.

The peak luminosities of most SNe are between \(10^{42} \, \mathrm{erg\,s^{-1}}\) and \(10^{43} \, \mathrm{erg\,s^{-1}}\). The peak luminosity of SN 2016iog is approximately \(L \sim 6.31 \times 10^{42} \, \mathrm{erg\,s^{-1}}\), which is slightly lower than that of SN 1979C (\(L \sim 9.75 \times 10^{42} \, \mathrm{erg\,s^{-1}}\)) and SN~1998S (\(L \sim 1.50 \times 10^{43} \, \mathrm{erg\,s^{-1}}\)), but higher than the luminosities of other SNe in the sample. 
SN 2016iog thus occupies a relatively bright region within the typical luminosity range. 
In contrast, SN 1999em (\(L \sim 1.23 \times 10^{42} \, \mathrm{erg\,s^{-1}}\)) and SN 2009kr (\(L \sim 8.30 \times 10^{41} \, \mathrm{erg\,s^{-1}}\)) are among the dimmest events.

The pseudo-bolometric LC of SN 2016iog compared through blackbody fits with those of other Type II SNe whose characteristics are summarised in Table~\ref{tab:s2}, is shown in Fig. \ref{fig:Pseudobolometric_light_curves}. 
The LC of SN 2016iog is similar to that of SN 2017ahn, which exhibits a comparable peak luminosity (\(L \sim 5.56 \times 10^{42} \, \mathrm{erg\,s^{-1}}\)), although SN 2016iog has a faster decline rate.

Following the initial peak, SN 2016iog shows a rapid luminosity decline, which is steeper than that of other SNe in the sample. After approximately 100 days post-explosion, the decline rate flattens, approaching the theoretical slope attributed to the radioactive decay of $^{56}\mathrm{Co} \to \,^{56}\mathrm{Fe}$ (0.98 mag/100d).  However, its late-time luminosity is fainter than that of other SNe in the sample, consistent with a lower mass of $^{56}$Ni in the explosion of SN 2016iog (see the following Sect. \ref{sect:Ni}). 

\subsection{The $^{56}$Ni mass} 
\label{sect:Ni}
The quantity of $^{56}$Ni formed during explosive nucleosynthesis can be estimated from  the radioactive LC tail observed when the nebular phase becomes optically thin.
Radioactive isotope decay produces $\gamma$-rays and positrons, which are then thermalized within the ejecta and re-emitted at optical wavelengths.
If $\gamma$-rays are completely confined within the ejecta (as observed in SN\,1987A), the $^{56}$Ni mass produced during the explosion can be inferred by comparing the pseudo-bolometric luminosity of the SN at nebular epochs with that of the well-characterized SN\,1987A \citep{Catchpole1988MNRAS.231P..75C,Catchpole1989MNRAS.237P..55C,Whitelock1988MNRAS.234P...5W}, under the assumption that the spectral energy distributions (SEDs) of both SNe are comparable.
This relation is given by the following equation \citep{Spiro14}:

\begin{equation}
 M(^{56}Ni) = M(^{56}Ni)_{\text{SN 1987A}} \times \frac{L_\mathrm{SN}(t)}{L_{\text{SN 1987A}}(t)} M_{\odot}
\end{equation}

where $M(^{56}\text{Ni})_{\text{SN~1987A}} = 0.075 \pm 0.005 M_{\odot}$ is the mass of $^{56}$Ni ejected by SN\,1987A \citep{Arnett1996SupernovaeAN}, and $L_\mathrm{SN}(t)$ and $L_{\text{SN~1987A}}(t)$ represent the pseudo-bolometric luminosities of the studied SN and SN\,1987A at epoch $t$, respectively, derived through identical filter sets. In this analysis, we perform a comparison between the pseudo-bolometric luminosities of SN\,2016iog (\textit{uBgVri}) and SN\,1987A (\textit{UBVRI}) across an identical wavelength range, focusing on the interval from 100 to 200 days, when both SNe are situated in their radioactive tail phases.
Based on this comparison, the mass of $^{56}$Ni for SN\,2016iog is estimated to be $M(^{56}Ni) = 0.014 \pm 0.007~ M_{\odot}$.  
We note that the late-time decline is derived solely from the $r$- and $i$-band data, and thus remains somewhat uncertain. While the shallower decline at late times could plausibly be powered by $^{56}$Ni decay, this is not guaranteed. If the decline is instead driven by some form of CSM interaction, the derived $^{56}$Ni mass should be regarded as an upper limit.

\subsection{Modelling the multi-band light curves with the MOSFiT framework}
\label{sec:MOSFiT}
\begin{figure}[ht]
    \centering
    \includegraphics[width=\columnwidth]{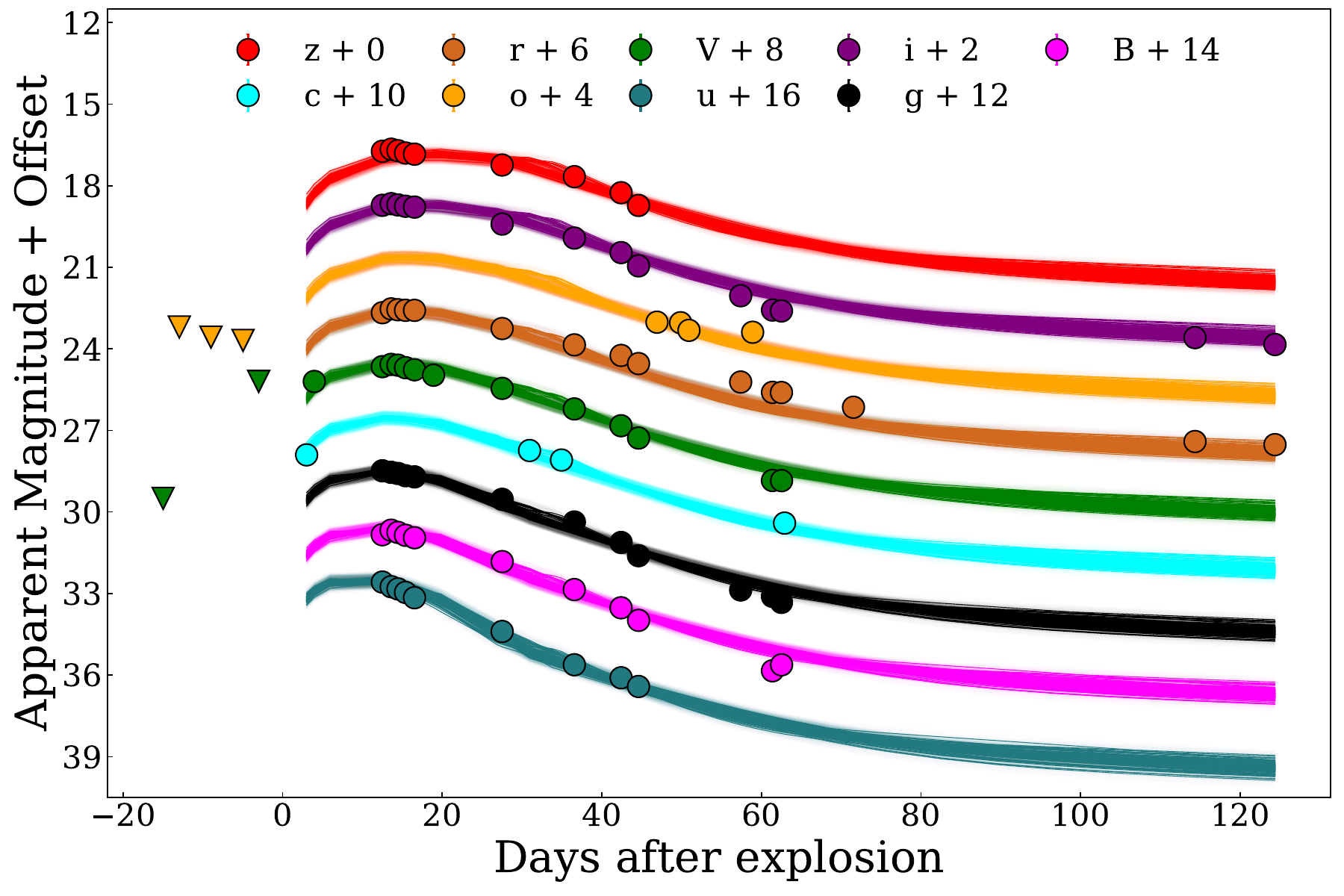} 
    \caption{Fits to the multi-band light curve of SN~2016iog using the \texttt{csmni} model in \texttt{MOSFiT}. The relevant parameters are listed in Table \ref{tab:MOSFiT}.
}
    \label{fig:mosfit}
\end{figure}

\begin{table}
\caption{Priors and marginalised posteriors for the \texttt{MOSFiT} \texttt{csmni} model of SN 2016iog.}
\label{tab:MOSFiT}
\centering
  \begin{tabular}{cccc}
  \hline\hline
  Parameter & Prior & Posterior & Units\\
  \hline
$\log{(f_{\mathrm{Ni}})}$ & $[-4, 0]$ & $-2.42^{+0.18}_{-0.18} $  \\
$\log{(\kappa_{\rm \gamma})}$ & $[-2, 1]$ & $-0.56^{+0.54}_{-0.50}$ & cm$^2$\,g$^{-1}$ \\
$\log{(E_{\rm k})}$ & $[1,50]$ & $0.24^{+0.13}_{-0.15}$ & 10$^{51}$ erg \\
$\log{(M_{\rm CSM})}$ & $[-2, 1.5]$ & $-1.22^{+0.13}_{-0.18}$ & M$_\odot$ \\
$\log{(M_{\rm ej})}$ & $[-1, 1.0]$ & $0.57^{+0.13}_{-0.15}$ & M$_\odot$ \\
$\log{(n_{\rm H,\mathrm{host}})}$ & $[16,23]$ & $16.95^{+1.07}_{-0.65}$ & cm$^{-2}$ \\
$\log{(R_0)}$ & $[-1, 2.4]$ & $0.66^{+0.26}_{-0.23}$ & AU \\
$\log{(\rho_{0})}$ & $[-15, -7]$ & $-10.02^{+0.43}_{-0.41}$ & g\,cm$^{-3}$ \\
$\log{(T_{\rm min})}$ & $[3.7, 3.9]$ & $ 3.76^{+0.02}_{-0.01}$ & K  \\
$\log{(\sigma)} $ & $[-4, 2] $ & $ -0.61^{+0.04}_{-0.03}$  &   \\
$v_{\rm {ej}}$ & $-$ & $6.9^{+1.7}_{-1.4}$ &  $10^{3}~$km s$^{-1}$ \\
$\kappa$ & 0.34 & - & cm$^2$\,g$^{-1}$ \\
$n$ & 12 & - & \\
$\delta$ & 0 & - & \\
$s$ & $2$ & - & \\
\hline
\hline
\end{tabular}
\end{table}

We used the publicly available Modular Open Source Fitter for Transients (\texttt{MOSFiT}\footnote{\url{https://mosfit.readthedocs.io/en/latest/index.html}}; \citealt{Guillochon2018ApJS..236....6G}) to fit the multi-band LCs of SN~2016iog. This tool accepts multi-band photometry as input and allows specifying prior distributions for the model parameters.
We employed the built-in model \texttt{csmni}, which incorporates luminosity contributions from both the radioactive decay of $^{56}$Ni and the additional luminosity due to CSM interaction. In this framework, a fraction of the kinetic energy from the SN ejecta is converted into radiative energy due to collisions with the CSM. The $^{56}$Ni decay model is based on \cite{Nadyozhin1994ApJS...92..527N}, while the CSM interaction model follows the semi-analytic approach outlined by \cite{Chatzopoulos2013ApJ...773...76C}.
The model is built upon the following physical assumptions: the onset of CSM interaction is determined by the time $t_{\rm int} = \frac{R_0}{v_{\rm ej}}$, where $R_0$ is the inner radius of the CSM shell and $v_{\rm ej}$  denotes the average velocity of the SN ejecta. Assuming that $v_{\rm ej}$ represents the mean photospheric velocity of the ejecta, it is derived from the free parameters $M_{\rm ej}$ (the ejecta mass) and $E_{\rm k}$ (the kinetic energy of the ejecta), under the assumption of constant density, using the relation $E_{\rm k} \approx \frac{3}{10} M_{\rm ej} v_{\rm ej}^2$ \citep{Arnett1982ApJ...253..785A}.
The model incorporates 10 free parameters: the mass fraction of $^{56}$Ni ($f_{\rm Ni} \equiv \frac{M_{\rm Ni}}{M_{\rm ej}}$), the $\gamma$-ray opacity ($\kappa_\gamma$),  the kinetic energy of the ejecta ($E_{\rm k}$), the CSM shell mass ($M_{\rm CSM}$), the total ejecta mass ($M_{\rm ej}$), the hydrogen column density of the host galaxy ($n_{\rm H,\mathrm{host}}$), the inner radius of the CSM shell ($R_0$), the density of the CSM at $R_0$ ($\rho_0$), the minimum temperature ($T_{\rm min}$) reached by the expanding and cooling photosphere,  and a white-noise variance term ($\sigma$) to account for additional uncertainties (in magnitudes) ensuring that the reduced $\chi^2 = 1$.
A power-law density profile for the CSM shell is assumed, expressed as $\rho(r) = q r^{-s}$, where $q = \rho_0 R_0^s$ \citep{Chatzopoulos2012ApJ...746..121C}. The power-law index is set to $s = 2$, which corresponds to a steady-wind CSM model \citep{Chevalier2011ApJ...729L...6C}. Additionally, three parameters are fixed: the Thomson scattering opacity ($\kappa = 0.34 \, \text{cm}^2 \, \text{g}^{-1}$), a typical value for H-rich ejecta, as suggested by \cite{Nagy2018ApJ...862..143N}, and the density power-law parameters for the inner ($\rho_{\rm ej} \propto r^{-\delta}$) and outer ($\rho_{\rm ej} \propto r^{-n}$) ejecta, with $\delta = 0$ and $n = 12$, respectively, typical for H-rich ejecta \citep{Chatzopoulos2013ApJ...773...76C}.

We adopt simple uniform or log-uniform prior distributions for all free parameters involved in the model. The comparison between the \texttt{MOSFiT} model LC and the observed data of SN~2016iog is shown in Fig.~\ref{fig:mosfit}, along with the best-fit parameters listed in Table~\ref{tab:MOSFiT}. 
The corner plot, which shows the posterior probability distributions and the correlations between the parameters from the model fits, is presented in Appendix~\ref{appendix:corner}, as depicted in Fig.~\ref{fig:modfit_corner}.
The model successfully reproduces the multi-band LCs of SN~2016iog, with the fitting converging and the parameters well constrained. Some key explosion parameters include: $M(^{56}Ni)$ = $0.0139^{+0.012}_{-0.009}$ M$_\odot$, which is consistent with the $^{56}$Ni mass of SN~2016iog derived by comparing its tail luminosity with that of SN~1987A;  $M_{\rm ej}$ = $3.691^{+1.265}_{-1.092}$ M$_\odot$. 
Nevertheless, the lower mass is consistent with the rapid decline observed in the LC and the absence of typical metal lines in the spectra of SN~2016iog (see Sect. \ref{section:spectrasequence}). Regarding the CSM, the key parameters are: $M_{\rm CSM}$ = $0.060^{+0.022}_{-0.020}$ M$_\odot$, with an inner radius $R_{\rm 0} = 4.57^{+1.82}_{-1.70}$ AU ($\sim$ 983 R$_\odot$) and a density ${\rho_{\rm 0} = 9.51^{+15.95}_{-5.83}\times10^{-11}}$ g cm$^{-3}$.  To reproduce the high luminosity and rapid evolution of SN~2016iog in the context of Type II SNe, the model favoured a low-density, low-mass CSM surrounding the progenitor star, which was struck by the low-mass ejecta. 

\subsection{Correlations of parameters}
\label{sect:para}

\begin{figure*}[ht]
    \centering
    \includegraphics[width=\textwidth]{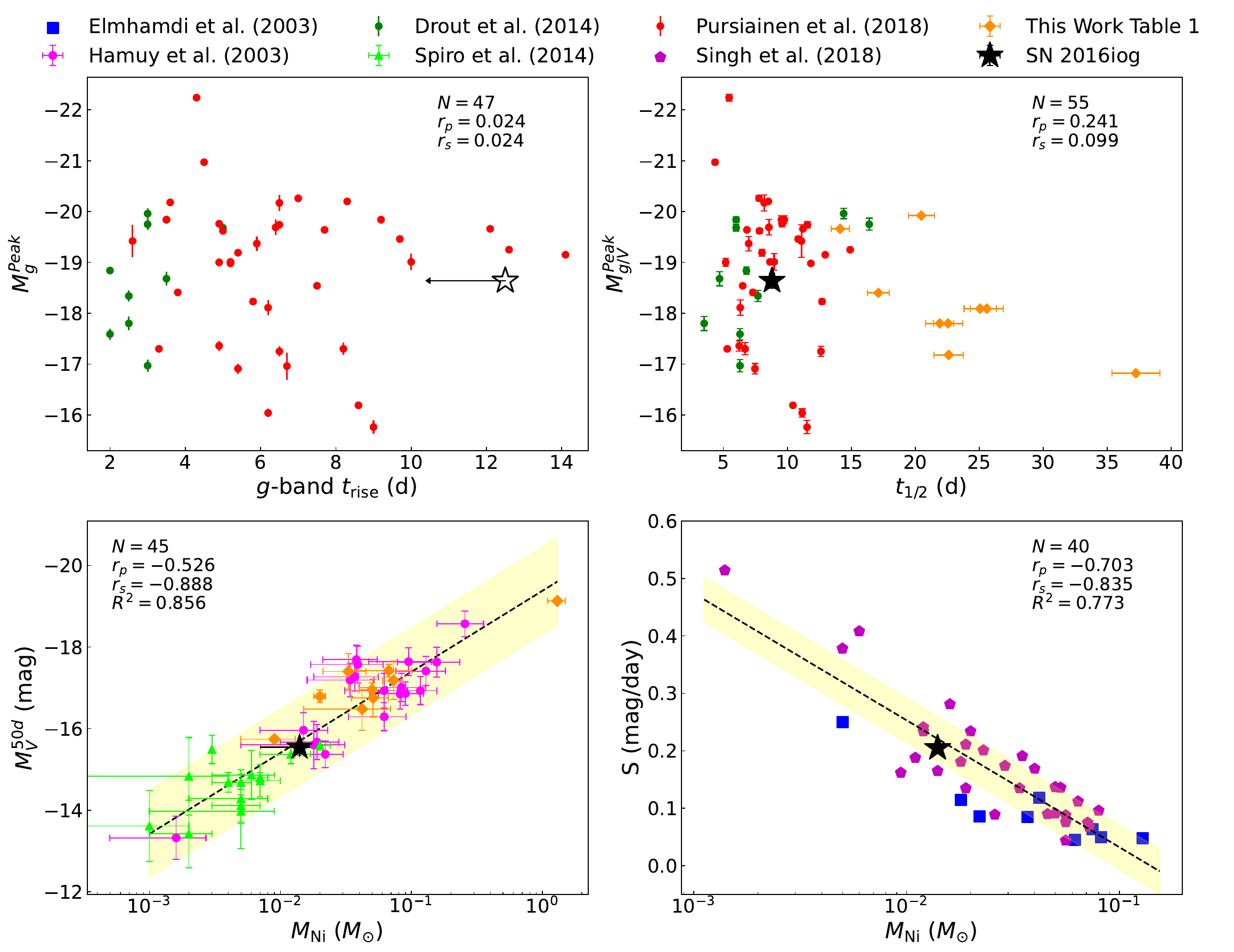} 
    \caption{Correlations between parameters of Type II SNe. 
    \textit{Top left}:  Absolute $g$-band peak magnitude versus $g$-band rise time, with the first $g$-band point of SN~2016iog taken as a upper limit owing to the lack of rising-phase data; 
    \textit{Top right}: Absolute $g$/$V$-band peak magnitude versus the time above half-maximum ($t_{1/2}$), with the data from \cite{Drout2014ApJ...794...23D} in the $g$-band, and those from \cite{Pursiainen2018MNRAS.481..894P}, this work Table~\ref{tab:s2}, and SN~2016iog in the $V$-band.; 
    \textit{Bottom left}: The absolute \(V\)-band magnitude at \(t = 50\) days ($M^{50d}_{V}$) versus the ejected \(^{56}\)Ni masses (M$_{\text{Ni}}$); 
    \textit{Bottom right}: The $V$-band LC maximum decline rate (steepness parameter, $S$)   versus the ejected \(^{56}\)Ni masses (M$_{\text{Ni}}$).  The weighted Pearson correlation coefficient ($r_p$), the Spearman rank correlation coefficient ($r_s$), and the number of SNe (N), are provided for each parameters pair. The black dashed line represents the linear relationship between the two parameters, and the yellow region indicates the error range. \(R^2\) represents the coefficient of determination. }
    \label{fig:combined}
\end{figure*}

Comparing the properties of SN~2016iog with those of other SNe helps to characterize its physical properties more precisely.
\citet{Hamuy2003ApJ...582..905H, Spiro14} reported a correlation between the absolute $V$-band magnitude at $t = 50$~days ($M_V^{\text{50d}}$) and the $^{56}$Ni mass ($M_{Ni}$), while \citet{Elmhamdi2002PhotometryAS, Singh2018MNRAS.480.2475S} noted a correlation between the steepness parameter ($S$) and the $^{56}$Ni mass.  The $S$ is defined as the maximum decline rate of the $V$-band LC during the transition phase: $S = -{dM_V}/{dt}$  \citep{Elmhamdi2002PhotometryAS,Singh2018MNRAS.480.2475S}. From the calculation, we obtain $S_{\mathrm{SN2016iog}} = 0.205$ mag/d. 
\citet{Drout2014ApJ...794...23D, Pursiainen2018MNRAS.481..894P} discussed numerous rapidly evolving and highly luminous transients, some of which show similarities to SN~2016iog. We compared the parameters of absolute $g/V$-band peak magnitude ($M_{g/V}^{Peak}$), $g$-band rise time, and the time above half-maximum ($t_{1/2}$). 

Specifically, we show the absolute $g$-band peak magnitude as a function of the $g$-band rise-to-peak time (for SN~2016iog, the first detection is taken as an upper limit on the rise time and a lower limit on the peak luminosity due to the lack of early $g$-band data), the absolute $g/V$-band peak magnitude against the time above half-maximum, and both the absolute $V$-band magnitude at $t=50$ days and the steepness parameter in relation to the ejected $^{56}$Ni mass, as illustrated in Fig.~\ref{fig:combined}.
To assess the pairwise correlations among these parameters, we calculated the weighted Pearson coefficient ($r_p$) and the Spearman rank coefficient ($r_s$), as shown in Fig.~\ref{fig:combined}.

In the upper-left panel, there is no obvious correlation between the absolute $g$-band peak magnitude and the $g$-band rise-to-peak time. SN~2016iog is most similar to a subset of the rapidly evolving and highly luminous transients presented by \cite{Pursiainen2018MNRAS.481..894P}, some unclassified events of which may correspond to H-rich SNe similar to SN~2016iog. In the upper-right panel, no clear correlation is apparent between $M_{g/V}^{\rm Peak}$ and $t_{1/2}$. For SN 2016iog, we calculate $t_{1/2} = 8.8~\rm d$. SN~2016iog lies within the range of events presented by \cite{Drout2014ApJ...794...23D, Pursiainen2018MNRAS.481..894P}, suggesting that a subset of these events may share similar physical mechanisms with SN~2016iog.

\begin{figure*}[ht]
    \centering
    \includegraphics[width=0.8\textwidth]{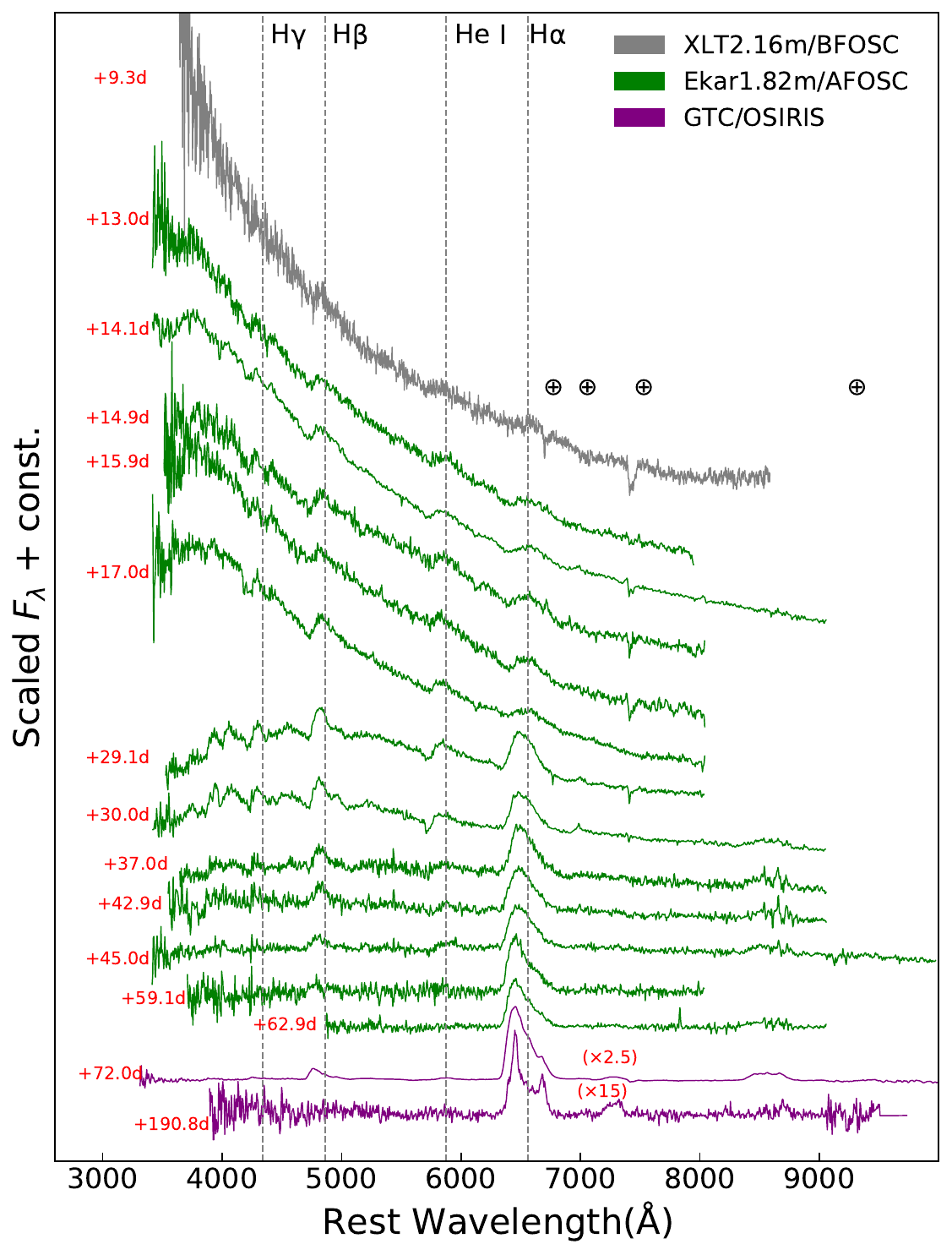} 
    \caption{Optical spectral evolution of SN 2016iog from +9.3 days to +190.8 days since the explosion. The spectra have been corrected for reddening and redshift, and vertically shifted for better visualization. The last two spectra have been amplified by different intensity factors to make them more prominent. The epochs used are indicated to the left of each spectrum. Different colours are employed to distinguish between the various telescopes, with the corresponding telescope labels shown in the upper right corner. The positions of major telluric absorption lines are denoted by the $\bigoplus$ symbol.}
    \label{fig:sp_ev}
\end{figure*}

The lower-left panel shows the absolute \(V\)-band magnitude at \(t = 50\) days as a function of the \(^{56}\)Ni mass for the SNe~II sample, based on the data from \cite{Hamuy2003ApJ...582..905H}, \cite{Spiro14}, and other SNe reported in Table~\ref{tab:s2}. We perform a linear regression on $\log(M_{\mathrm{Ni}})$ and $M^{50d}_{V}$, resulting in the following equation:
\begin{align}
\label{eq:2}
    M_V^{\text{50d}} = -1.988 \log(M_{\mathrm{Ni}}) - 19.380
\end{align}
with \(R^2 = 0.856\), where \(R^2\) represents the coefficient of determination, indicating that approximately 85.6\% of the variance in \(M_V^{\text{50d}}\) is explained by $\log(M_{{Ni}})$, showing a strong correlation (R = 0.925).
This indicates that SN 2016iog follows the typical core-collapse (CC) SNe energy mechanism driven by  radioactive decay.
In the lower-right panel, the estimated \(^{56}\)Ni mass is plotted vs. the steepness parameter (\(S\)). 
We perform a linear regression on $\log(M_{\mathrm{Ni}})$ and \(S\), resulting in the following equation:
\begin{align}
\label{eq:3}
    S = -0.221 \log(M_{\mathrm{Ni}}) - 0.188
\end{align}
with \(R^2 = 0.773\),  where \(R^2\) indicates that approximately 77.3\% of the variance in \(S\) is explained by $\log(M_{{Ni}})$, showing a strong correlation (R = 0.879). These equations show that the \(^{56}\)Ni mass of SN 2016iog are consistent with those of typical Type II SNe, indicating the reliability of our \(^{56}\)Ni mass estimation.

\section{Spectroscopy}
\label{section:spect}
\subsection{Spectral sequence}
\label{section:spectrasequence}
Our spectroscopic observations of SN 2016iog span a period from +9.3 days to +72.0 days after the explosion, followed by a large observational gap, before a final spectrum obtained at a phase of +190.8 days after the explosion. 
Details of the spectroscopic observations are provided in Table~\ref{tab:spec} in the Appendix \ref{SpecInfo}, while our sequence of spectra for SN 2016iog is shown in Fig.~\ref{fig:sp_ev}. Our earliest spectrum (+9.3 days) is nearly featureless. 

Classical high-ionization features such as \ion{C}{IV}, \ion{N}{IV}, \ion{N}{III} and \ion{He}{II}, are not unequivocally detected in our early spectrum of SN~2016iog. 
The lack of high-ionization lines in our +9.3 day spectrum of SN 2016iog is not surprising given it was probably taken too late to exhibit these early-stage features.
Weak emission features attributed to H$\alpha$ and H$\beta$ are observed in this spectrum of SN 2016iog. 
After $\sim$ 10 days, broad and shallow spectral line profiles begin to form and become increasingly prominent as the continuous spectrum cools.
From +13.0 days to +30.0 days, broad P-Cygni profiles are observed for H$\beta$ $\lambda$4861 and \ion{He}{I} $\lambda$5876.
Due to the spectral line broadening effect, the P-Cygni profile of \ion{He}{I} $\lambda$5876 may be blended with the \ion{Na}{I} doublet $\lambda\lambda$ 5890, 5896. In the spectra at +29.1 days and +30.0 days, we also observed a prominent \ion{He}{I} $\lambda$7065 emission line.
While the H$\alpha$ emission line is prominent, its broad absorption counterpart is barely distinguished, and it is less prominent than what is usually observed in Type II SNe. 
This weaker H$\alpha$ absorption profile may indicate the presence of CSM surrounding SN~2016iog \citep{Hillier2019A&A...631A...8H}, and is also associated with the faster luminosity decline observed in Type II SNe \citep{Guti2014ApJ...786L..15G},  with similar weak H$\alpha$ absorption detected in both SN~1998S \citep{Pozzo2004MNRAS.352..457P, Mauerhan2012MNRAS.424.2659M, Luc2025} and PTF11iqb \citep{Smith2015MNRAS.449.1876S}, which exhibit a prominent multi-component H$\alpha$ at late times.

The following five spectra, from $+$37.0 days to $+$62.9 days, exhibit a broad emission of the H$\alpha$ line, and the absorption component is no longer visible.  A detailed discussion can be found in Sect. \ref{Spectra}.
The Balmer lines also exhibit a broad emission spectrum, while the metal lines are very weak or even non-existent, such as \ion{Sc}{II} $\lambda$5527 and \ion{Fe}{II} $\lambda$5169. 
At $+$62.9 days, a double-peak structure begins to emerge in the H$\alpha$ emission line, resembling that of SN 1998S. At +72.0 days, the spectrum primarily shows H$\alpha$ and H$\beta$, with the \ion{Ca}{II} NIR  $\lambda\lambda\lambda$ 8498, 8542, 8662  being very weak. In the last spectrum we observed at $+$190.8 days, H$\alpha$ clearly presents an asymmetric double-peak structure. In addition, an asymmetric emission feature near 7300 \AA\ is observed in the emission line, which is attributed to [\ion{Ca}{II}] $\lambda\lambda$7291,7323.

The temperature was derived by fitting the continuum of each spectrum to a blackbody function. The temperature measurement results for SN~2016iog are summarised in Table \ref{tab:specT}, and the complete temperature evolution is presented in the upper panel of Fig. \ref{fig:lin_temp}, where we compare it with other representative Type II SNe. 
Fig.~\ref{fig:lin_temp} also shows the results for SN~2016iog based on photometric blackbody fits.
The error estimates were computed using the Monte Carlo method, by repeating the fitting process 100 times, with the errors considered as the standard deviation of the fitting parameters.
In the early phases, the initial temperature of most SNe is close to 20,000 K. At +9.3 days, the temperature of SN 2016iog is $T_{\mathrm{bb}}$ = 18,100 $\pm$ 2,180 K, which is similar to the temperature of SN 1998S at +11.3 days ($T_{\mathrm{bb}} \sim 18,000$ K). 
Subsequently, the temperature of SN 2016iog rapidly decreased, from $T_{\mathrm{bb}} = 11,290 \pm 670$ K at +13.0 days  to $T_{\mathrm{bb}} = 5,720 \pm 540$ K at +37.0 days, with the temperature of SN 2016iog during this phase being similar to that of SNe 1999em, 2014G, and 2019nyk.
Between +42.9 d and +45.0~d, the blackbody temperature of SN 2016iog remained stable at $T_{\mathrm{bb}} = 5,940 \pm 500$ K and $T_{\mathrm{bb}} = 5,510 \pm 620$ K respectively, maintaining a plateau around 5,500 K that aligns with SNe 1999em, 2019nyk.
However, SNe 1998S, 2014G, and ASAS-SN15nx are hotter  with $T_{\mathrm{bb}}\sim$ 6,500 K than SN 2016iog, while SNe 2009kr and 2013ej are colder with $T_{\mathrm{bb}}\sim$ 4,000 K.
This difference may be related to the envelope mass, the explosion energy, and the degree of interaction with the CSM for each SN. For example, events with strong CSM interaction such as SN~1998S usually exhibit higher temperatures at this stage, whereas low-energy explosions (e.g. SNe~2009kr and 2013ej) tend to show lower temperatures. 
The blackbody radiation fit is only applicable during the photospheric phase. During the nebular phase, when emission lines dominate the SED, the blackbody fit loses its physical relevance.  Therefore, we did not measure the temperature from the spectrum of SN~2016iog at phases later than  +45.0 days.

\subsection{Line velocity} 
\begin{figure}
    \includegraphics[width=\columnwidth]{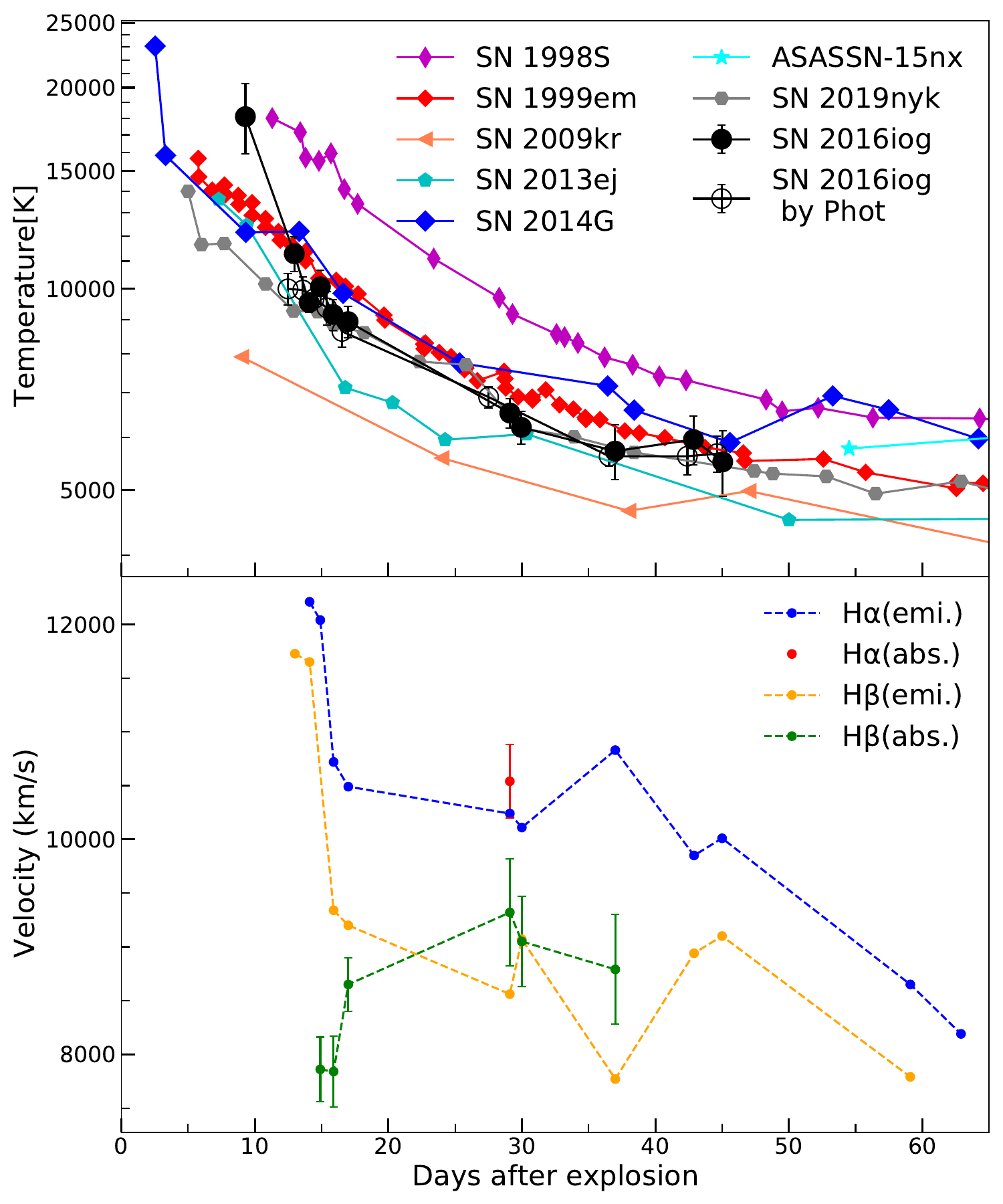}
    \caption{Spectroscopic measurements of SN 2016iog compared with those of other Type II SNe. \textit{Top panel}: The temperature evolution of SN 2016iog, compared with other SNe. The black solid dots represent the temperatures of SN~2016iog derived from spectroscopic blackbody fits, while the black hollow dots represent the temperatures from blackbody fits to the photometry. The values reported for SN 2009kr were derived from the spectra presented in \cite{Elias2010ApJ...714L.254E}, those for ASASSN-15nx were obtained from the spectra shown in \cite{Bose2018ASASSN15nxAL}, while for the other SNe, the values from the literature were used.
    \textit{Bottom panel}: The velocity evolution of the Balmer lines of SN 2016iog. The suffix abs. denotes velocities measured from the minimum of the H absorption trough, while emi. refers to velocities derived from the FWHM of the emission component. }
\label{fig:lin_temp}
\end{figure}

\begin{figure}
    \includegraphics[width=0.95\columnwidth]{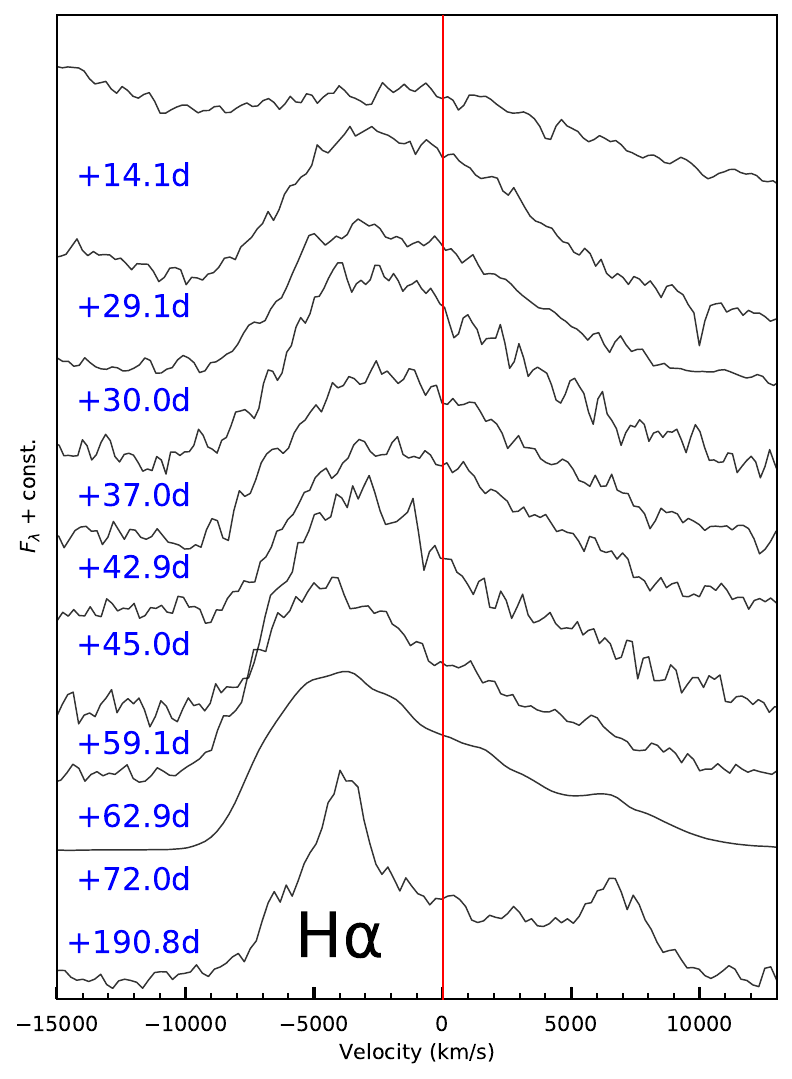}
    \caption{Evolution of the \(\mathrm{H\alpha}\) profiles in velocity space for SN 2016iog.}
\label{fig:Ha_ev}
\end{figure}

\begin{table}
\caption{Spectroscopic measurements of the blackbody temperature and the H line velocities of SN 2016iog.}
\label{tab:specT}
\centering
\resizebox{\columnwidth}{!}{
\begin{tabular}{cccccc}
\hline \hline 
Date&Phase$^a$ &  Temp. & H$\alpha$ (emi.)  & H$\beta$ (emi.) & H$\beta$ (abs.)  \\
  &[days]&  [K] & [km s$^{-1}$] & [km s$^{-1}$] & [km s$^{-1}$] \\
\hline
2016-12-02 &+9.3 &  $18100\pm2180$ & $-$ & $-$ & $-$   \\
2016-12-06 & +13.0&  $11290\pm670$ & $-$ & $11730$ & $-$  \\
2016-12-07&+14.1 &  $9540\pm330$ & $12210$  & $11650$ & $-$   \\
2016-12-08&+14.9 &  $10070\pm600$ & $12040$ & $-$ & $ 7860\pm300$    \\
2016-12-09 &+15.9&  $9160\pm480$ & $10720$ & $9340$ & $7840\pm330$  \\
2016-12-10 & +17.0&  $8930\pm490$  & $10490$ & $9200$ & $8650\pm250$ \\
2016-12-22& +29.1&  $6520\pm330$ & $10240$$^b$ & $8560$ & $ 9320\pm500$  \\
2016-12-23 &+30.0&  $6210\pm350$ & $10110$ & $9070$ & $9050\pm420$  \\
2016-12-30 &+37.0 &  $5720\pm540$ & $10830$ & $7770$ & $8790\pm510$  \\
2017-01-04 &+42.9&  $5950\pm500$ & $9850$ & $8940$ & $-$  \\ 
2017-01-07& +45.0&  $5510\pm620$ & $10010$ & $9100$ & $-$ \\
2017-01-21 &+59.1 &  $-$ & $8650$ & $7790$ & $-$ \\
2017-01-25& +62.9&  $-$ &  $8190$ & $-$ & $-$ \\
\hline\hline \\ 
\end{tabular}
}
\begin{flushleft}
    $^a$ Phases are calculated relative to the explosion epoch (MJD = 57715.6) in the reference frame of the observer. \newline
    $^b$ The H$\alpha$ absorption line velocity at +29.1 days is $10540 \pm 340$ km s$^{-1}$.
    \end{flushleft}
\end{table}

\begin{figure}
    \includegraphics[width=\columnwidth]{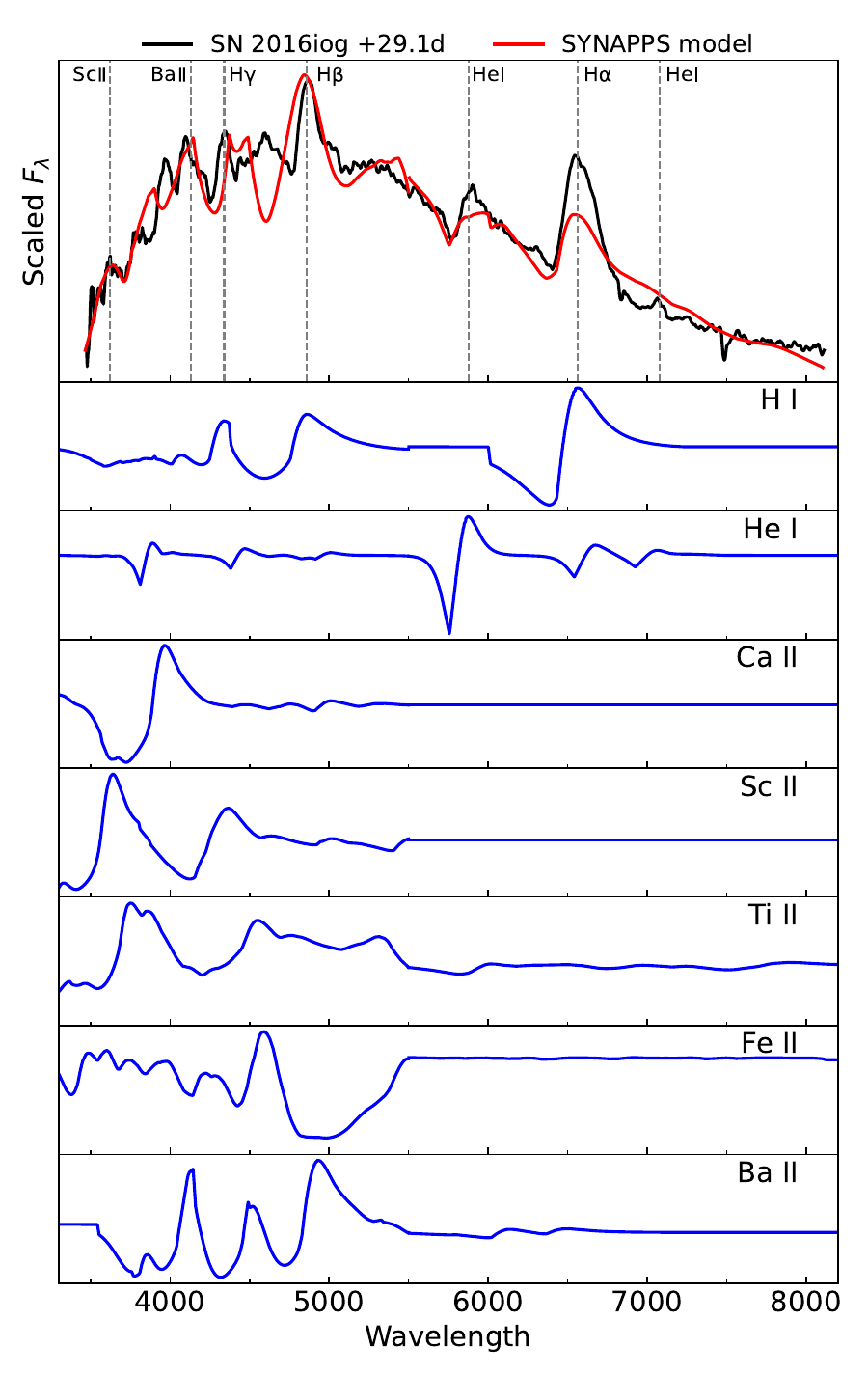}
    \caption{Best-fit SYNAPPS model to the +29.1d spectrum of SN 2016iog. The seven panels at the bottom display the individual contribution of each ion to the synthetic spectrum shown in the top panel.} 
\label{fig:syn}
\end{figure}

Since SN 2016iog only exhibits a small number of identifiable P-Cygni features in the spectral lines, we only measured the velocities for H$\alpha$ $\lambda$6563 and H$\beta$ $\lambda$4861.
We calculated the velocities of H$\alpha$ $\lambda$6563 and H$\beta$ $\lambda$4861 from the minima of their absorption features (abs.), with the uncertainties estimated using the Monte Carlo method. We measured the full width at half maximum (FWHM) of the emission components (emi.) of H$\alpha$ $\lambda$6563 and H$\beta$ $\lambda$4861 from the line profiles.
The velocity evolution of H$\alpha$ $\lambda$6563, H$\beta$ $\lambda$4861 is shown at the lower panal of Fig.~\ref{fig:lin_temp}, with the specific measurement results listed in Table~\ref{tab:specT}.

The velocity of the H$\beta$ line ($\lambda$4861) is 7,860~$\pm$~300 km~s$^{-1}$ at +14.9 days, increases to 9,320~$\pm$~500 km~s$^{-1}$ at +29.1 days, and then decreases to 8,790~$\pm$~510 km~s$^{-1}$ at +37.0 days. 
Due to the weaker H$\alpha$ $\lambda$6563 P-Cygni features, we only measured the velocity at +29.1 days, which is  $10,540 \pm 340$ km s$^{-1}$. The emission of both H$\alpha$ and H$\beta$ is more pronounced, and the velocity trends derived from the FWHM are also consistent.  
The H$\alpha$ emission velocity declined from 12,210 km s$^{-1}$ at +14.1 days to 10,110 km s$^{-1}$ at +30.0 days, exhibited a sudden increase to 10,830 km s$^{-1}$ at +37.0 days, and then decreased to 8,190 km s$^{-1}$ by +62.9 days. Similarly, the H$\beta$ emission velocity dropped from 11,730 km s$^{-1}$ at +13.0 days to 7,770 km s$^{-1}$ at +37.0 days, rose again to 9,100 km s$^{-1}$ at +45.0 days, and subsequently declined to 7,790 km s$^{-1}$ at +59.1 days.

The evolution of the H$\alpha$  profiles (with velocities in the rest frame) is presented in detail in Fig.~\ref{fig:Ha_ev}. 
During the +14.1 day spectrum, the H$\alpha$ emission shows a blueshift of approximately $-$3,000 km s$^{-1}$. 
In the early spectra of Type II SNe, a blueshifted and broad H$\alpha$ line is expected, as the redshifted receding side of the line is obscured by the optically thick hydrogen envelope \citep{Anderson2014bMNRAS.441..671A}. 
In subsequent spectra of SN 2016iog, the H$\alpha$ profile consistently shows a similar blueshift of approximately $-$4,000 km s$^{-1}$, and gradually begins to exhibit asymmetry.
After the photospheric phase ends, the peak of the H$\alpha$ line is expected to move to zero velocity \citep{Anderson2014bMNRAS.441..671A}. 
However, in the +190.8 day spectrum of SN 2016iog, the asymmetry of the H$\alpha$ profile becomes more pronounced, ultimately resulting in a double-peaked profile in the final spectrum. 
This may be due to either the presence of dust or asymmetric CSM around SN 2016iog, with further details discussed in Sect.~\ref{section:190.8} and Sect.~\ref{sect:PTF}, respectively.

To determine the elements responsible for the photospheric spectrum of SN 2016iog, we generated a synthetic spectrum similar to the one observed at +29.1 days using SYNAPPS\footnote{\url{https://github.com/rcthomas/es/}} \citep{SYN, Thomas2013ascl.soft08008T}. 
The spectrum from +29.1 days was selected for modelling because it has a relatively high S/N and clearly displays many strong features. 
The upper panel of Fig.~\ref{fig:syn} compares the synthetic spectra generated with SYNAPPS to the observed spectra, while the seven lower panels sequentially activate individual ions to illustrate the contribution of each of the seven ions used in the modeling. The modeling indicates that the blue part of the spectrum ($< 5000$~\AA) is primarily dominated by  \ion{H}{I}, \ion{Ca}{II}, \ion{Sc}{II}, \ion{Ti}{II}, \ion{Fe}{II}, and \ion{Ba}{II}.
The dip in the $7450 - 7750$ \AA\ region is due to a residual telluric absorption feature.

\subsection{Comparison of Type II SNe spectra}
\label{Spectra}

\begin{figure}
    \includegraphics[width=\columnwidth]{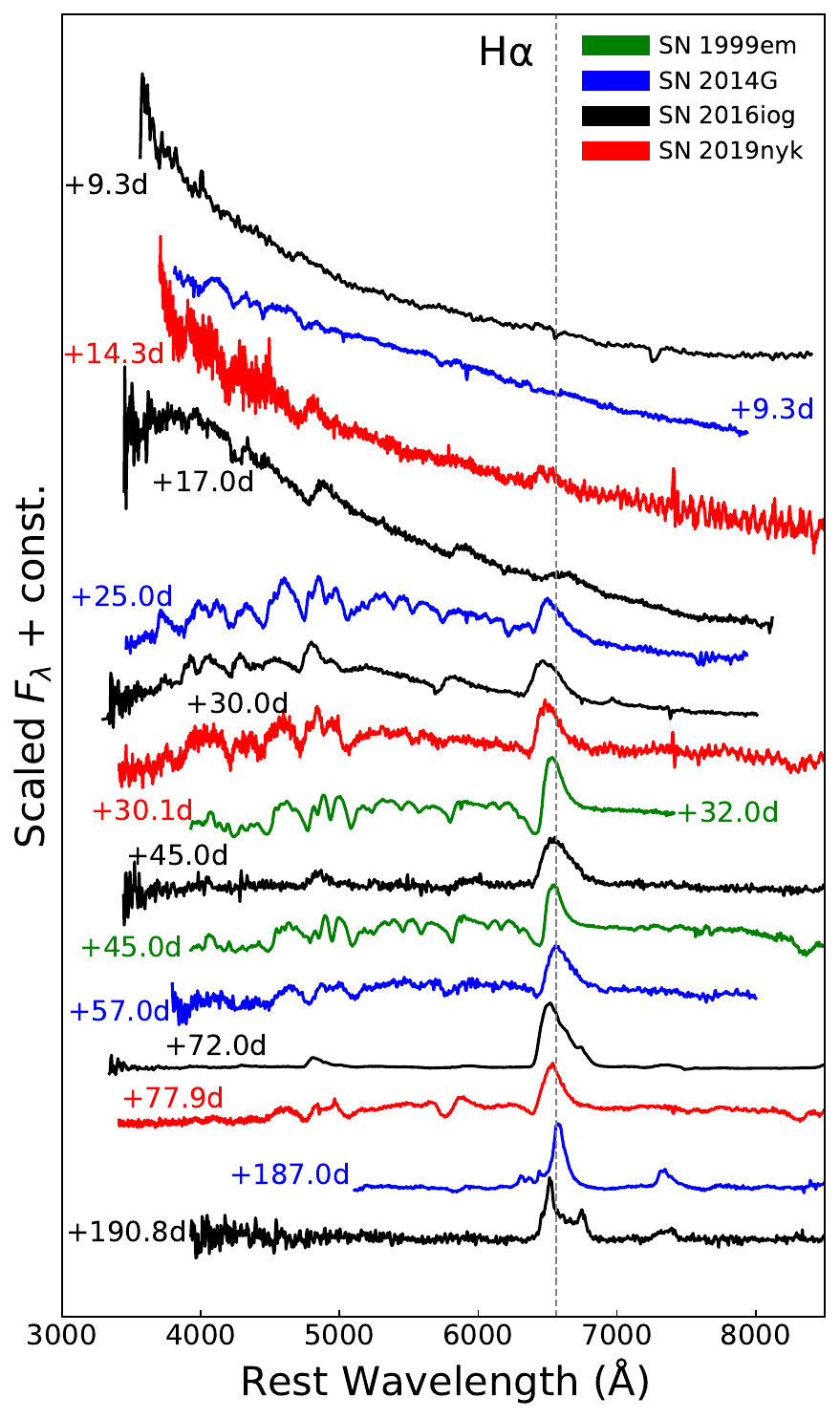}
    \caption{Comparison of the spectral evolution of SN~2016iog with Type~II SNe exhibiting different light-curve decline rates.}
    \label{fig:spec_evl}
\end{figure}

\begin{figure*}[ht]
    \centering
    \includegraphics[width=0.98\textwidth]{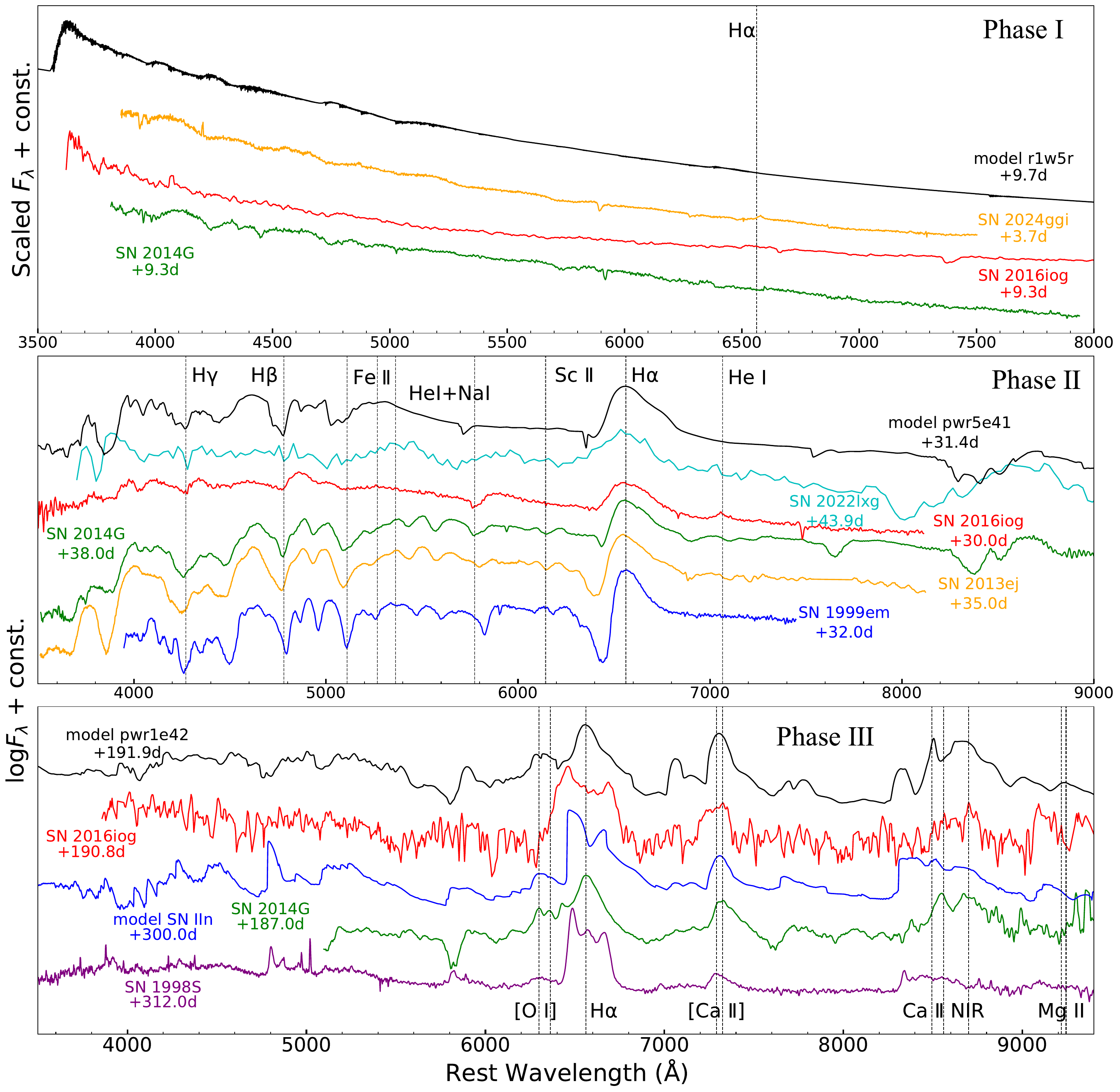} 
    \caption{Comparison of the spectra of SN 2016iog at different phases with those of other Type II SNe and model spectra.
    \textit{Top panel}: The early-phase spectra of SN 2016iog at +9.3 days are compared with those of SNe 2014G, 2024ggi, and the \texttt{r1w5r} model at similar phases.
    \textit{Middle panel}: The recombination-phase spectra of SN 2016iog at +30.0 days are compared with those of SNe 1999em, 2013ej, 2014G, 2022lxg and the  \texttt{pwr5e41} model at similar phases.
    \textit{Bottom panel}: The nebular-phase spectra of SN 2016iog at +190.8 days are compared with those of SNe 1998S, 2014G, the \texttt{SN IIn} model, and the \texttt{pwr1e42} model at similar phases. All observed spectra are in the rest frame and have been corrected for reddening. The spectra have been scaled for improved comparison. }
    \label{fig:spectrum_comparison}
\end{figure*}

\begin{figure*}[ht]
    \centering
    \includegraphics[width=0.98\textwidth]{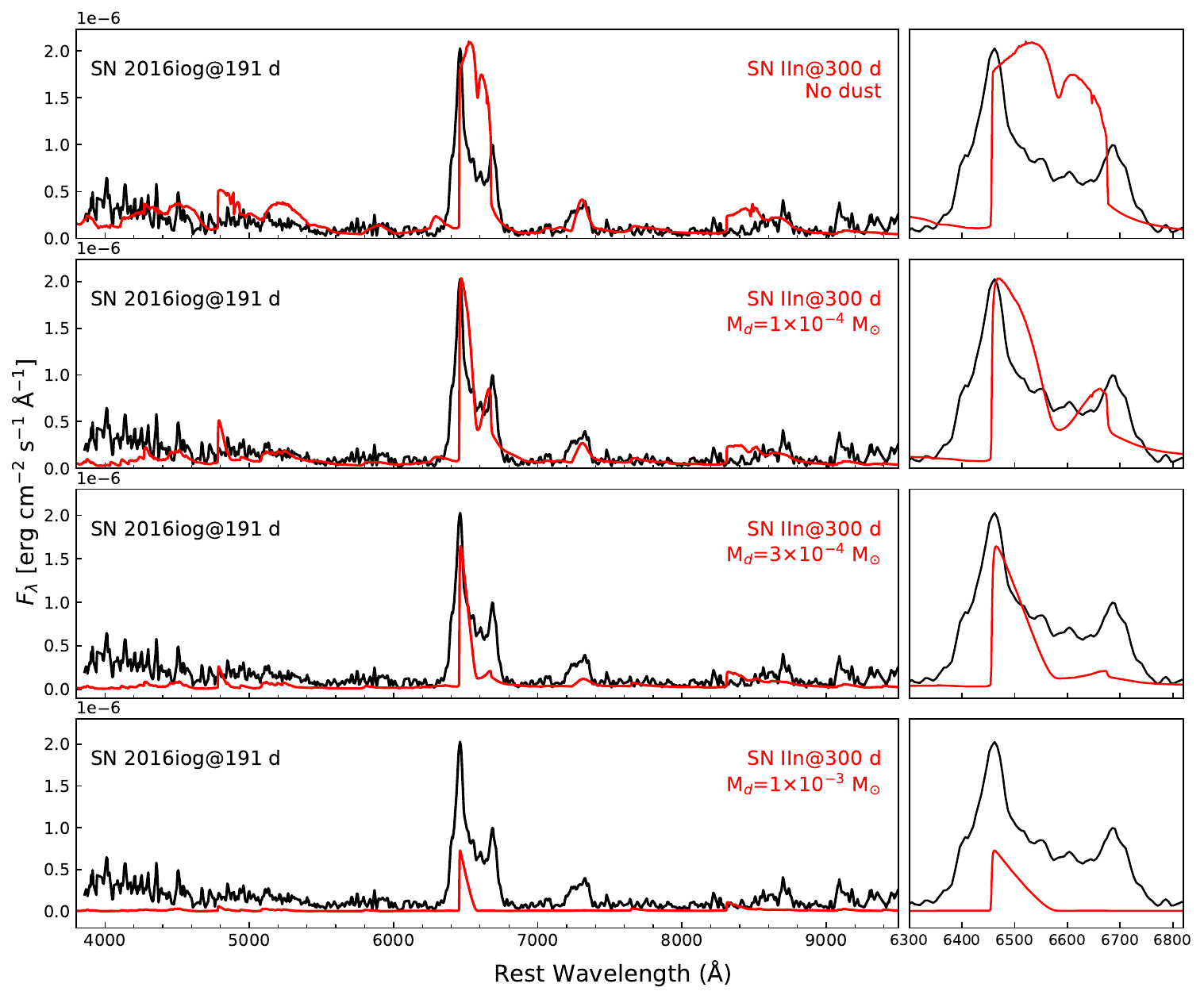} 
    \caption{Comparison of the spectra of SN 2016iog at +191 days with the \texttt{SN IIn} model spectrum for different dust mass scenarios. Each subplot on the  shows a zoom-in of the 6300-6800~\AA~region, focussing on the H$\alpha$ line details. The spectrum of SN 2016iog has been corrected for redshift and extinction, and scaled for improved comparison.}
    \label{fig:dust}
\end{figure*}

SN~2016iog exhibits an unusually high decline rate among Type~II SNe. To explore the similarities and differences in spectral evolution across SNe with varying decline rates, we compare it with Type~II SNe 1999em ($s_2 = 0.3 \pm 0.02$ mag/100d), 2014G ($s_2 = 2.87 \pm 0.05$ mag/100d), and 2019nyk ($s_2 = 2.84 \pm 0.03$ mag/100d). The spectral evolution of these SNe in comparison with SN~2016iog is shown in Fig.~\ref{fig:spec_evl}. We find that, at roughly the same epochs, SN~2016iog exhibits weaker metal lines and H$\alpha$ absorption compared to the other SNe. We also performed a comparison of different models with observed spectra of SN~2016iog and the Type~II SNe introduced above during the early, recombination, and nebular phases, as shown in Fig.~\ref{fig:spectrum_comparison}.
The model spectra used in our comparison were non-local thermodynamic equilibrium radiative-transfer simulations computed using \texttt{CMFGEN} \citep{Hillier2012MNRAS.424..252H}. Some of the comparison spectra are obtained from WISeREP\footnote{\url{https://www.wiserep.org/}} \citep{Yaron2012PASP..124..668Y, Goldwasser2022TNSAN.191....1G}.

Due to observational limitations, earlier spectra were not obtained. Therefore, the +9.3 day spectrum of SN 2016iog is compared with those of other Type II SNe observed at similar phases, including SNe 2014G, 2024ggi, and the \texttt{r1w5r} model \citep{Luc2017}, as shown in the top panel of Fig.~\ref{fig:spectrum_comparison}.
The \texttt{r1w5r} model has the following parameters: \( R = 501 \, R_\odot \), a mass-loss rate of \( 5 \times 10^{-3} \, M_\odot \, \text{yr}^{-1} \), and a transition to a lower mass-loss rate of \( 10^{-6} \, M_\odot \, \text{yr}^{-1} \) at a distance of \( 2 \times 10^{14} \, \text{cm} \). This \texttt{r1w5r} model represents the influence of CSM interaction on the spectral profile.
SNe~2014G and 2024ggi both exhibited flash spectra at very early times, indicating interaction, and also showed nearly featureless spectra similar to that of SN~2016iog at +9.3\,days.
After a period of evolution, as seen in SNe 2014G, 2023ixf, and 2024ggi, interaction between the SN and the surrounding medium leads to a gradual evolution in spectral features. As the explosion continues, the SN shock slows down as it propagates outwards, hindered by the surrounding CSM. The interaction between the ejecta and the CSM causes the ejecta to decelerate due to the resistance from the medium, while the medium is accelerated by the ejecta, ultimately forming a cool dense shell (CDS) structure. After the shock has swept through all the dense CSM, the spectrum originates entirely from the CDS. This moment, the CDS possesses a high optical depth, effectively blocking radiation from the inner ejecta \citep{Luc2017, Luc2024interactingsupernovae}. The nearly featureless spectrum observed in SN 2016iog at +9.3 days likely forms within this shell. Similarly, SNe 2014G, 2024ggi, and the \texttt{r1w5r} model, which are known to have early interactions, also nearly featureless spectra at similar phases.

In the middle panel of Fig. \ref{fig:spectrum_comparison}, we present the spectrum of SN 2016iog obtained at +30.0 days during the recombination phase, and compare it with spectra from SNe 1999em, 2013ej, 2014G, 2022lxg, and the \texttt{pwr5e41} model \citep{Luc2022} at similar phases.
The \texttt{pwr5e41} model has the following parameters: an initial mass of 15 M$_\odot$, kinetic energy of \( 1.3 \times 10^{51} \, \text{erg} \), 0.03 M$_\odot$ of \(^{56}\)Ni, shock power of \( 5 \times 10^{41} \, \text{erg\,s}^{-1} \), and a mass-loss rate of approximately \( 3.1 \times 10^{-4} \, M_{\odot}\,\text{yr}^{-1} \).
The \texttt{pwr5e41} model suggests that, under a certain shock power, the H$\alpha$ absorption is filled, leading to the appearance of a broad H$\alpha$ profile.
The sequence from SN 1999em to SN 2016iog shows a P-Cygni absorption of H$\alpha$, which progressively less pronounced. As proposed by \cite{Guti2014ApJ...786L..15G}, the weaker absorption for SN 2016iog may correspond to a lower envelope mass, a steeper density gradient, and the presence of CSM interaction, with additional emission from the collision between the ejecta and the CSM filling in the absorption.

As described in \cite{Guti2014ApJ...786L..15G}, we can estimate the equivalent width ratio ($a/e$) of the absorption to emission components in the H$\alpha$ P-Cygni profile.
SNe with a smaller $a/e$ ratio  typically exhibit a faster decline in the LC (larger $s_1$, $s_2$, $s_3$), exemplified by  SNe 2016iog ($a/e$ $\sim$ 0); 2014G ($\sim$ 0.2); 2013ej ($\sim$ 0.4); 1999em ($\sim$ 0.8). Specifically, $s_2$ for SN 2016iog is approximately $8.85~\text{mag}/100\text{d}$, which is greater than those estimated for SN 2014G ($s_2$ $\sim 2.87~\text{mag}/100\text{d}$), SN 2013ej ($s_2$ $\sim 2.04~\text{mag}/100\text{d}$) and the classical Type IIP SN 1999em ($s_2$ $\sim 0.30~\text{mag}/100\text{d}$). These SNe with a smaller $a/e$ ratio also exhibit a higher maximum luminosity: the peak luminosity of SN 2016iog is approximately $6.31 \times 10^{42} \, \mathrm{erg\,s^{-1}}$, which is greater than the luminosities of SN 2014G ($\sim 4.14 \times 10^{42} \, \mathrm{erg\,s^{-1}}$), and SN 2013ej ($\sim 2.89 \times 10^{42} \, \mathrm{erg\,s^{-1}}$). 
The comparison shows that the spectrum of SN~2016iog exhibits only a weak P-Cygni absorption feature in H$\alpha$. According to \citet{Luc2022}, a strong CSM interaction associated with higher shock power can suppress the formation of H$\alpha$ absorption features.
This is consistent with the H$\alpha$ profile observed in the +30.0 day spectrum of SN~2016iog.
The \texttt{pwr5e41} model exhibits stronger emission below 5000~\AA ~compared to SN~2016iog, and this difference may be related to variations in shock power.
Apart from this discrepancy, the overall spectral match between SN~2016iog and the model is quite good.

We compare the nebular phase spectra of SN 2016iog with those of SNe 1998S, 2014G, the \texttt{pwr1e42} model \citep{Luc2022}, and the \texttt{SN IIn} model \citep{Luc2025}, as shown in the bottom panel of Fig.~\ref{fig:spectrum_comparison}. 
The \texttt{pwr1e42} model shares the same basic parameters as the \texttt{pwr5e41} model, with the only difference being a reduction in shock power to $1 \times 10^{41} \, \text{erg\,s}^{-1}$, which results in more typical Type II SN characteristics. The \texttt{SN IIn} model includes dust with a mass of $10^{-4} \, \text{M}_{\odot}$, and the detailed discussion of this model can be found in Sect. \ref{section:190.8}.
At this stage, the spectrum of SN 2016iog exhibits striking resemblance to that of SN 1998S, particularly in the weak strength of the [\ion{O}{I}]~$\lambda\lambda$6300, 6363 doublet and the presence of an asymmetric H$\alpha$ emission line, characterized by a stronger blue wing and a double-peaked profile. Furthermore, compared to SN 2014G, SN 2016iog exhibits a broader [\ion{Ca}{II}]~$\lambda\lambda$7291, 7323 emission line, but the \ion{Ca}{II} NIR $\lambda\lambda\lambda$8498, 8542, 8662 feature is relatively weaker.

SNe 2016iog, 1998S, and the \texttt{SN IIn} model all show asymmetric double peaks in H$\alpha$, in contrast to more typical Type II SNe, such as SN 2014G and the \texttt{pwr1e42} model, which exhibit a single symmetric peak.  
\cite{Dessart2025A&A...696L..12D} points out that invoking dust in a spherical model can reproduce the observations.
However, if dust is present in a spherically symmetric dense shell in the outer ejecta, it would generate a fully asymmetric spectral feature, affecting observations from any distant observer \citep{Luc2025}. In the absence of dust, the H$\alpha$ line shows a broad and symmetric `boxy' profile. If the shell is uniform and spherically symmetric, and there is no dust obscuration, the strengths of the redshift and blueshift components should be symmetric. Radiation from an optically thin shell moving at a constant velocity produces a box, flat-topped profile, where the radiation intensity in each velocity interval is independent of the projected velocity. The central dip arises from optical depth effects within the dense shell (see Sect. \ref{section:190.8}).
When dust is present, photons from the redshifted side (the direction away from the observer) must pass through more dust, resulting in significant attenuation, whereas photons from the blueshift side (the direction towards the observer) experience less attenuation due to shorter travel paths. As a result, the spectrum exhibits blue-red asymmetry. The larger the dust mass, the stronger the attenuation on the red side \citep{Jerkstrand2017hsn..book..795J, Luc2025}. This could explain the observed spectra of SN 2016iog. A detailed discussion can be found in Sect. \ref{section:190.8}.

We also propose that the asymmetric H$\alpha$ double peak observed in SN 2016iog results from its surrounding asymmetric CSM.
\cite{Jerkstrand2017hsn..book..795J} suggest that the two peaks of H$\alpha$ are caused by the geometry of the ejecta (e.g., a disk with a hole). Additionally, the asymmetric H$\alpha$ profile requires the ring to have a higher material density along certain directions. A detailed discussion can be found in Sect. \ref{sect:PTF}.

\subsection{Late time spectrum and dust formation}
\label{section:190.8}

Dust formed in the metal-rich ejecta can explain the systematic skewness in the emission line profiles  as well as the excessive optical light dimming observed in the late phases post-explosion \citep{Lucy1989LNP...350..164L}. By comparing the degree of skewness in the emission profiles with model predictions, we can infer the dust mass. The presence or absence of dust in a large sample of SNe can also be inferred from the excess infrared radiation \citep[e.g. SNe 1980K, 2004et, 2005ip, 2017eaw;~][]{Zsiros2023SerendipitousDO,  Shahbandeh2023MNRAS.523.6048S, Shahbandeh2025ApJ...985..262S}.
Theoretically, in standard Type II SNe, dust is expected to form suddenly in the metal-rich inner ejecta around 500 days after the explosion, primarily in the form of silicates, and it reaches a mass of about 0.01 M$_\odot$ after approximately 3 to 5 years \citep{Sarangi2022A&A...668A..57S}. For ejecta strongly interacting with the CSM, dust is expected to form about a year after the explosion, but at this point, it is located in the compressed dense shell formed at the interface between the ejecta and the CSM \citep{Sarangi2022ApJ...933...89S}. In interacting SNe, dust will eventually exist both in the inner ejecta and in the dense shell \citep{Luc2025}.
The dust in the inner ejecta has a minimal impact on the spectrum. When most dust resides in the low-velocity regions (v < 3,000 km s$^{-1}$) of the dense shell, it only slightly attenuates the intensity of metal lines such as [\ion{O}{I}] $\lambda\lambda$6300, 6364. At the 3,000 km s$^{-1}$ region, it begins to affect the H$\alpha$ profile. At the 5,000 km s$^{-1}$ region, the spectrum undergoes significant and pronounced wavelength-dependent changes. The flux at shorter wavelengths is strongly attenuated, while the continuous spectrum at longer wavelengths (e.g. $>8000$ \AA) is only weakly affected. However, the effect of dust on the H$\alpha$ profile is significant. As shown in Fig. \ref{fig:dust}, the intensities of [\ion{O}{I}] \(\lambda\lambda 6300, 6364\) and [\ion{Ca}{II}] \(\lambda\lambda 7291, 7323\) gradually decrease with increasing dust mass \citep{Luc2025}.

To explain the double-peaked structure and asymmetry observed in the H$\alpha$ profile of the SN~2016iog spectrum at +190.8 days, we compare different dust mass models at +300 days using the \texttt{SN IIn} model from \cite{Luc2025}, shown on the left side in Fig.~\ref{fig:dust}, and magnify the H$\alpha$ profiles for each of the different dust mass models, shown on the right side in Fig.~\ref{fig:dust}.
The \texttt{SN IIn} model parameters are as follows: the model age is 300 days, with an ejecta mass (\(M_{\rm ej}\)) of 12 M$_\odot$, kinetic energy (\(E_{\rm kin}\)) of \(1.5 \times 10^{51}\) erg, \(^{56}\)Ni mass (\(M_{^{56}\mathrm{Ni}}\)) of 0.032 M$_\odot$, and CSM velocity (\(V_{\rm CDS}\)) of 5,000 km s$^{-1}$. In the \texttt{SN IIn} model, dust is assumed to be composed of 0.1 micron silicate grains. 
In the \texttt{SN IIn} simulation, most of the radiation originates from the outer ejecta dense shell at \(V_{\rm CDS} = 5,000 \, \mathrm{km \, s}^{-1}\).  

From the spectra of the \texttt{SN IIn} model with varying dust mass, we observe that as the dust mass increases, the asymmetry of H$\alpha$ increases significantly, while the intensities of [\ion{O}{I}] $\lambda\lambda 6300, 6364$ and [\ion{Ca}{II}] $\lambda\lambda 7291, 7323$ gradually weak.
The H$\alpha$ profile of SN 2016iog at +190.8 days qualitatively matches the \texttt{SN IIn} model obtained with 10$^{-4}$ M$_{\odot}$ of dust. The H$\alpha$ emission originates from a dense shell in the outer ejecta. In the model, the dense shell is located at 8,000 km/s, and if the velocity were higher (e.g. 10,000 km/s or more), the fit could improve. In the model, the dense shell is very narrow (with a velocity width of about 100 km/s), and this is the reason why the flux drops sharply at velocities greater than 8,000 km/s and less than $-$8,000 km/s (redshifted and blueshifted peaks). Asymmetry or disruption of the shell would cause the emission to spread over a broader velocity range, as seen in the SN 2016iog +190.8 days spectrum \citep{Luc2025}.
Compared to the \texttt{SN IIn} model, it can be seen that the asymmetry of H$\alpha$ in SN~2016iog is similar to the model, but [\ion{O}{I}] $\lambda\lambda 6300, 6364$ is weaker.
Therefore, we estimate that the dust mass surrounding SN~2016iog is approximately $10^{-4}$ M$_{\odot}$ at 190.8 days. Although the \texttt{SN IIn} model lacks spectra at 200~d, it can qualitatively explain the observations of SN~2016iog, where the spectral features are primarily related to varying dust mass.

Compared to the prominent [\ion{O}{I}]~$\lambda\lambda$6300, 6364 doublet typically observed in SN~2014G (see the bottom panel of Fig.~\ref{fig:spectrum_comparison}), SN~2016iog exhibits little or no evidence of this feature.
While forbidden lines such as [\ion{O}{I}] generally take time to emerge, they are commonly seen in a large number of SNe II by $\sim$200 days after explosion. At later epochs (e.g. 300--400 days), as the outer ejecta become transparent, [\ion{O}{I}] emission is almost universally present in non-interacting Type II events (e.g. SN 2014G).
Adopting the explanation from the \texttt{SN IIn} model, dust is present around SN 2016iog at +190.8 days, with the spectrum at this time appearing to be dominated by continued interaction between the SN ejecta and a dense circumstellar shell \citep{Luc2025}.
In such conditions, emission from deeper layers of the ejecta---particularly the [\ion{O}{I}] region---may be obscured by newly formed dust within the dense shell. This dust can cause substantial optical extinction, possibly on the order of $\sim$2 magnitudes \citep{Luc2025}, thereby suppressing the visibility of nebular [\ion{O}{I}] lines. A similar effect is seen in SN~1998S at +312 days and may account for the weak [\ion{O}{I}] emission observed in SN~2016iog at +190.8 days. 
Another possibility is that the progenitor star of SN 2016iog had a relatively low initial mass, resulting in a smaller amount of oxygen production in the core. This would lead to weaker [\ion{O}{I}] emission in the ejecta, as the amount of oxygen synthesized would be limited. This behaviour is also seen in some low-luminosity Type II SNe, such as SN 2005cs, 2018is, 2021gmi \citep{Pastorello2009MNRAS.394.2266P, Dastidar2025A&A...694A.260D, Meza-Retamal2024ApJ...971..141M}, where the [\ion{O}{I}] emission is notably weaker compared to normal Type IIP SNe like SN 1999em.

\subsection{Comparison with SN~1998S and PTF11iqb}
\label{sect:PTF}
\begin{figure}
    \includegraphics[width=\columnwidth]{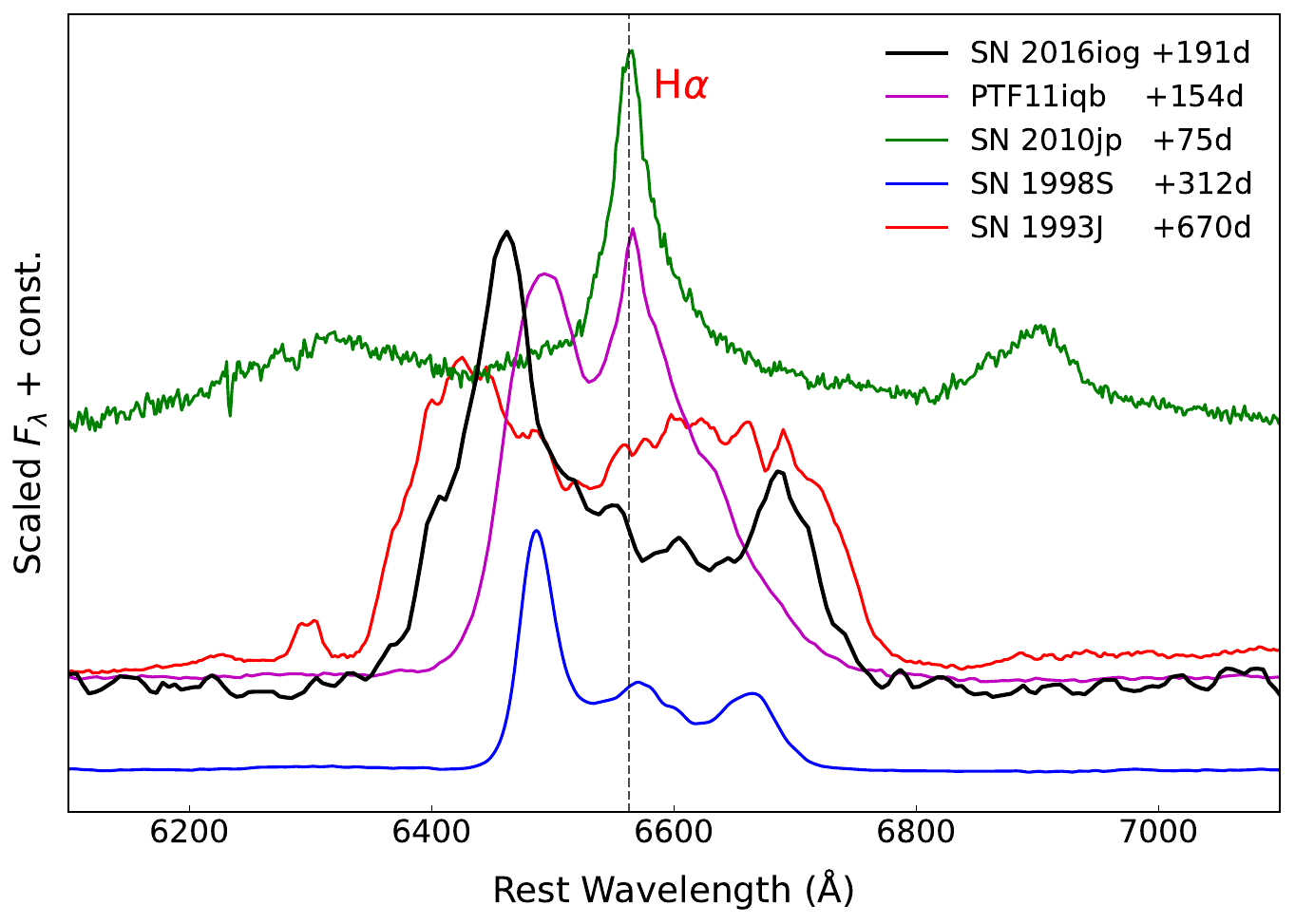}
    \caption{Comparison of the H$\alpha$ profile of SN 2016iog with other SNe that have exhibited multi-peaked H$\alpha$ features. The spectra have been scaled for improved comparison. }
\label{fig:compreHa}
\end{figure}

\begin{figure}
    \includegraphics[width=\columnwidth]{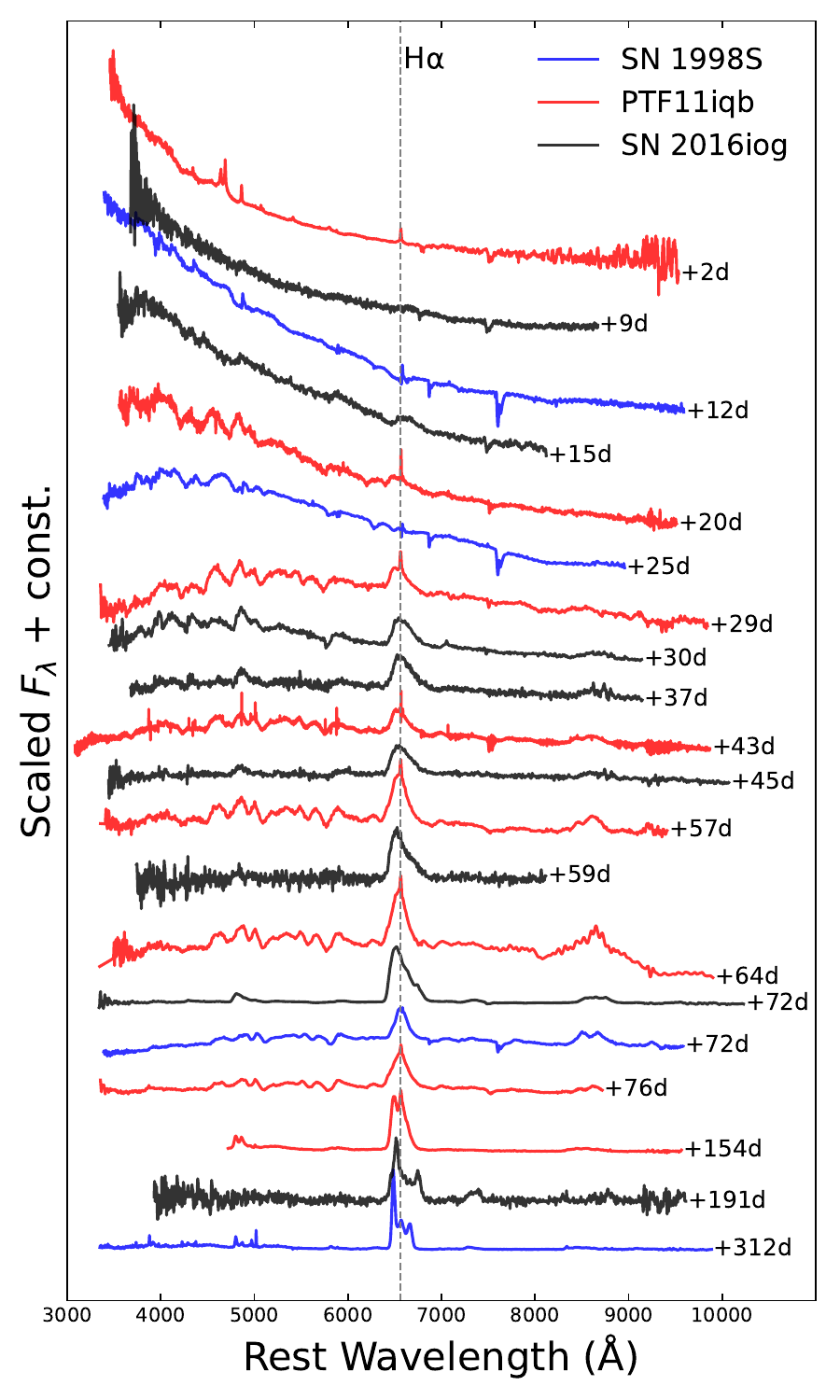}
    \caption{Spectral evolution of SN~2016iog is compared with those of SN~1998S and PTF11iqb. All spectra of SN 2016iog are plotted in black, those of PTF11iqb in red, and those of SN 1998S in blue. The spectrum of PTF11iqb obtained at +2 days is taken from \cite{Smith2015MNRAS.449.1876S}, while the remaining spectra of PTF11iqb are retrieved from the Padova-Asiago Spectra Archive and will be presented in a forthcoming paper on PTF11iqb.}
\label{fig:PTF}
\end{figure}

Due to observational limitations, such as the lack of infrared observations, although the spectrum of SN 2016iog matches well with the model spectrum assuming the presence of dust, it cannot be regarded as the only possible explanation. We compare the multi-peaked features observed in the H$\alpha$ profile of SN 2016iog with those in other SNe with similar features, such as SNe 1993J, 1998S, 2010jp and PTF11iqb \citep{Matheson2000AJ....120.1487M, Dessart2025A&A...696L..12D, Smith2012MNRAS.420.1135S, Smith2015MNRAS.449.1876S}, as shown in Fig. \ref{fig:compreHa}. 
The multi-horned structure of the H$\alpha$ line in SN 2010jp, has been interpreted as the result of the interplay between a non-spherical geometric structure and a jet-driven explosion \citep{Smith2012MNRAS.420.1135S}. In this scenario, the central narrow peak is produced by the interaction with the CSM, where the narrow lines originate from shock heating in the dense CSM. The jet breaks through the stellar envelope along the polar axis, forming a bipolar high-velocity outflow. When the line of sight is inclined relative to the jet axis, both blue-shifted (towards) and red-shifted (away) components are observed.
However, this spectrum shows significant differences compared to SN 2016iog. In SN 2010jp, the multi-horned line profile is also present in the H$\beta$ line across multiple epochs, but SN 2016iog exhibits a relatively weak H$\beta$ profile, making this interpretation less applicable to SN 2016iog.

The H$\alpha$ double-horned profile of SN 1993J, is not as pronounced as in SN 2016iog, but the broad features match best. \cite{Matheson2000AJ....120.1499M} suggests that the double-horned phenomenon in SN 1993J is attributed to the combined effect of an aspherical geometric structure (such as a flattened or disk-like structure) and the material distribution along the line of sight. This interpretation builds upon the earlier model proposed to explain the spectra of SN 1987A \citep{Leonard2000ApJ...536..239L, Gerardy2000AJ....119.2968G}. SN 2016iog exhibits a more pronounced H$\alpha$ profile asymmetry compared to SN 1993J. We therefore infer that if SN 2016iog is surrounded by a more pronounced aspherical geometric structure, stronger than that of SN 1993J, combined with material distribution along the line of sight, this could explain the observed H$\alpha$ profile of SN 2016iog.  The edge of a disk-like structure, where the projection velocity is maximal, would produce high-velocity emission components, while the central region, where the velocity is more aligned with the line of sight, would exhibit weaker emission, resulting in a dip between the peaks. The weakening of the redshifted peak could be due to either dust extinction or optical depth effects, where emission from the receding side is partially absorbed \citep{Matheson2000AJ....120.1499M}.  

However, the emergence of the H$\alpha$ double-horned profile in SN 1993J occurred much later than in SN 2016iog. PTF11iqb has been interpreted as exhibiting an H$\alpha$ double-horned profile due to an asymmetric CSM, and the appearance of its double-peaked structure occurs at a phase comparable to that of SN 2016iog. We present the spectral evolution of SN 1998S, PTF11iqb, and SN 2016iog, arranged according to the time since explosion, in Fig.~\ref{fig:PTF}. Overall, the spectra of the three SNe in Fig.~\ref{fig:PTF} exhibit remarkable similarities, undergoing comparable spectral evolution at similar phases. However, we note two differences among these three SNe. Since SN 1998S ($M_{V} = -19.66$ mag) exhibits a brighter peak luminosity than PTF11iqb ($M_{r} = -18.32$ mag) and SN 2016iog ($M_{V} = -18.64$ mag; see Fig.~\ref{fig:Absolute_magnitude}), the early spectra of SN 1998S also display stronger signatures of interaction. Subsequently, the LC of PTF11iqb enters a plateau phase ($s_1 = 1.61$ mag/100\,d $> s_2 = 0.83$ mag/100\,d), whereas SNe 1998S ($s_1 \approx s_2 = 3.57$ mag/100\,d) and 2016iog ($s_1 = s_2 = 8.85$ mag/100\,d) do not exhibit a plateau phase.

At later times, the H$\alpha$ line profile of SN 2016iog shows significant differences from that of PTF11iqb, but exhibits a striking similarity to the spectrum of SN 1998S at +312\,d. All three SNe develop qualitatively similar asymmetric and multi-peaked H$\alpha$ profiles at late times. However, the H$\alpha$ profile of PTF11iqb is blueshifted from +120\,d to +200\,d and becomes strongly redshifted after +500\,d, whereas SN 1998S shows a persistent blueshift.
For SN 2016iog, no later-time spectra were obtained due to observational limitations. The case of SN 1998S is discussed in detail in Sect.~\ref{section:190.8}, where the persistent blueshift is interpreted as being caused by dust formation that obscures the receding parts of the system \citep{Leonard2000ApJ...536..239L, Pozzo2004MNRAS.352..457P, Luc2025}. The spectral evolution of the H$\alpha$ profile in PTF11iqb, from a blueshifted peak to a stronger redshifted peak, may be attributed to an asymmetric CSM combined with viewing-angle effects \citep{Smith2015MNRAS.449.1876S}. A flattened CSM structure with one side significantly denser than the other is difficult to produce with a single star. While axisymmetric configurations in the CSM, such as discs or bipolar nebulae, may result from rapid stellar rotation, achieving strong azimuthal asymmetry is unlikely for a single star. This may be achieved through mass-loss in a binary system with non-zero eccentricity or by unsteady mass-loss.

RY~Scuti is a rare massive eclipsing binary observed during an active phase of mass transfer, in which one component is being stripped of its hydrogen envelope while evolving toward the Wolf–Rayet stage. It remains the only known system of this type with a spatially resolved toroidal circumstellar nebula \citep{Smith2002ASPC..279..325S, Grundstrom2007ApJ...667..505G, Smith2015MNRAS.449.1876S}. The velocity-resolved observations reveal that the torus exhibits pronounced azimuthal asymmetry, likely resulting from episodic mass-loss events that occurred within the past few hundred years \citep{Grundstrom2007ApJ...667..505G}. The integrated emission-line profile of [\ion{N}{ii}]~$\lambda6583$ from the nebula is strongly asymmetric and multi-peaked, displaying a brighter red component and bearing a close resemblance to the late-time H$\alpha$ profile of PTF11iqb \citep{Smith2015MNRAS.449.1876S}. Therefore, the case of RY~Scuti provides direct evidence that such an asymmetric toroidal CSM configuration can exist in nature, reinforcing the plausibility of similar environments inferred for interacting SNe.

When the CSM has an asymmetric density distribution, the H$\alpha$ profile varies with the viewing angle. Observing from the side with higher CSM density yields a blueshifted peak, as in the spectra of SNe 1998S (+312\,d) and 2016iog (+191\,d), whereas viewing from the side with lower density produces a redshifted peak, as in the spectra of PTF11iqb after +500\,d (see Fig.~7 of \citealt{Smith2015MNRAS.449.1876S}). At viewing angles near the interface between the high-density and low-density regions, more symmetric multi-peaked H$\alpha$ profile can be seen, as observed in SN~1993J at +670~d (see Fig. \ref{fig:compreHa}).  Therefore, if SN 2016iog has an asymmetric CSM and is observed from the side with higher CSM density, it can naturally produce a spectrum with a clearly blueshifted H$\alpha$ peak, as seen at +191\,d \citep{Smith2015MNRAS.449.1876S}. 

The H$\alpha$ red peak of SN 2016iog is already visible at +72\,d, although it is weaker than in the later spectrum at +191\,d (see Fig. \ref{fig:sp_ev}). In the +72\,d spectrum of SN 2016iog, the red peak in the H$\alpha$ profile is relatively weak and appears as a secondary feature compared to the strong H$\alpha$ asymmetry observed at +191\,d. The asymmetry in the H$\alpha$ profile of SN 2016iog is present at all earlier times, but may be more apparent as a blueshifted peak. The overall asymmetry of its spectra is likely caused by optical-depth effects on emission from the outer ejecta and the dense shell, and does not necessarily require dust. We can also see emission profiles with a central dip in dust-free models (see Fig.~\ref{fig:dust}), which result from the larger column density in the mid-plane regions and produce the central dip in H$\alpha$ \citep{Luc2025}. The +72\,d spectrum of SN~2016iog suggests that asymmetry may also contribute, since mass loss in supergiants is inherently asymmetric.  Compared with SN~1998S and PTF11iqb, the earlier emergence of multi-component H$\alpha$ lines in SN~2016iog is likely a consequence of its more rapid photometric evolution.

In summary, the multi-horned H$\alpha$ feature observed in the spectrum of SN~2016iog may result from either a small amount of dust or a non-spherical geometry, such as a flattened or disk-like structure.  The asymmetry of the double-horned profile is likely caused by dust extinction, optical depth effects, or azimuthal asymmetry in the circumstellar medium.

\section{Discussion}
\label{section:sum}

In this paper, we present a detailed analysis of the photometric and spectroscopic properties of the Type IIL SN\,2016iog. The main findings of our study are summarised as follows:
\begin{itemize}
\item 
The $V$-band LC of SN~2016iog rises rapidly to its peak absolute magnitude ($M_V = -18.64 \pm 0.15$ mag) at +13.9 days after the explosion, placing it among the most luminous Type~II SNe (see Fig.~\ref{fig:Absolute_magnitude}).
The peak is followed by a steep decline, with a rate ($s_1 = s_2 = 8.85 \pm 0.15$ mag/100d) that exceeds those of most Type II SNe, while it is closer to those expected in SNe IIb. In contrast, at late times, SN\,2016iog is less luminous than most Type II SNe.
By comparing the late-time evolution of SN~2016iog, we determine an ejected mass of $^{56}$Ni of $0.014 \pm 0.007\,M_{\odot}$.
\item 
Based on LC modelling, we estimate an ejecta mass of 3.7\,M$_\odot$ and an initial radius of approximately 983\,R$_\odot$. 
In addition, our model suggests that SN~2016iog is surrounded by a low-density ($\sim 9.51 \times 10^{-11}$ g cm$^{-3}$), low-mass ($\sim 0.060~M_\odot$) CSM around its progenitor. Most massive stars likely experience enhanced mass loss towards the end of their lives, leading to a reduction of the progenitor mass at the time of the SN explosion \citep{Ofek2014ApJ...789..104O, Bruch2021ApJ...912...46B, Strotjohann2021ApJ...907...99S}. 
The presence of pre-SN mass-loss suggests that circumstellar interaction may be a common feature to some extent in CC SNe.
Therefore, the progenitor of SN 2016iog is likely to have been experienced  mass loss.
\item 
The spectrum of SN~2016iog at +9.3 days exhibits a nearly featureless continuum. At later epochs, the spectra show low-contrast P-Cygni profiles in H$\beta$ and \ion{He}{I}~$\lambda$5876. Throughout its evolution, H$\alpha$ consistently appears as a broad line dominated by the emission component, while the absorption component remains weak, fading to completely disappear at phases later than about one month. At very late times (more than six months after the explosion), H$\alpha$ develops an unusual asymmetric double-peaked profile.
\item 
The rapid LC rise to a luminous peak, combined with the quasi-featureless early-time spectra, suggests that SN~2016iog likely experienced interaction with CSM in its early phases.
During the recombination phase, the rapid decline of the LC indicates that the ejecta have become optically thin. 
In the late-time spectra, we infer the presence of approximately $10^{-4}$\,M$_{\odot}$ of dust in the surrounding region, or an asymmetric CSM may be present, which may have formed during this phase.
\end{itemize}

\subsection{On the rapid decline of the light curve}
\label{section:RD}

\begin{figure}
    \includegraphics[width=\columnwidth]{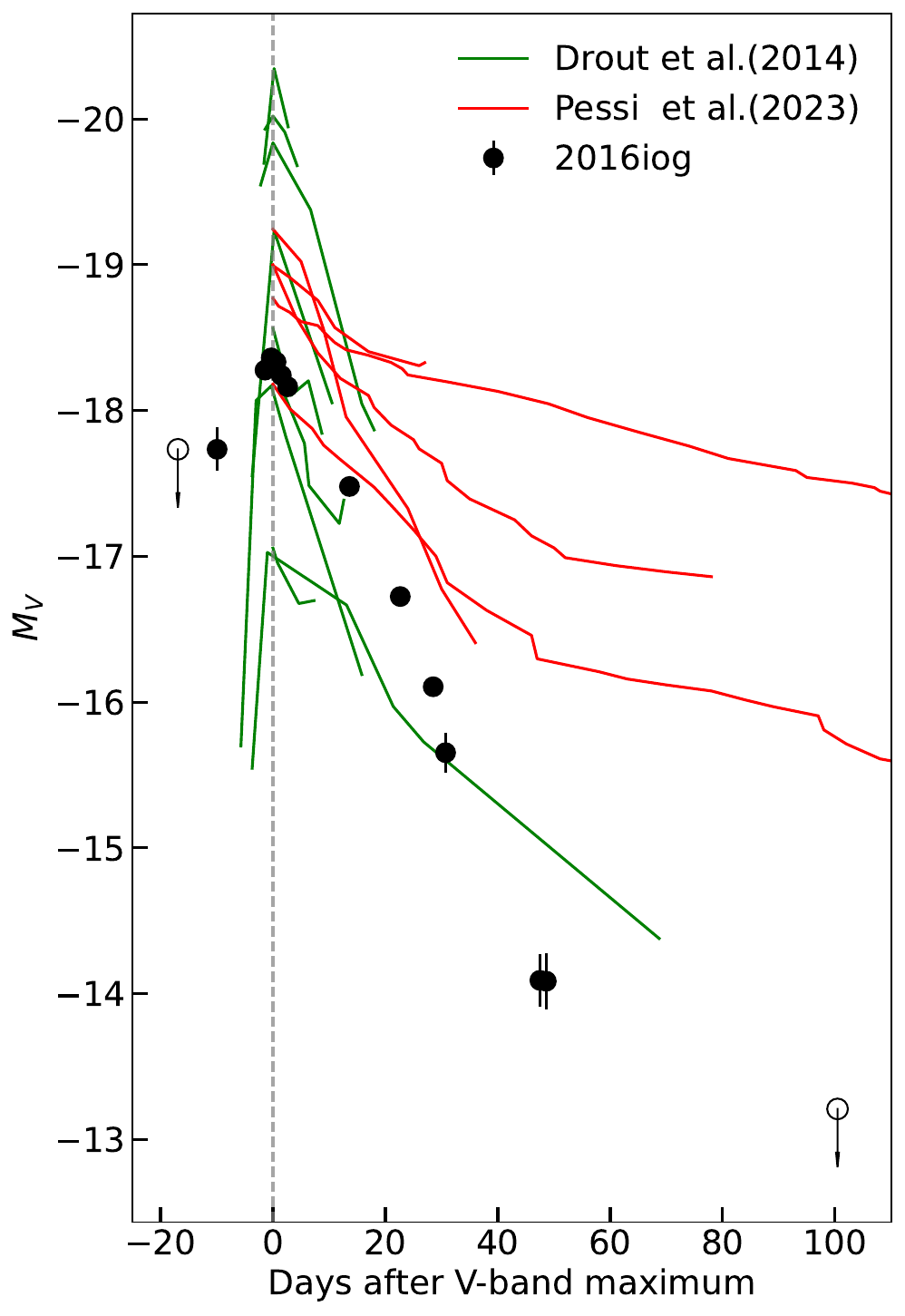}
    \caption{Comparison of the absolute $V$-band light curve of SN~2016iog with other rapidly evolving transients.  The $V$-band data from \cite{Pessi2023MNRAS.523.5315P} are shown as red lines, while the sample from \cite{Drout2014ApJ...794...23D}, which lacks $V$-band observations, is represented by the $r/z$-band data shown as green lines.}
\label{fig:fast}
\end{figure}

SN 2016iog was bright in the early phases ($M^{\text{peak}}_V <-18.5$ mag), but over time, it became dimmer than nearly all Type II SNe (see the upper-right panel of Fig.~\ref{fig:Absolute_magnitude}). This indicates that the ejected mass was relatively small ($\sim$ $3.7 ~ M_{\odot}$), as otherwise, the LC would have maintained a plateau phase around 50 to 100 days, similar to SN 1999em (see Fig. \ref{fig:Absolute_magnitude}). The decline rate of the LC of SN~2016iog ($s_2 = 8.85 \pm 0.15$ mag/100d) shows some similarity to those of rapidly evolving transients reported by \cite{Drout2014ApJ...794...23D, Pursiainen2018MNRAS.481..894P} and \cite{Pessi2023MNRAS.523.5315P}. We compare the LC of SN~2016iog with those of rapidly evolving transients \citep{Drout2014ApJ...794...23D} and fast-evolving Type~II SNe \citep{Pessi2023MNRAS.523.5315P} in Fig.~\ref{fig:fast}. In Sect.~\ref{sect:para} (see Fig.~\ref{fig:combined}), we compare some LC-related parameters (e.g. $M^{Peak}_g$) of SN~2016iog with those of the samples from \cite{Drout2014ApJ...794...23D, Pursiainen2018MNRAS.481..894P}.

A common characteristic shared by these rapidly evolving transients and SN~2016iog is a $t_{1/2}$ shorter than 12~days and an absolute magnitude in the range $-20 < M < -16.5$~mag. For many rapidly evolving transients, the early-time radiation is not powered by radioactive $^{56}$Ni \citep{Drout2014ApJ...794...23D}. A growing body of evidence indicates that interaction with the surrounding CSM can significantly affect both the spectral features and the peak luminosity of these events \citep{Hillier2019A&A...631A...8H}. Compared to the samples presented in \cite{Drout2014ApJ...794...23D, Pessi2023MNRAS.523.5315P}, the peak luminosity of SN~2016iog lies at an intermediate level. The spectra of the \cite{Drout2014ApJ...794...23D} sample are poorly constrained, making it difficult to determine whether the events are H-rich or H-poor SNe. In contrast, the sample of \cite{Pessi2023MNRAS.523.5315P} has been identified as H-rich Type~II SNe, and the match with SN~2016iog in Fig.~\ref{fig:fast} is correspondingly better. By comparing these LCs, it is possible that some of the SNe in the \cite{Drout2014ApJ...794...23D} sample may also belong to the H-rich population.

For the rapid evolution of Type~II SNe, \cite{Pessi2023MNRAS.523.5315P} proposed a possible explanation: these objects may be surrounded by CSM that is denser than in ordinary SNe~II, but still not sufficiently thick to become optically opaque to electron scattering on large scales. 
\cite{Morozova2017ApJ...838...28M} modelled the LCs of rapidly declining SNe II, and proposed that the dense CSM embedding RSGs serves as a viable scenario for these events. A dense CSM is known to be relatively common in H-rich CC SNe. This may explain the rapid brightness decline observed in SN~2016iog. An alternative explanation for the late-time dimming of the optical luminosity could be dust formation. This scenario would also be supported by late spectroscopic clues (see Sect. \ref{section:190.8}). 

In summary, a rapid decline in the LC generally indicates that the ejecta have become optically thin, with the power source steeply diminishing due to multiple factors such as $\gamma$-rays escape, low \(^{56}\)Ni mass, and dust extinction.

\subsection{Interaction exhibited in the spectrum} 
\label{Sec:inter-spec}
In the early phase, the spectrum exhibits a nearly smooth continuum. According to \cite{Luc2024interactingsupernovae}, 1-2 weeks later, the shock wave from the SN propagates outwards through the CSM and undergoes deceleration. Material from both the ejecta and the CSM accumulates into a CDS. As a result, the spectrum shows a nearly featureless continuum with weak blue-shifted absorption \citep{Luc2022}, which is consistent with the spectrum of SN 2016iog at $+$9.3 days. 
At the same time, the rapid luminosity rise, discussed in Sect. \ref{section:RD}, is probably caused by the interaction between the ejecta and the CSM. Therefore, we propose that the characteristic early spectrum of SN 2016iog is likely the result of its interaction with the CSM.

During the recombination phase, H$\alpha$ displays a very weak absorption component (that later vanishes) and a broad emission. 
In a normal SN, the H$\alpha$ and other lines narrow with time, and eventually one sees narrower lines from [\ion{O}{I}] or [\ion{Ca}{II}] (e.g. SN 2019nyk). However, in the spectra of SN 2016iog between +13.0 days and +62.9 days, we observe broad lines, with H$\alpha$ persisting at all epochs. This suggests continued power injection from the outer ejecta interacting with the CSM \citep{Luc2022}. The evolution of the line profile can only be explained by interaction \citep{Guti2014ApJ...786L..15G,Hillier2019A&A...631A...8H, Luc2022}.

In the nebular phase, the spectra of SN 2016iog showed a strong asymmetry in its  H$\alpha$ profile. In particular, at +190.8 days, it shows two peaks, one (more prominent) blue-shifted by 4,000 km s$^{-1}$, and the weaker one red-shifted by 7,000 km s$^{-1}$. This peculiar line profile and the rapid optical luminosity decline can be explained by the formation of dust in a CDS, producing a significant flux extinction in the optical bands as well as modifying the observed line profile. In Sect. \ref{section:190.8}, we suppose  that this could be attributed to the presence of around 10$^{-4}$~M$_{\odot}$ of dust, which is likely distributed across both the inner ejecta and the dense shell. The dense shell is formed due to the high-velocity collision between the SN ejecta ($v > 5,000$ km $s^{-1}$) and the CSM, which generates both a reverse shock and a forward shock, leading to compression between the two \citep{Luc2025}. Thus, the high-velocity H$\alpha$ emission emerging from the dense shell is influenced by dust, resulting in the formation of the double-peaked structure, which serves as evidence of the interaction. If the double-peaked H$\alpha$ structure in the spectrum of SN 2016iog at +190.8 days is caused by the interaction between the ejecta and CSM, leading to an aspherical geometric structure, then it is unquestionable that interaction is occurring at this phase.

To summarise, in the recombination and nebular phases of SN 2016iog, we observe evidence of interaction, suggesting that CSM was already present. Therefore, it is highly likely that interaction also occurred during the early phase of SN 2016iog.

\subsection{The progenitor scenario}
\label{section:progenitor}
For SN~2016iog, we cannot constrain the progenitor mass from the [\ion{O}{I}] $\lambda\lambda$6300, 6364 features in the spectrum at +190.8 days. 
\cite{Fransson1987ApJ...322L..15F, Fransson1989ApJ...343..323F} found that the ratio (R) of the nebular-phase spectral luminosities of [\ion{Ca}{II}] $\lambda\lambda$7291, 7324 and [\ion{O}{I}] $\lambda\lambda$6300, 6364 remains nearly constant in the late stages. Objects with a larger R are expected to have smaller main-sequence masses. Although the spectrum of SN 2016iog at +190.8 days may not be considered a very late-phase spectrum, the [\ion{O}{I}] $\lambda\lambda$6300, 6364 lines in SN~2016iog are weak to the point of being undetectable, while the [\ion{Ca}{II}] $\lambda\lambda$7291, 7324 lines show a relatively noticeable feature. Considering this, the value of R for SN~2016iog is expected to be relatively large, suggesting that it likely originated from a progenitor with a smaller main-sequence mass. However, if we accept the assumption made in Sect. \ref{section:190.8} that dust exists around SN~2016iog, the weakness of [\ion{O}{I}] could be due to dust extinction, which would lead to an underestimation of the progenitor mass \citep{Luc2025}.

Many progenitors of Type II SNe experience substantial mass loss before explosion \citep{Bruch2021ApJ...912...46B}, suggesting that such mass loss likely contributes to the formation of the CSM that subsequently interacts with the SN ejecta.
Most Type II SNe show evidence of early interaction with CSM \citep{Forster2018NatAs...2..808F, Morozova2020ApJ...891L..32M}. The presence of CSM could enhance the SN continuum, which may account for the absence of metal spectral lines \citep{Branch2000PASP..112..217B}.  
\cite{Pessi2023MNRAS.523.5315P} compared six luminous SNe II with fast-declining LCs. They suggested that the mechanism behind the luminosity of these SNe II is likely the interaction between the SN ejecta and the CSM, which converts kinetic energy into radiative energy.
The low-density CSM serves as the energy source for these luminous SN II samples.
From the multi-band LCs modelled using MOSFIT, it is inferred that there exists a low-density CSM (${\rho_{\rm 0} = 9.51^{+15.95}_{-5.83}\times10^{-11}}$ g cm$^{-3}$) around SN 2016iog. Since SN~2016iog exhibits similar LCs and spectra to the sample compared in \cite{Pessi2023MNRAS.523.5315P} (e.g. SNe 2017cfo, 2017gpp), and both show the presence of low-density CSM, these SNe possibly have the same progenitor star type.

The LC morphology of SN~2016iog reveals a high luminosity peak, followed by a rapid decline. Based on the \texttt{MOSFIT} modelling, the ejecta mass of SN~2016iog is found to be relatively small ($\sim$ 3.7 M$_\odot$), a large radius ($\sim$ 983 R$_\odot$) and a low \(^{56}\)Ni mass ($\sim$ 0.014 M$_\odot$). However, due to the faint appearance of [\ion{O}{I}] $\lambda\lambda$6300, 6364 and [\ion{Ca}{II}] $\lambda\lambda$7291, 7324 in the nebular-phase spectrum, it suggests that most, if not all, of the energy likely comes from the interaction (see Sect. \ref{Sec:inter-spec}), with the \(^{56}\)Ni mass remaining highly uncertain. 
Consequently, the $^{56}$Ni mass estimated for SN~2016iog ($0.014 \pm 0.007~M_{\odot}$), should be regarded as an upper limit.

Therefore, we consider a RSG progenitor with a low-mass hydrogen-rich envelope surrounded by a low-density CSM.
The density of this CSM is insufficient to produce significant optical depth effects via electron scattering on large scales (higher than that in other ordinary Type II SNe). 
If mass transfer occurs in a binary system to form CSM with the aforementioned characteristics, the resulting CSM may exhibit asymmetry. The progenitor of SN~2016iog may involve a lower-mass single or binary star system, where the outer H-rich envelope is partially removed by intense stellar winds or interaction with a companion star \citep{Podsiadlowski1992ApJ...391..246P, Folatelli2014ApJ...792....7F, Fox2014ApJ...790...17F, Ryder2018ApJ...856...83R}.
In both progenitor channels, the rapid increase to peak luminosity, colour behaviour, and spectral characteristics of SN~2016iog can be qualitatively explained.

\section{Conclusions} 
\label{section:conclusion}
The optical LC of SN 2016iog rises rapidly, with a bright peak, followed by a rapid decline. The spectra show broad and asymmetric H$\alpha$ emission lines, with very little H$\alpha$ absorption component observed.
SN~2016iog is most striking in the dominance of interaction as the main power source.
From the rapid rise of the early LC of SN~2016iog, the quasi-featureless spectra, to the spectra in the recombination and nebular phases, evidence for interaction is present. 
In the late-time spectra, the SN~2016iog exhibits a rare asymmetric double-peaked H$\alpha$ profile, likely influenced by dust formation or asymmetries in the ejecta or CSM. These observational characteristics make SN~2016iog an important case for understanding the diversity and complexity within the Type II subclass of CCSNe. We assume an RSG progenitor star with its outer H-rich envelope partially stripped, surrounded by low-density CSM, and the subsequent appearance of dust at later stages, in order to qualitatively explain the observed characteristics of SN~2016iog.

Future observational facilities, such as the Chinese Space Station Telescope\footnote{\url{http://nao.cas.cn/csst/}} \citep[CSST; ][]{CSST2025} and the Vera C.\ Rubin Observatory\footnote{\url{https://www.lsst.org/}} \citep{Hambleton2023PASP..135j5002H}, will provide higher-cadence monitoring of Type II SNe. These data will greatly improve our ability to detect and characterise fast-evolving events, refine existing theoretical models, and advance our understanding of this enigmatic subclass of CCSNe.

\section{Data availability}
Optical photometric measurements of SN 2016iog are available at the CDS via  \url{https://cdsarc.cds.unistra.fr/viz-bin/cat/J/A+A/705/A104}

Our observations of the specta are available via the Weizmann Interactive SN Data Repository (WISeREP; \citealt{Yaron2012PASP..124..668Y}).

\bibliographystyle{aa}
\bibliography{pzhref.bib} 
\onecolumn 
\begin{appendix}
\section{Supplementary figures}
\label{appendix:corner}
\begin{figure*}[ht]
    \centering
    \includegraphics[width=1\textwidth]{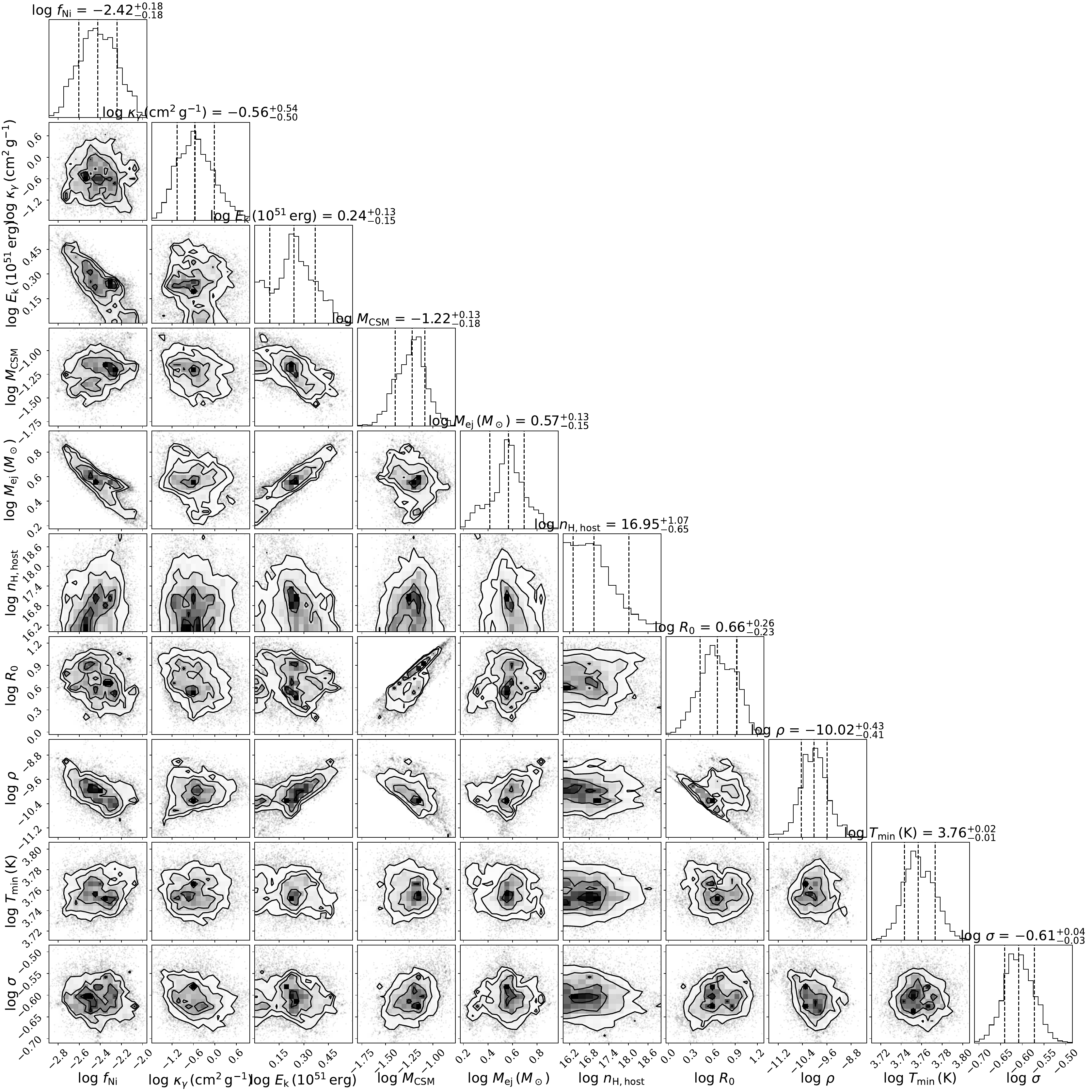} 
    \caption{Posterior probability density functions for the free parameters of the model light curves in Fig. \ref{fig:mosfit}. }
    \label{fig:modfit_corner}
\end{figure*}

\newpage
\section{Spectroscopic tables}
\label{SpecInfo}

\begin{table}[h]
\caption{Log of spectroscopic observations of SN 2016iog.}
\label{tab:spec}
\centering
\begin{tabular}{cccccccc}
\toprule
\hline  
Epoch &MJD & Phase$^a$&Telescope+Instrument & Grism/Grating+Slit & Spectral range & Resolution & Exp.time  \\
 &  & [days]& &  & [\AA]    & [\AA]           & [s] \\
\hline
2016-12-02&57724.890 & +9.3  & Xinglong2.16m+BFOSC & gm4 & $3740-8800$ & 19 & 1200   \\
2016-12-06&57728.555 & +13.0 &  Ekar1.82m+AFOSC & GR04+1.69" & $3400-8240$ & $13$ & $1800$   \\
2016-12-07&57729.662 & +14.1 & Ekar1.82m+AFOSC & VPH6+1.69" & $3150-9290$ & $15$ & $2400$    \\
2016-12-08&57730.508 & +14.9 &Ekar1.82m+AFOSC & GR04+1.69" & $3400-8240$ & $13$ & $2700$   \\
2016-12-09&57731.455 & +15.9 & Ekar1.82m+AFOSC & GR04+1.69" & $3400-8240$ & $13$ & $2700$ \\
2016-12-10&57732.584 & +17.0 & Ekar1.82m+AFOSC & GR04+1.69" & $3400-8240$ & $13$ & $2700$  \\
2016-12-22&57744.704 & +29.1 & Ekar1.82m+AFOSC & GR04+1.69"& $3400-8240$ & $13$ &$2700$  \\
2016-12-23&57745.566 & +30.0 & Ekar1.82m+AFOSC & VPH6+1.69" &$3150-9290$  & $15$ & $2700$    \\
2016-12-30&57752.586 & +37.0 & Ekar1.82m+AFOSC & VPH6+1.69" & $3150-9290$& $15$ &$2700$  \\
2017-01-04&57758.457 & +42.9 & Ekar1.82m+AFOSC & VPH6+1.69" & $3150-9290$ & $15$ & $3000$   \\
2017-01-07&57760.648 & +45.0 & Ekar1.82m+AFOSC & VPH6+1.69" & $3150-9290$& $15$ & $3000$  \\
2017-01-21&57774.665 & +59.1 & Ekar1.82m+AFOSC & GR04+1.69"& $3400-8240$ & $13$  & $3600$   \\
2017-01-25&57778.520 & +62.9& Ekar1.82m+AFOSC & VPH6+1.69" & $3150-9290$& $15$ & $3600$   \\
2017-02-03&57787.645 & +72.0 & GTC+OSIRIS & R300B+1.0" & $3400-10390$ & $17$ & $1800$   \\
2017-06-01&57906.432 & +190.8 & GTC+OSIRIS &  R300B+1.0" & $3400-9980$ & $17$ & $1500$  \\
\hline\hline 
\multicolumn{8}{l}{{$^a$ Phases are calculated relative to the explosion epoch (MJD = 57715.6) in the reference frame of the observer.}} 
\\

\end{tabular}
\end{table} 
\begin{table}[H]
\caption{Observed magnitudes of SN\,2016iog in the Johnson BV bands and the `G' band of Gaia.}
\label{2016BV} 
\resizebox{\textwidth}{!}{ 
\centering     
\begin{tabular}{c c c c c c c} 
\hline \hline
Date & MJD & Phase$^a$ &  B & V & G$^b$ & Instrument\\ 
\hline
20161108& 57700.613& -15.0 &  $-$ &     >21.5 & >21.5  &GAIA\\
20161120& 57712.590 &  -3.0&$-$&     >17.2  &$-$&ASASSN\\
20161127& 57719.560 &  $+$4.0  &$-$& 17.200(0.150)&   $-$ &ASASSN\\
20161206 &57728.085 &  $+$12.5 &16.838(0.022)& 16.657(0.013) &   $-$ &AFOSC \\
20161207 &57729.190&  $+$13.6  &16.671(0.038)&16.573(0.039)  &$-$&AFOSC \\
20161208 &57730.045  &  $+$14.4&16.752(0.021)& 16.601(0.031)  &$-$& AFOSC \\
20161208 &57730.990 &  $+$15.4&16.858(0.027) &16.693(0.016)  &   $-$ &AFOSC \\
20161210 &57732.120  &  $+$16.5&16.948(0.045)& 16.771(0.051)  &$-$& AFOSC\\
20161212 &57734.510  &  $+$18.9 &$-$&  16.970(0.200)   &$-$&GAIA\\
20161216 &57737.507 & $+$21.9 &    $-$&$-$& 16.99(0.20) &  GAIA \\
20161216 &57737.508 & $+$21.9 &    $-$&$-$& 17.04$^c$ &  GAIA \\
20161221 &57743.110  & $+$27.5&17.825(0.027)& 17.455(0.026) & $-$&  AFOSC \\
20161227 & 57749.013 &$+$33.4&    $-$&$-$& 17.77$^c$  & GAIA \\
20161227 & 57749.087 &$+$33.5&    $-$&$-$ &17.81$^c$  &  GAIA \\
20161230 &57752.145  &  $+$36.5&18.854(0.029)& 18.210(0.039) & $-$&  AFOSC \\
20161231 &57753.092& $+$37.5   &18.826(0.055) & $-$&$-$& 1\\
20161231 &57753.094& $+$37.5   &18.701(0.052) & $-$&$-$& 1\\
20161231 &57753.096& $+$37.5   &18.916(0.060) & $-$&$-$& 1\\
20170110 &57758.005 &  $+$42.4 &19.524(0.054)& 18.829(0.054) &$-$& AFOSC \\
20170107 &57760.205  &  $+$44.6&19.987(0.080) &19.281(0.137)&$-$&   AFOSC \\
20170123 &57776.985  &  $+$61.4&21.842(0.293) &20.843(0.178) &$-$& AFOSC \\
20170125 &57778.125  &  $+$62.5&21.622(0.232)& 20.849(0.195) &$-$& AFOSC \\
20170317 &57829.935  &  $+$114.3&>21.8 &>21.7 &$-$ &AFOSC \\
\hline\hline
\end{tabular}
}

\begin{minipage}{\textwidth}
\small
$^a$ Phases are calculated relative to the explosion epoch (MJD = 57715.6) in the reference frame of the observer.\\
$^b$ G band of Gaia consists of unfiltered white-light observations, with the passband defined by the response of the instrument.\\
$^c$ G band of Gaia is obtained from \url{http://gsaweb.ast.cam.ac.uk/alerts/alert/Gaia16cfv/}, marked as detection, but no error are displayed.\\
1 The observations were made by individual observers Bialkow and Z. Kolaczkow.
\end{minipage}
\end{table}
\begin{table}[H]
\centering
\caption{The observed magnitudes of SN\,2016iog in the \textit{oc} band were obtained using the ATLAS.} \label{2016oc}
\begin{tabular}{cccc|cccc}
\hline \hline
Date & MJD &  Filter & Mag & Date & MJD & Filter & Mag  \\ 
\hline
20170109&	57762.51	&o&	19.02(0.15)&20170408&	57851.35	&o&	>19.18 \\
20170112&	57765.47	&o&	19.04(0.29)& 20170411&	57854.35	&o&	>19.12\\
20170113&	57766.52	&o&	19.32(0.32)&20170412&	57855.34	&o&	>19.41\\
20170121&	57774.50	&o&	19.39(0.13)&20170413&	57856.31	&o&	>19.30\\
20160122&	57409.54	&o&	>18.53&20170415&	57858.36	&o&	>19.65\\
20160126&	57413.50	&o&	>19.10&20170416&	57859.30	&o&	>19.40\\
20160127&	57414.49	&o& >18.88&20170417&	57860.33	&o&	>19.21\\
20160131&	57418.49	&o&	>19.80&20170420&   57863.46	&o&	>19.30\\
20160201&	57419.53	&o&	>19.89&20170421&	57864.46	&o&	>19.61\\
20160222&	57440.46	&o&	>18.42&20170425&	57868.45&	o&	>18.53\\
20160223&	57441.43	&o&	>18.65&20170507&	57880.26	&o&	>19.25\\
20160323&	57470.37	&o& >17.24&20170509&	57882.33	&o&	>19.63\\
20160327&	57474.39&	o&	>19.05&20170514&	57887.25	&o&	>19.63\\
20160414&	57492.34&	o&	>19.37&20170610&	57914.26	&o&	>19.16\\
20160420&	57498.33&	o&	>18.62&20170614&	57918.27	&o&	>19.92\\
20160422&	57500.42&	o&	>18.61&20170621&   57925.26	    &o&	>20.21\\
20160515&	57523.29&	o&	>19.22&20170623&	57927.27	&o&	>20.10\\
20160527&	57535.27&	o&	>18.28&20161126&	57718.60	&c&	17.90(0.04)\\
20161017&	57678.64&	o&	>18.46&20161224&	57746.54	&c&	17.74(0.02)\\
20161110&	57702.64&	o&	>19.20&20161228&	57750.54	&c&	18.09(0.06)	\\
20161114&	57706.61&	o&	>19.57&20170125&	57778.50	&c&	20.41(0.28)	\\
20161118&	57710.63&	o&	>19.69&20160206&	57424.50	&c&	>20.22\\
20170210&	57794.46&	o&	>18.55&20160301&	57448.47	&c&	>19.63\\
20170214&	57798.46&	o&	>19.67&20160302&	57449.47	&c&	>20.41\\
20170215&	57799.51&	o&	>19.27&20160311&	57458.43	&c&	>19.77\\
20170217&	57801.50&	o&	>19.61&20160404&	57482.38	&c&	>19.84\\
20170218&	57802.50&	o&	>19.99&20160410&	57488.39	&c&	>19.63\\
20170219&	57803.47&	o&	>18.82&20160507&	57515.29	&c&	>19.18\\
20170221&	57805.47&	o&	>19.01&20160606&	57545.26	&c&	>19.83\\
20170223&	57807.49&	o&	>18.94&20161025&	57686.64	&c&	>19.12\\
20170305&	57817.46&	o&	>18.94&20170129&	57782.49	&c&	>20.67\\
20170306&	57818.46&	o&	>20.07&20170202&	57786.49	&c&	>20.52\\
20170315&	57827.44&	o&	>19.17&20170222&	57806.47	&c&	>19.92\\
20170317&	57829.42&	o&	>19.30&20170305&	57817.38	&c&	>20.11\\
20170318&	57830.43&	o&	>19.40&20170322&	57834.43	&c&	>19.57\\
20170319&	57831.39&	o&	>19.17&20170326&	57838.41	&c&	>20.01\\
20170323&	57835.40&	o&	>19.81&20170330&	57842.39	&c&	>19.77\\
20170327&	57839.37&	o&	>19.30&20170419&	57862.41	&c&	>20.11\\
20170328&	57840.38&	o&	>19.78&20170426&	57869.44	&c&	>20.57\\
20170331&	57843.36&	o&	>19.26&20170427&	57870.41	&c&	>20.73\\
20170401&	57844.37&	o&	>19.63&20170521&	57894.31	&c&	>20.73\\
20170404&	57847.36&	o&	>19.34&20170525&	57898.34	&c&	>20.01\\
20170405&	57848.35	&o&	>19.55&20170529&	57902.30	&c&	>20.33\\
20170407&	57850.39	&o&	>19.42&    $-$ &  $-$           &$-$&  $-$  \\
\hline
\hline
\end{tabular}
\end{table}
\begin{table}[htbp]
\caption{Observed magnitudes of SN 2016iog in the Sloan ugriz bands.}\label{2016ugriz}
\resizebox{\textwidth}{!}{ 
\centering                                      
\begin{tabular}{c c c c c c c c c}  
\hline \hline
 Date & MJD & Phase$^a$ & u & g & r & i & z & Instrument\\ 
\hline 
20161206 & 57728.095 & $+$12.5 & 16.579(0.013) & 16.486(0.021) & 16.672(0.019) & 16.720(0.016) & 16.734(0.019) & AFOSC\\ 
20161207 & 57729.200 & $+$13.6 & 16.749(0.038) & 16.543(0.032) & 16.531(0.027) & 16.661(0.027) & 16.658(0.014) & AFOSC\\
20161208 & 57730.045 & $+$14.4 & 16.830(0.023) & 16.584(0.027) & 16.563(0.052) & 16.706(0.033) & 16.714(0.024) & AFOSC\\
20161208 & 57730.995 & $+$15.4 & 16.959(0.038) & 16.667(0.025) & 16.583(0.072) & 16.756(0.073) & 16.798(0.018) & AFOSC\\
20161210 & 57732.120 & $+$16.5 & 17.155(0.043) & 16.709(0.034) & 16.585(0.018) & 16.787(0.068) & 16.840(0.016) & AFOSC\\
20161221 & 57743.105 & $+$27.5 & 18.395(0.065) & 17.534(0.025) & 17.254(0.016) & 17.412(0.061) & 17.240(0.021) & AFOSC\\
20161230 & 57752.155 & $+$36.6 & 19.620(0.064) & 18.369(0.031) & 17.854(0.030) & 17.926(0.035) & 17.668(0.019) & AFOSC\\
20170104 & 57758.017 & $+$42.4 & 20.098(0.124) & 19.123(0.058) & 18.242(0.032) & 18.461(0.052) & 18.258(0.041) & AFOSC\\
20170107 & 57760.197 & $+$44.6 & 20.419(0.109) & 19.609(0.036) & 18.542(0.034) & 18.949(0.053) & 18.726(0.040) & AFOSC\\
20170119 & 57773.000 & $+$57.4 & $-$ & 20.870(0.061) & 19.222(0.034) & 20.049(0.060) & $-$ & AFOSC\\
20170123 & 57776.990 & $+$61.4 & $-$ & 21.106(0.119) & 19.601(0.066) & 20.573(0.115) &  $-$& AFOSC\\
20170125 & 57778.110 & $+$62.5 &  $-$& 21.325(0.054) & 19.607(0.038) & 20.607(0.084) &  $-$& AFOSC\\
20170203 & 57787.140 & $+$71.5 & $-$ &  $-$& 20.148(0.032) &  $-$&  $-$& OSIRIS\\
20170317 & 57829.945 & $+$114.3 &  $-$& >22.1 & 21.409(0.166) & 21.589(0.230) & >20.3 & AFOSC\\
20170327 & 57839.965 & $+$124.4 & $-$ & >22.8 & 21.524(0.131) & 21.842(0.175) &  $-$& AFOSC\\
20170601 & 57905.930 & $+$190.3 &  $-$&  $-$& 22.306(0.178) & $-$ & $-$ & OSIRIS \\
\hline\hline

\multicolumn{9}{l}{{$^a$ Phases are calculated relative to the explosion epoch (MJD = 57715.6) in the reference frame of the observer.}} 
\\
\end{tabular}
}
\end{table}

\section{Acknowledgements}
We gratefully thank the anonymous referee for his/her insightful comments and suggestions that improved the paper.
We thank Luc Dessart for his invaluable contributions to the interpretation of  LCs, spectra, and model comparisons, as well as his insightful comments and revisions that significantly improved this work.
We thank  Subhash Bose, Giuliano Pignata and Raya Dastidar, Panos Charalampopoulos  for providing the spectra and measurement data of ASASSN-15nx, SN 2019nyk, SN 2022lxg, Ofer Yaron for supplying some of the comparison spectra of other SNe, and Eddie Baron for providing installation assistance for SYNAPPS. We thank Jujia Zhang, Zeyi Zhao, and Luhan Li for their helpful discussions. 

This work is supported by the National Natural Science Foundation of China (Nos. 12105032, 12288102, 12303054, 12225304, 12473031), the Natural Science Foundation of Chongqing (No. cstc2021jcyj-msxmX0481), the National Key Research and Development Program of China  (Grant No. 2021YFA1600404, 2024YFA1611603), the Yunnan Fundamental Research Projects (Grant Nos. 202401AU070063, 202501AS070078),the Yunnan Science and Technology Program (Nos. 202501AS070005, 202201BC070003), the International Centre of Supernovae, Yunnan Key Laboratory (No. 202302AN360001).
AP, AR, SB, EC, NER, PO, LT, GV acknowledge support from the PRIN-INAF 2022, "Shedding light on the nature of gap transients: from the observations to the models".
AR also acknowledges financial support from the GRAWITA Large Program Grant (PI P. D'Avanzo).
NER also acknowledges support from the Spanish Ministerio de Ciencia e Innovaci\'on (MCIN) and the Agencia Estatal de Investigaci\'on (AEI) 10.13039/501100011033 under the program Unidad de Excelencia Mar\'ia de Maeztu CEX2020-001058-M.
AF acknowledges funding by the European Union - NextGenerationEU RFF M4C2 1.1 PRIN 2022 project "2022RJLWHN URKA" and by INAF 2023 Theory Grant ObFu 1.05.23.06.06 "Understanding R-process \& Kilonovae Aspects (URKA)".
JI acknowledges funding by Spanish program Unidad de Excelencia Mar\'ia de Maeztu CEX2020-001058-M, PID2023-149918NB-I00, and 2021-SGR-1526 (Generalitat de Catalunya).
RR acknowledges support from Grant RYC2021-030837-I, funded by MCIN/AEI/ 10.13039/501100011033 and by "European Union NextGeneration EU/PRTR", along with partial support by the Spanish MINECO grant, PID2023-148661NB-I00.
S.-P. Pei is supported by the  Science and Technology Foundation of Guizhou Province (QKHJC-ZK[2023]442).
XFW is supported by the National Natural Science Foundation of China (NSFC grants 12288102, 12033003, and 11633002) and the Tencent Xplorer Prize.

Based on observations obtained with the Xinglong 2.16-m telescope (XLT), located at the Xinglong Observatory, China.
Based on observations obtained with the Cima Ekar 1.82 m Telescopio Copernico, installed at the INAF (Istituto Nazionale di Astrofisica) - Astronomical Observatory of Padova, Italy.
Based on observations obtained with the Gran Telescopio Canarias (GTC), installed in the Spanish Observatorio del Roque de los Muchachos of the Instituto de Astrofisica de Canarias, on the island of La Palma.
This work has made use of data from the Asteroid Terrestrial-impact Last Alert System (ATLAS) project. The Asteroid Terrestrial-impact Last Alert System (ATLAS) project is primarily funded to search for near earth asteroids through NASA grants NN12AR55G, 80NSSC18K0284, and 80NSSC18K1575; byproducts of the NEO search include images and catalogs from the survey area. This work was partially funded by Kepler/K2 grant J1944/80NSSC19K0112 and HST GO-15889, and STFC grants ST/T000198/1 and ST/S006109/1. The ATLAS science products have been made possible through the contributions of the University of Hawaii Institute for Astronomy, the Queen's University Belfast, the Space Telescope Science Institute, the South African Astronomical Observatory, and The Millennium Institute of Astrophysics (MAS), Chile.

We thank Las Cumbres Observatory and its staff for their continued support of ASAS-SN. ASAS-SN is funded in part by the Gordon and Betty Moore Foundation through grants GBMF5490 and GBMF10501 to the Ohio State University, and also funded in part by the Alfred P. Sloan Foundation grant G-2021-14192. Development of ASAS-SN has been supported by NSF grant AST-0908816, the Mt. Cuba Astronomical Foundation, the Center for Cosmology and AstroParticle Physics at the Ohio State University, the Chinese Academy of Sciences South America Center for Astronomy (CAS-SACA), and the Villum Foundation.
This research has made use of the NASA/IPAC Extragalactic Database (NED), which is operated by the Jet Propulsion Laboratory, California Institute of Technology, under contract with the National Aeronautics and Space Administration.
{\sc iraf} was distributed by the National Optical Astronomy Observatory, which was managed by the Association of Universities for Research in Astronomy (AURA), Inc., under a cooperative agreement with the U.S. NSF.

\end{appendix}

\end{document}